\documentclass[pdflatex,sn-mathphys-num]{sn-jnl}% Math and Physical Sciences Numbered Reference Style 
\usepackage{graphicx}%
\usepackage{multirow}%
\usepackage{amsmath,amssymb,amsfonts}%
\usepackage{amsthm}%
\usepackage{mathrsfs}%
\usepackage[title]{appendix}%
\usepackage{xcolor}%
\usepackage{textcomp}%
\usepackage{manyfoot}%
\usepackage{booktabs}%
\usepackage{algorithm}%
\usepackage{algorithmicx}%
\usepackage{algpseudocode}%
\usepackage{listings}%

\begin{document}

\title[Article Title]{Benchmarking of Geant4 simulations for the COSI Anticoincidence System}


\author*[1,2]{\fnm{Alex} \sur{Ciabattoni}}\email{alex.ciabattoni@inaf.it}

\author[2]{\fnm{Valentina} \sur{Fioretti}}

\author[3]{\fnm{John A.} \sur{Tomsick}}

\author[3]{\fnm{Andreas} \sur{Zoglauer}}

\author[4]{\fnm{Parshad} \sur{Patel}}

\author[5]{\fnm{Lee} \sur{Mitchell}}

\author[2]{\fnm{Andrea} \sur{Bulgarelli}}

\author[6]{\fnm{Pierre} \sur{Jean}}

\author[1,2]{\fnm{Gabriele} \sur{Panebianco}}

\author[2]{\fnm{Nicolò} \sur{Parmiggiani}}

\author[1,2]{\fnm{Cristian} \sur{Vignali}}

\author[6]{\fnm{Peter} \sur{von Ballmoos}}

\author[5]{\fnm{Eric} \sur{Wulf}}

\affil[1]{\orgdiv{Department of Physics and Astronomy ``Augusto Righi"}, \orgname{University of Bologna}, \orgaddress{\street{Via P. Gobetti 93/2}, \city{Bologna}, \postcode{40129}, \country{Italy}}}

\affil[2]{\orgdiv{INAF/OAS}, \orgaddress{\street{Via Gobetti 93/2}, \city{Bologna}, \postcode{40129}, \country{Italy}}}

\affil[3]{\orgdiv{Space Sciences Laboratory}, \orgname{University of California}, \orgaddress{\street{7 Gauss Way}, \city{Berkeley}, \postcode{94720}, \state{CA}, \country{USA}}}

\affil[4]{\orgdiv{George Mason University}, \orgaddress{\street{4400 University Dr}, \city{Fairfax}, \postcode{22030}, \state{VA}, \country{USA}}}

\affil[5]{\orgdiv{U.S. Naval Research Laboratory}, \orgaddress{\street{4555 Overlook Ave. SW}, \city{Washington}, \postcode{20375}, \state{DC}, \country{USA}}}

\affil[6]{\orgdiv{IRAP}, \orgname{University of Toulouse, CNRS, CNES, UPS}, \orgaddress{\street{9 avenue du colonel Roche}, \city{Toulouse}, \postcode{31028}, \country{France}}}


\abstract{The Compton Spectrometer and Imager (COSI) is an upcoming NASA Small Explorer satellite mission, designed for all-sky observations in the soft gamma-ray domain with the use of germanium detectors (GeDs). An active Anticoincidence System (ACS) of BGO scintillators surrounds the GeDs to reduce the background and contribute to the detection of transient events. Accurately modeling the ACS performance requires simulating the intricate scintillation processes within the shields, which significantly increases the computational cost. We have encoded these effects into a correction matrix derived from dedicated Geant4 simulations with the inclusion of the optical physics. For this purpose, we use laboratory measurements for the energy and spatial response of the ACS lateral wall to benchmark the simulation and define instrument parameters, including the BGO absorption length and the electronic noise. We demonstrate that the simulations replicate the experimental energy resolution and light collection uniformity along the BGO crystal, with maximum discrepancies of 20\% and 10\%, respectively. The validated simulations are then used to develop the correction matrix for the lateral wall, accounting for the light collection efficiency and energy resolution based on the position within the crystal. The gamma-ray quantum detection efficiency is also position-dependent via the inclusion of the optical physics. It is enhanced by $\sim$8\% close to the SiPMs and suppressed by $\sim$2\% in the adjacent corners with respect to the average value. Finally, we explore the energy threshold and resolution of the bottom ACS, considering the impact of its smaller crystals compared with the lateral walls.}




\maketitle

\section{Introduction} \label{sec:intro}
\begin{figure}
\centering
\includegraphics[width=0.45\textwidth]{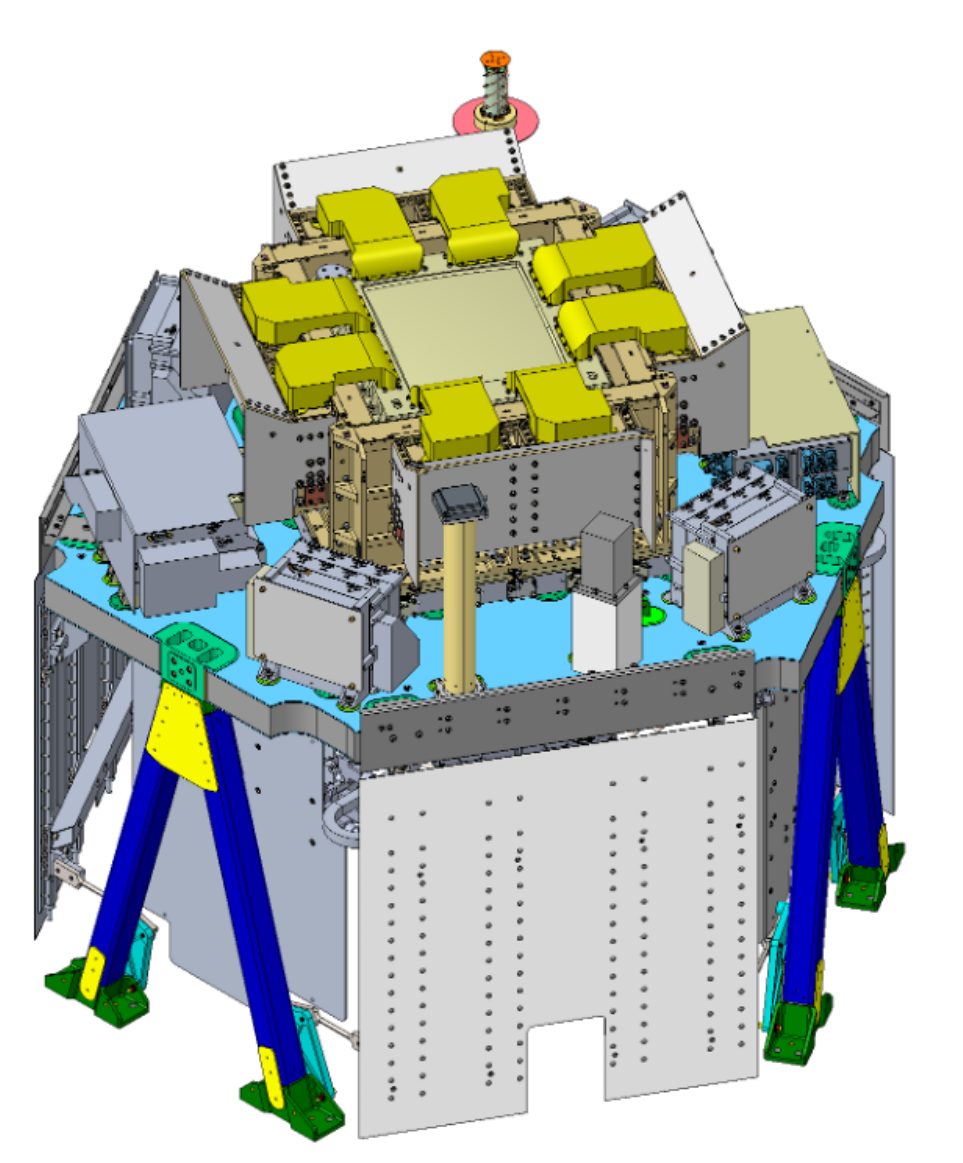}\includegraphics[width=0.45\textwidth]{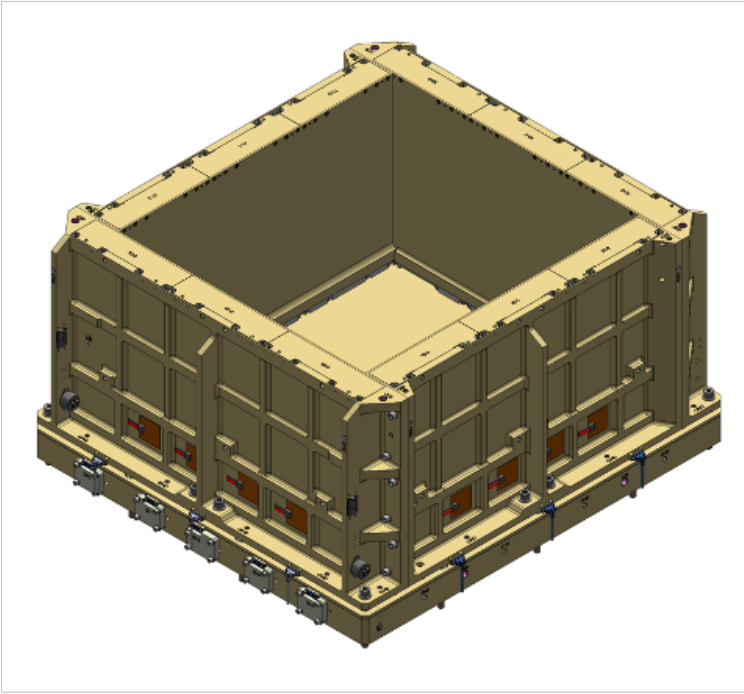}
\caption{COSI payload design (left) and ACS design (right)}\label{fig:payload}
\end{figure}

The Compton Spectrometer and Imager (COSI) \citep{2024icrc.confE.745T} is an approved NASA Small-Explorer (SMEX) satellite mission scheduled for launch in 2027. It is a wide-field gamma-ray telescope aimed to survey the entire sky at 0.2-5 MeV, and will operate on a low-Earth near-equatorial orbit. Its main science goals include mapping the 511 keV positron emission in our Galaxy, detecting diffuse emission from nuclear lines like $^{26}$Al and $^{60}$Fe, gaining insights into extreme environments through polarization measurements of compact objects, and contributing to multi-messenger astrophysics by detecting gamma-ray counterparts of gravitational wave sources. The payload (Fig. \ref{fig:payload}) consists of semiconductor germanium detectors (GeDs), that exploit Compton scatterings of the incoming photon to retrieve its properties, surrounded on four lateral sides and underneath by Anticoincidence System (ACS) shields, defining an instantaneous field of view (FoV) for the GeDs of $\sim$25\% of the sky. The ACS is composed of bismuth germanate (BGO) panels that emit scintillation light in the optical range from interaction with ionizing radiation. The scintillation light is read out by silicon photomultipliers (SiPMs), coupled with each BGO panel, with a required energy threshold of $\sim$80 keV. The ACS offers both passive attenuation and active vetoing of the background events in the GeDs, improving the overall sensitivity of the mission. A fundamental function of the ACS is to serve as a monitor for gamma-ray transient events, such as Gamma-Ray Bursts (GRBs). By monitoring the ACS count rate and analyzing the relative rates among different panels, it is possible not only to detect GRBs but also to constrain their localization. The use of systems with multiple scintillators for transient detection has been successfully implemented in past and present space missions, including INTEGRAL-SPI \citep{2003A&A...411L.299V}, AGILE \citep{PEROTTI2006228}, Fermi-GBM \citep{2015ApJS..216...32C}, and BurstCube \citep{2019ICRC...36..604S}, making significant contributions to transient science and multi-messenger astronomy.

Calibration campaigns are essential for accurately assessing the detector performance. Laboratory measurements of the GeDs have been conducted to investigate the complexities of the Compton measurement process. These measurements enable the conversion of detector parameters into physical quantities, such as energy and position, which are critical for effective Compton reconstruction \citep{2022NIMPA103166510B}. Calibration measurements are also crucial for simulation benchmarking. Validating simulations against laboratory data ensures that key detector characteristics, such as energy resolution and threshold, are realistically modeled. Standard COSI simulations are performed using the Cosima software, which is built on the Geant4 toolkit \citep{AGOSTINELLI2003250, 1610988, ALLISON2016186} and the MEGAlib framework \citep{2006NewAR..50..629Z}. The implementation of realistic effects into simulated data will be handled by the software Nuclearizer through the so-called Detector Effects Engine (DEE). The DEE is constructed and tuned using calibration data. A dedicated DEE for the GeDs has been developed to include effects like cross-talk, strip pairing, energy and depth calibrations, and to ensure that the simulated data closely replicates real measurements \citep{2019NIMPA.94662643S}. Currently, MEGAlib ACS simulations do not account for the complex scintillation processes occurring within the BGO shields. While incorporating the generation and tracking of optical photons would enhance the accuracy of the simulated response, it would also significantly increase the computational time in standard COSI simulations, as thousands of optical photons are generated for each keV of deposited energy. To address this, we built a dedicated Geant4 simulation of the calibration experimental apparatus and the optical physics to model the scintillation light transmission and signal read-out. Once benchmarked against laboratory measurements, we use the simulation of the ACS lateral wall to encode the energy dispersion as a function of deposited energy and BGO interaction position into a correction matrix. This correction will be applied to ACS MEGAlib simulations using the DEE module in Nuclearizer, to obtain a realistic model of the ACS both as vetoing and stand-alone detector.

In this work, we present the calibrations conducted for the ACS lateral wall, the simulation benchmarking, and the development of the correction matrix. We also show an application of the benchmarked simulation to the bottom ACS to assert the impact of the different crystal sizes on its performance. The paper is organized as follows. In Sec. \ref{sec:cosi_acs} we provide an overview of the COSI ACS design and role. In Sec. \ref{sec:calibration_and_benchmarking} we present the calibration campaigns conducted for a lateral ACS wall, the Geant4 simulation of the experimental setups and the results on the benchmarking. In Sec. \ref{sec:dee} we show the development of the correction matrix for the lateral ACS. Finally, in Sec. \ref{sec:bottom} we present the application of the benchmarked simulations to the bottom ACS.

\section{The COSI ACS} \label{sec:cosi_acs}
The COSI ACS comprises a total of 22 BGO crystals, 3 on each lateral side and 10 underneath the GeDs. Photons entering the instrument from outside the GeD field of view are effectively detected by the BGO shields, due to their high stopping power. In addition, the ACS is used to reject photons that are not fully absorbed in the GeDs and are therefore difficult to reconstruct: such photons typically interact with both the GeDs and the ACS, triggering a veto signal that causes the event to be discarded. The ACS also plays a key role in the detection and localization of GRBs and other transients outside the FoV of the GeDs. By analyzing the relative count rates among different panels, it is possible to constrain the localization of GRBs \citep{Parmiggiani_prep}, allowing for quick follow-up observations by other telescopes. During the flight, the COSI ACS data will be collected into 6 different channels. Each of them will be divided into two channels, below and above 2 MeV, for a total of 12 channels of BGO data. When an excess in the signal-to-noise ratio is detected in at least two channels, a trigger is raised and the data is downlinked via the Tracking and Data Relay Satellite System (TDRSS). The downlinked ACS data passes through the fast transient pipeline to get a first source type identification, significance and localization of the detected source, and a GCN notice is generated within 60 minutes. Subsequently, a comprehensive transient pipeline performs a reanalysis of the transient with all data downlinked and generates a GCN within a few hours from the trigger. 

The BGO emits scintillation light in the 375-650 nm band, peaking at 480 nm\footnote{\href{https://gammadata.se/wp-content/uploads/2024/01/BGO-data-sheet-v2.pdf}{https://gammadata.se/wp-content/uploads/2024/01/BGO-data-sheet-v2.pdf}}. Each BGO crystal is coupled with a $3\times 3$ array of SiPMs. The SiPMs are single-photon-sensitive devices aimed at collecting the optical light generated inside the BGO and the consequent generation of a proportional electric signal. Compared with other readout devices, the SiPMs offer substantial reductions in mass, volume, power consumption, and cost. They are composed of thousands of single-photon avalanche diodes (SPADs) operating in Geiger mode, that is when the applied voltage is above the breakdown voltage. In this regime, when a SPAD is hit by a photon, a large electric signal is generated from internal avalanche multiplication, which depends on the gain of the SiPM. The produced avalanche is eventually quenched and, after a characteristic recovery time, the SPAD is restored and ready for the detection of another photon. Each SPAD detects photons independently, and the final output is given by the sum of the photocurrents from each SPAD. Another key quantity related to the SiPM performance is the photon detection efficiency (PDE), that is the probability of the SiPM to produce a signal in response to an incident photon. It is a function of the overvoltage (defined as the difference between the applied voltage and the breakdown voltage), and the wavelength of the incident light. It includes the geometrical fill factor (i.e. the fraction of the total area occupied by the microcells), the quantum efficiency (i.e. the probability for an incident photon to produce an electron/hole pair) and the probability of Geiger discharge (i.e. the probability for the generated electron/hole to trigger the avalanche process). To maximize the light collection of the SiPMs, the BGO is wrapped with reflective layers in order to keep the scintillation photons inside the crystal.

\section{ACS calibration and simulation benchmarking} \label{sec:calibration_and_benchmarking}
\subsection{Calibration measurements} \label{sec:calibrations}
\begin{figure}
\centering
\includegraphics[width=\textwidth]{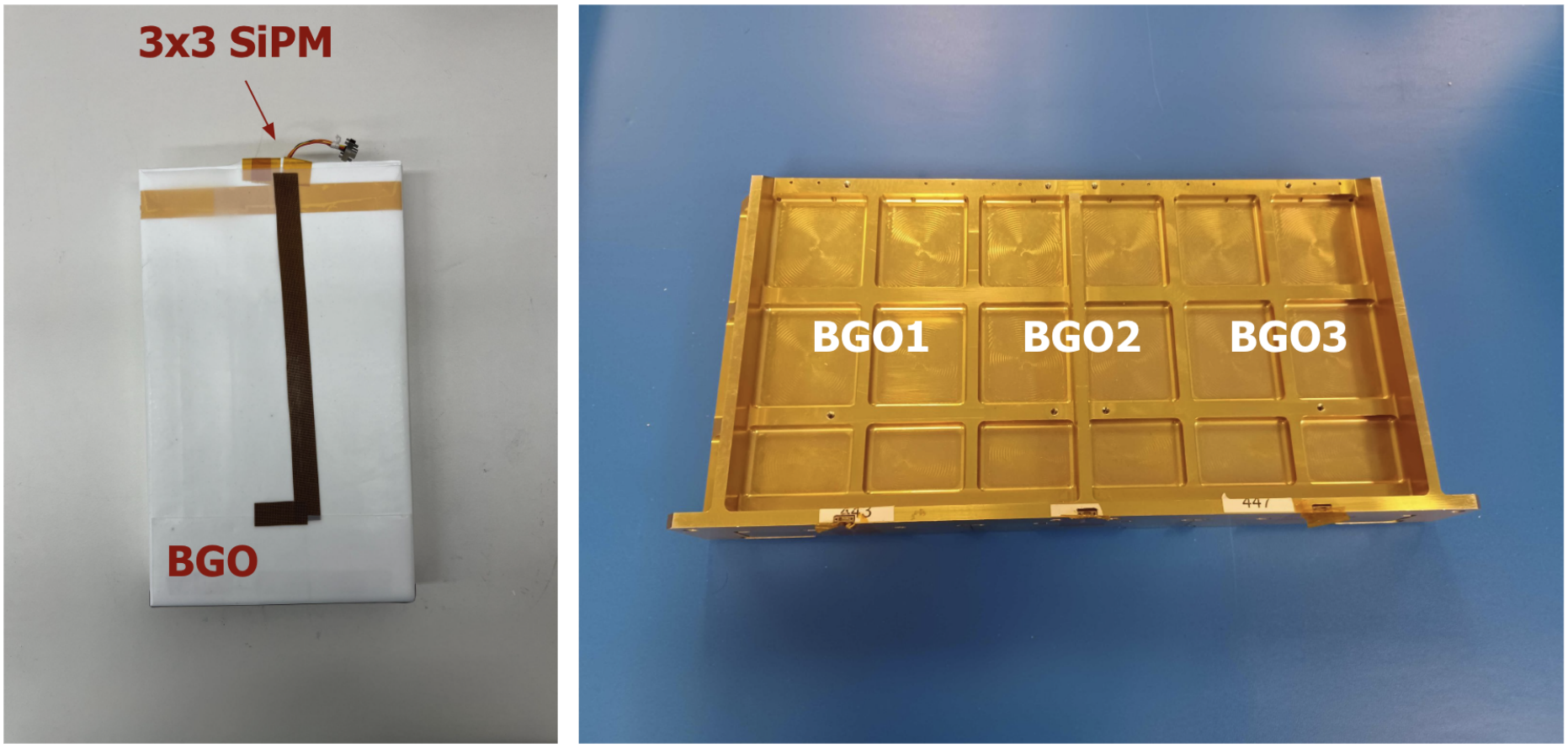}
\caption{Wrapped BGO crystal with coupled SiPM array on the left (picture taken at NRL) and the EM X-wall with the three BGO crystals on the right (picture taken at SSL)}\label{fig:exp_pictures}
\end{figure}

We use two calibration campaigns for the COSI ACS, one conducted at the Naval Research Laboratory (NRL, Washington DC) and one at the Space Sciences Laboratory (SSL, Berkeley). The measurements involve the Engineering Model (EM) X-wall, i.e. a lateral ACS panel consisting of 3 BGO crystals and read-out each by SiPM arrays (Fig. \ref{fig:exp_pictures}). Each BGO crystal consists of a $198\times 118 \times 23$ mm scintillator block, and a $3\times 3$ array of SiPMs (onsemi J-series 60035) is attached to the $118\times 23$ mm face with the use of a silicone optical coupler. The BGO is wrapped with two layers of VM2000, one layer of Tetratex and one layer of Teflon PTFE. The face where the SiPMs are attached is instead covered only with two layers of Tetratex. These reflective layers prevent scintillation light from escaping the BGO and maximize the SiPM response. The measurements at NRL are taken at room temperature using three radioactive sources: $^{241}$Am, $^{22}$Na and $^{137}$Cs. The $^{241}$Am and $^{137}$Cs sources are measured with and without collimation, while $^{22}$Na is used only without collimation. At SSL, we use a larger number of radioactive sources spanning a broader range of energy, and only measurements without collimation are taken. The full list of radionuclides used, along with their line energies and the corresponding calibration campaigns, is shown in Table \ref{tab:sources}.
The dataset of each calibration campaign consists of a measured ADC channel spectrum for each source and configuration (collimated or uncollimated), and one or more background spectra. The ADC channels are proportional to the charge/voltage collected by the SiPMs coupled with the BGO and hence to the energy deposit in the crystal. Each spectrum exhibits one or more photopeaks where the photons are fully absorbed in the BGO, and a continuum tail at lower energy which arises from the Compton scattering of the source photons within the crystal or the surrounding environment. We present here the measurements for the central BGO crystal.

\subsubsection{NRL campaign} \label{sec:x-wall_NRL}
\begin{figure}
\centering
\includegraphics[width=0.5\textwidth]{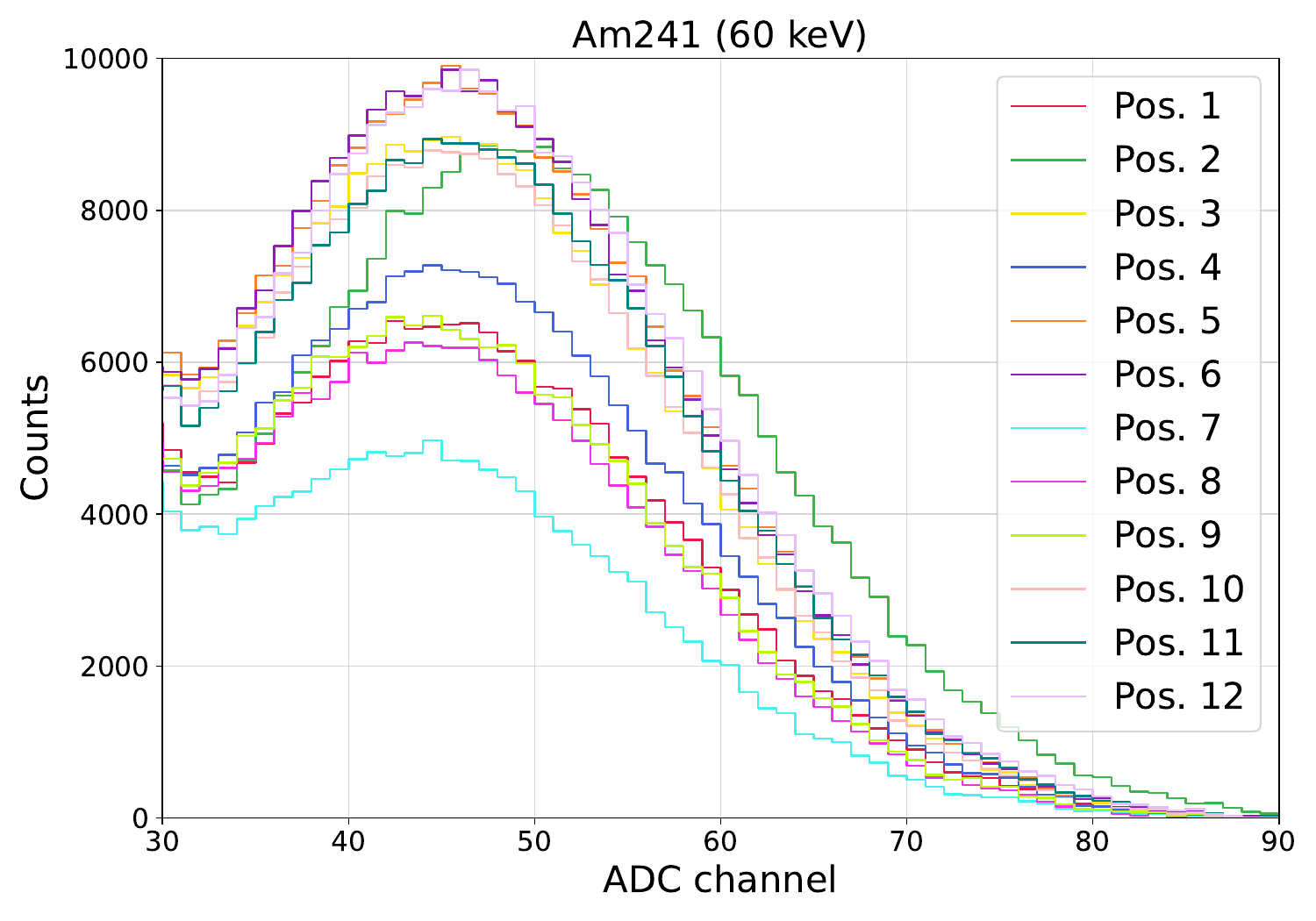}\includegraphics[width=0.5\textwidth]{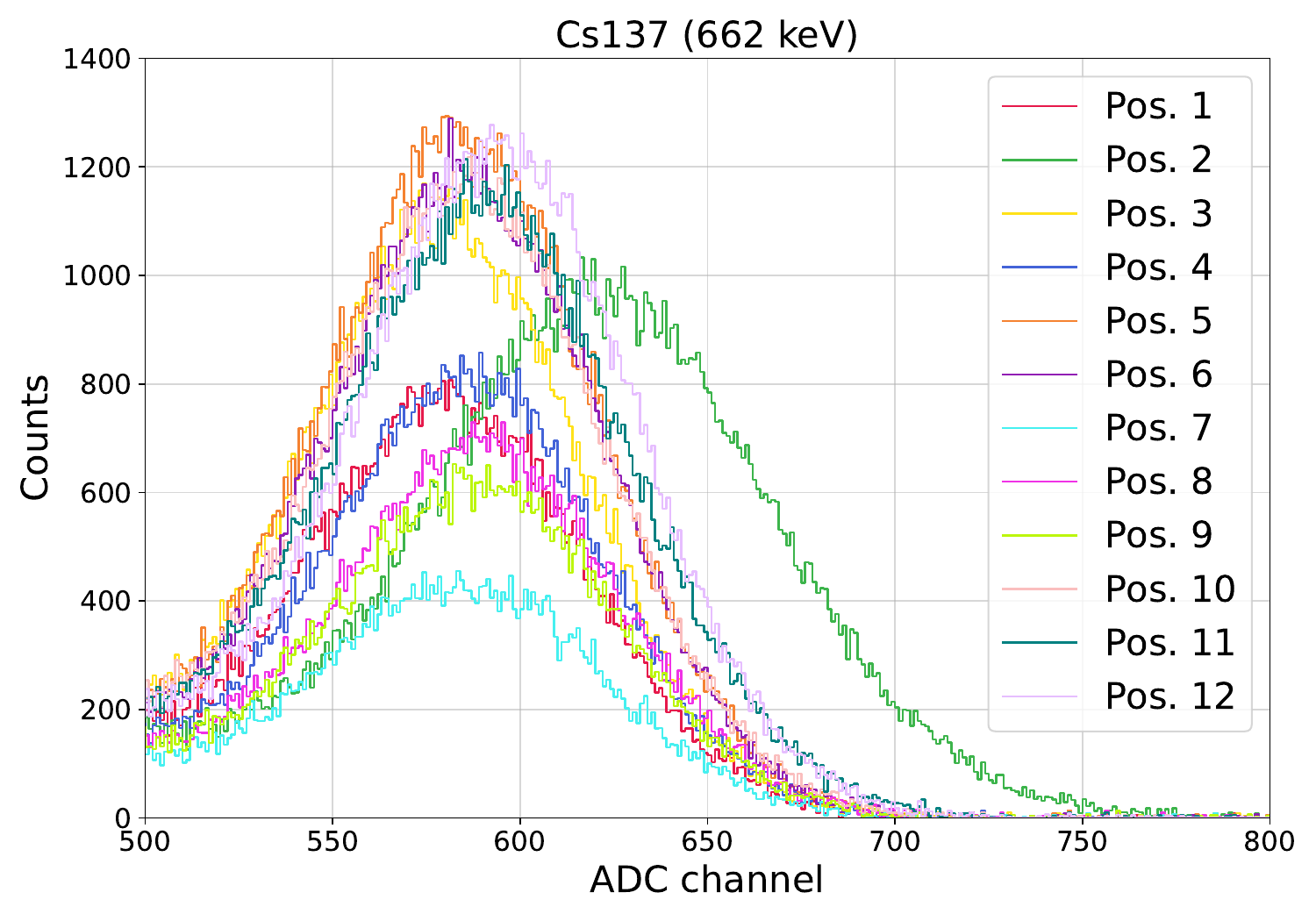}
\caption{Background-subtracted spectra for $^{241}$Am (left) and $^{137}$Cs (right) for each collimated position (see the positions in Fig. \ref{fig:rel_resp})}\label{fig:spectra_pos_NRL2}
\end{figure}

The measurements at NRL are taken with a multi-channel analyzer and an amplifier. The X-wall is placed on a metal table and covered with a black cloth to prevent light leakage. The uncollimated spectra are taken with $^{241}$Am and $^{137}$Cs placed on top of a foam block that sits on an aluminum frame, with the sources at $\sim$32 cm above the central BGO crystal. The $^{22}$Na source is instead positioned directly on top of the aluminum housing and centered on the crystal. The collimated measurements are performed using a lead cylinder with a 16 mm diameter hole and a 3D printed aligner in order to place the collimator at 12 different positions to scan the BGO surface. The background spectrum is taken only once for both collimated and uncollimated measurements and subsequently subtracted from the source spectra (Fig. \ref{fig:spectra_pos_NRL2}).

\subsubsection{SSL campaign} \label{sec:x-wall_SSL}
At SSL, we conducted additional calibration measurements for the same EM X-wall, but using a larger number of radioactive sources which are listed in Table \ref{tab:sources}. With respect to the NRL experiment, we have the same lowest energy line (60 keV from $^{241}$Am) which is just below the required threshold for the ACS (80 keV), but we extended the measurements up to 1836 keV from the $^{88}$Y emission. The spectra are acquired using a Picoscope.

\begin{figure}
\centering
\includegraphics[width=0.5\textwidth]{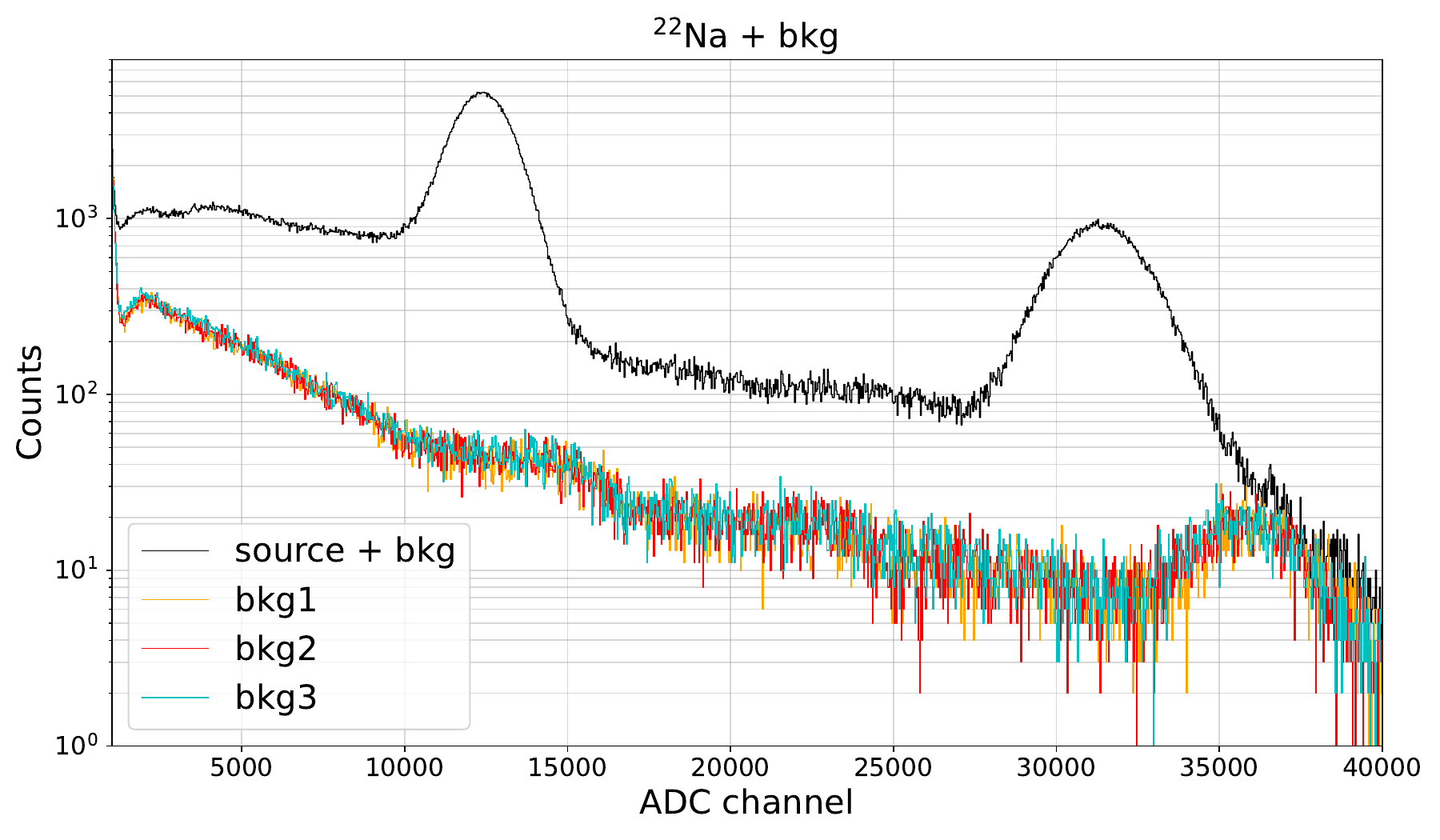}\includegraphics[width=0.5\textwidth]{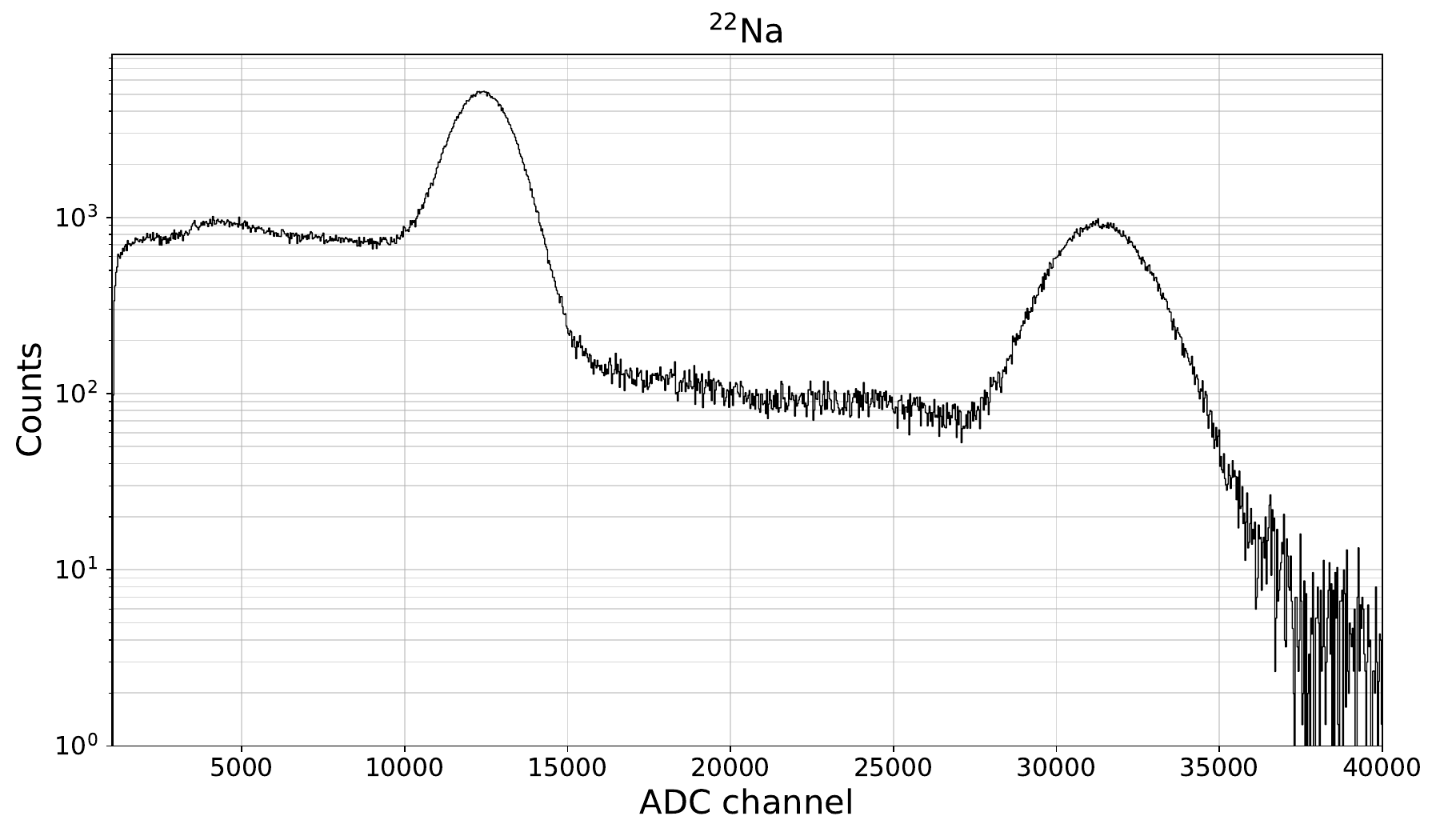}
\caption{Source + background spectra (on the left) and background-subtracted spectra (on the right) for $^{22}$Na, measured at SSL. In the caption of the left plot, `bkg1', `bkg2' and `bkg3' refer to the background measured before, in between and after the data acquisition with the sources}\label{fig:spectra_SSL}
\end{figure}

The X-wall is placed on a wooden table and anti-static mat and covered with a black cloth. A stand is used to hold the sources at $\sim$60 cm from the central BGO crystal, without collimation. We measure the background spectra three times, i.e. before, in between, and after the measurements with the sources, in order to check for possible time variations. We subtract from the source spectra the average of the three background spectra, and include their difference as systematic uncertainty in the estimated error. An example of the source spectrum, along with the background, and a background-subtracted spectrum is shown in Fig. \ref{fig:spectra_SSL}, for the $^{22}$Na. In App. \ref{secA1} we show all the spectra measured at NRL and SSL.

\begin{table}
\caption{Radioactive sources used for COSI ACS calibration measurements, including their most relevant line energies (neglecting lines below 60 keV), source activity, the campaigns in which they were used and the configuration (uncollimated or collimated)}\label{tab:sources}%
\begin{tabular*}{\textwidth}{@{\extracolsep\fill}lllll@{}}
\toprule
\textbf{Radionuclide} & \textbf{Line energies [keV]} & \textbf{Activity [$\mu$Ci]}\footnotemark[1] & \textbf{NRL} & \textbf{SSL}\\
\midrule
$^{241}$Am    & 60 & 98.03 (103) & Yes (uncoll. \& coll.)  & Yes (uncoll.)  \\
$^{109}$Cd    & 88 & 1000.0 & No & Yes (uncoll.)  \\
$^{57}$Co    & 122, 136 & 7.65 & No  & Yes (uncoll.)  \\
$^{133}$Ba    &  81, 276, 302, 356, 383 & 0.05289 & No & Yes (uncoll.)  \\
$^{22}$Na    & 511, 1275 & 11.83 (0.7) & Yes (uncoll.) & Yes (uncoll.)  \\
$^{137}$Cs    & 662 & 75.19 (9) & Yes (uncoll. \& coll.)  & Yes (uncoll.)  \\
$^{60}$Co    & 1173, 1333 & 55.69 & No  & Yes (uncoll.)  \\
$^{88}$Y    & 898, 1836 & 52.7 & No & Yes (uncoll.)  \\
\botrule
\end{tabular*}
\footnotetext[1]{The source activities reported here are based on the most recent measurements available at the time of the experiments. Values are given for the eight SSL sources, and the values in parentheses are for the three NRL sources.}
\end{table}

\subsubsection{Energy calibration} \label{sec:ene_cal_exp}
\begin{figure}
\centering
\includegraphics[width=0.5\textwidth]{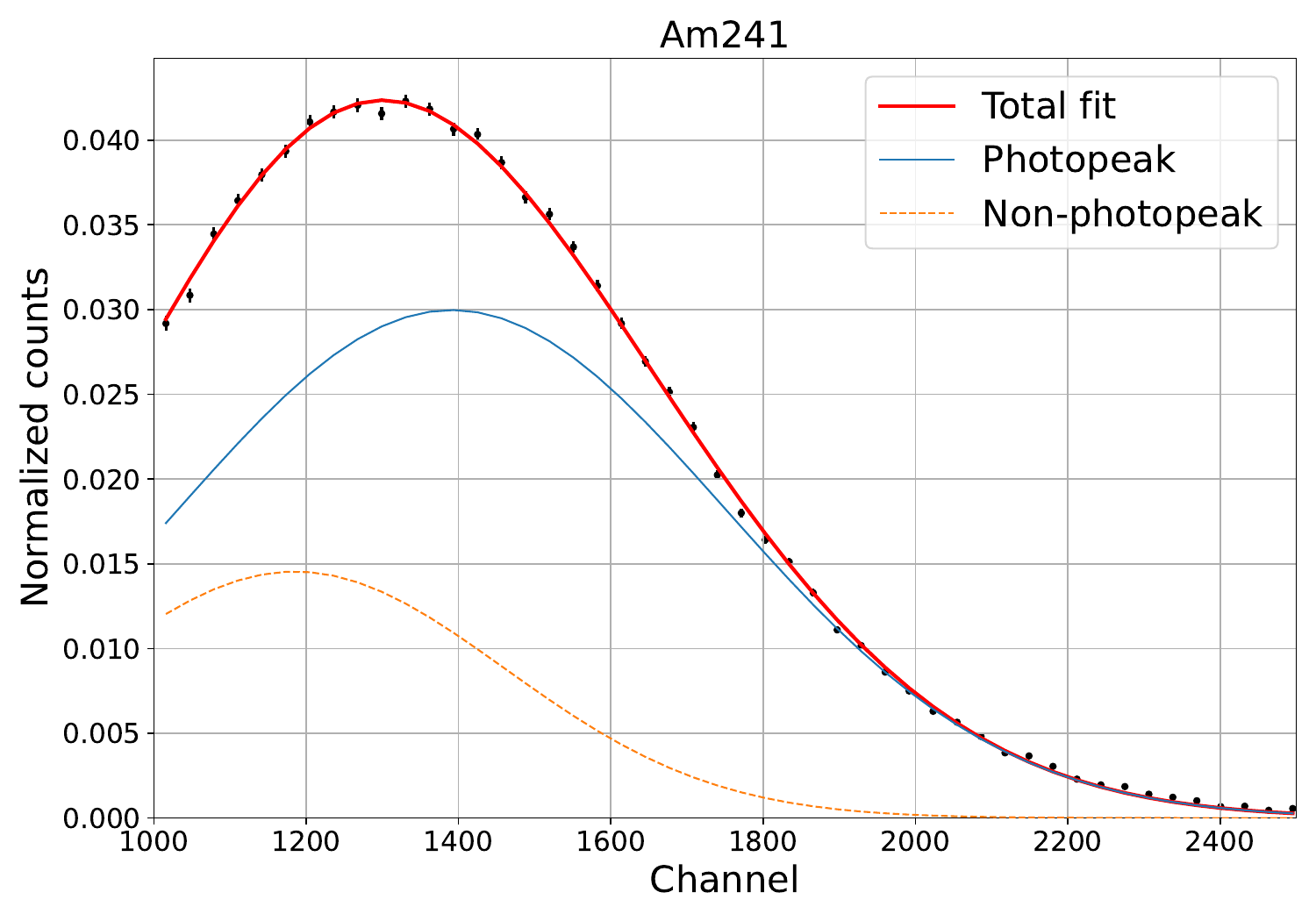}\includegraphics[width=0.5\textwidth]{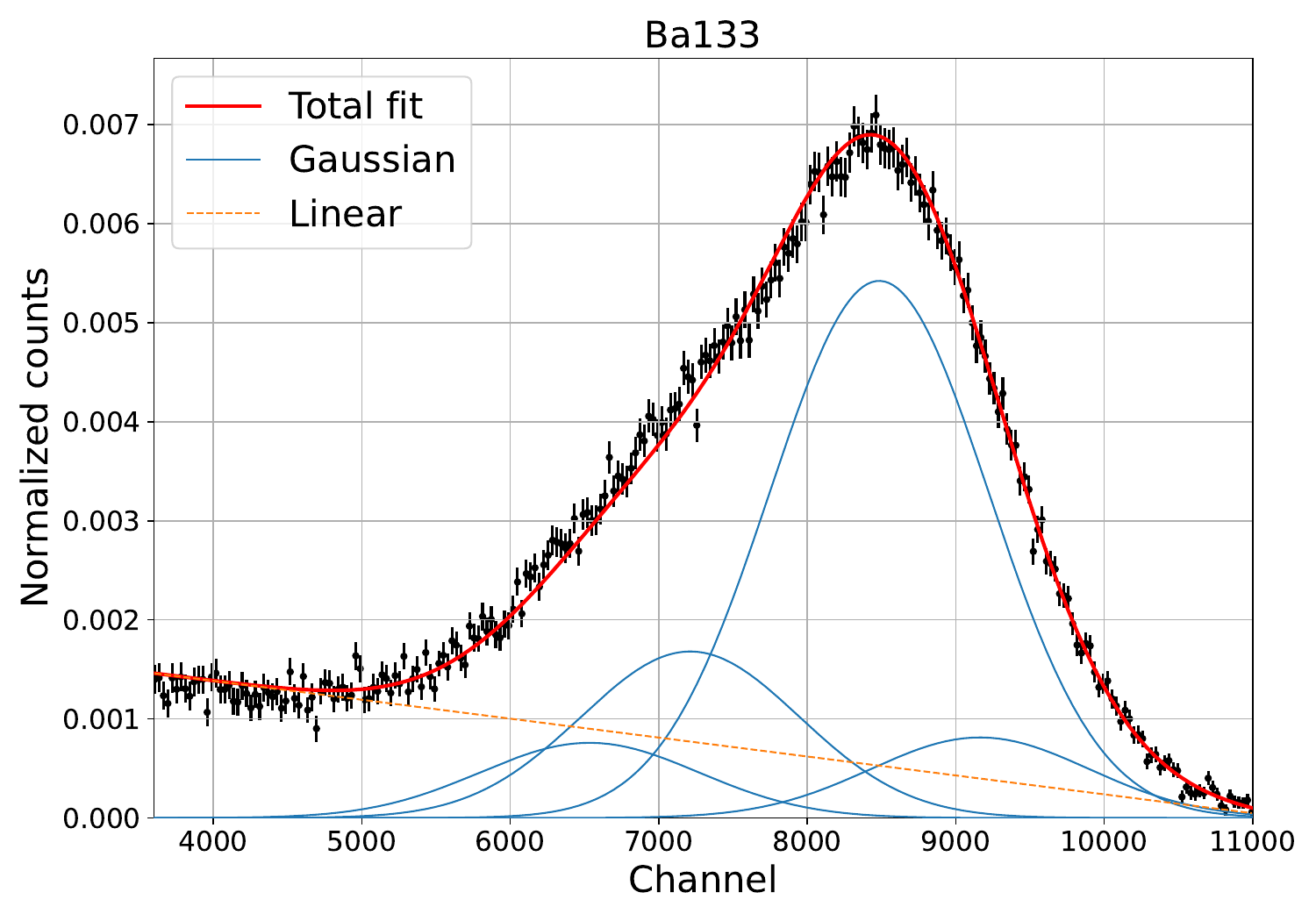}
\caption{Fit of the photopeaks for $^{241}$Am (left) and $^{133}$Ba (right) measured at SSL. For $^{241}$Am we use an extra Gaussian for the non-photopeak contribution, which is not fully resolved and therefore contaminates the photopeak. $^{133}$Ba has 4 overlapping photopeaks (276, 302, 356, 383 keV) and we consider the dominant one at 356 keV for the energy calibration. The linear model takes into account the continuum component of the spectra}\label{fig:fit_Am241_Ba133}
\end{figure}

\begin{figure}
\centering
\includegraphics[width=0.5\textwidth]{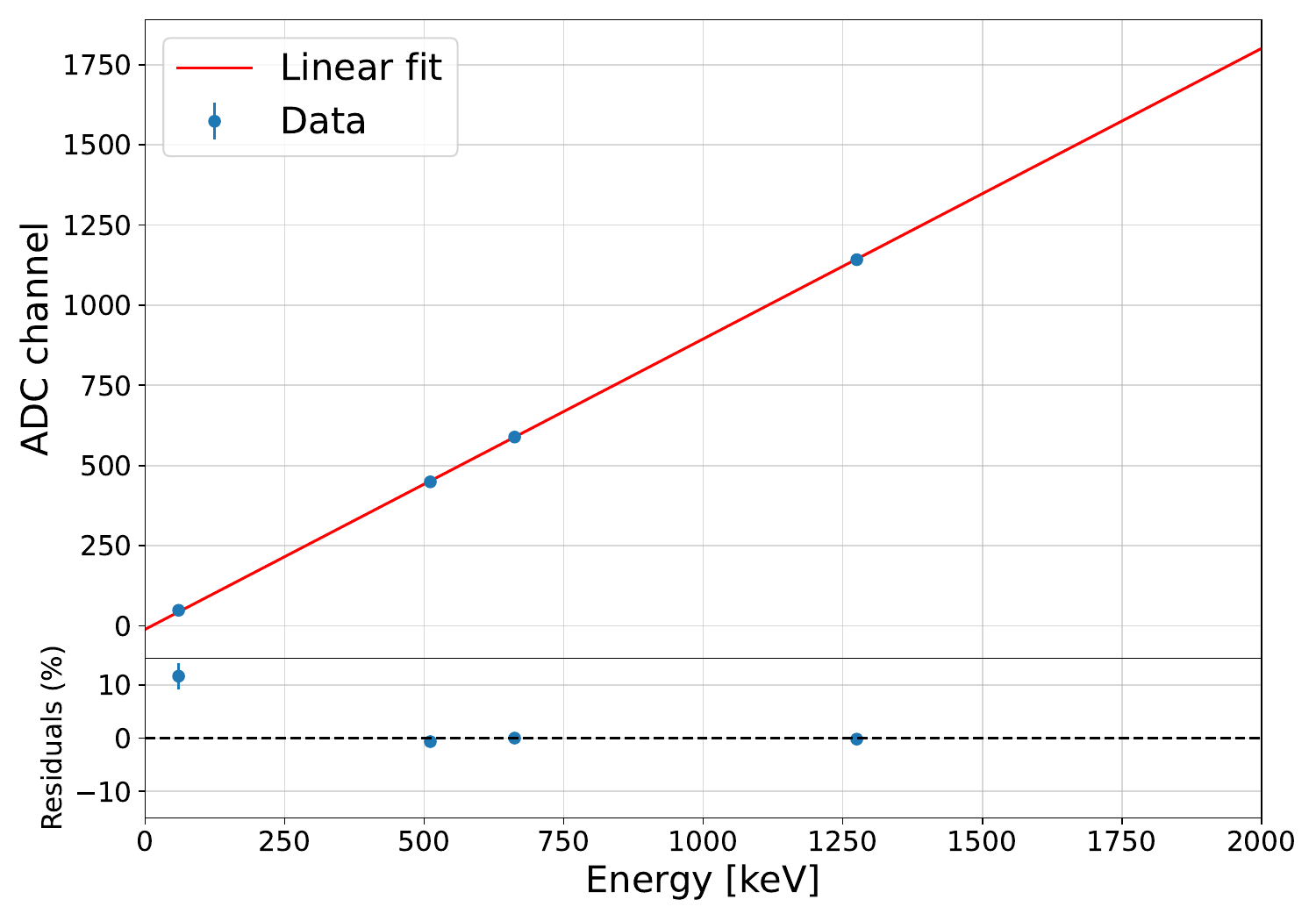}\includegraphics[width=0.5\textwidth]{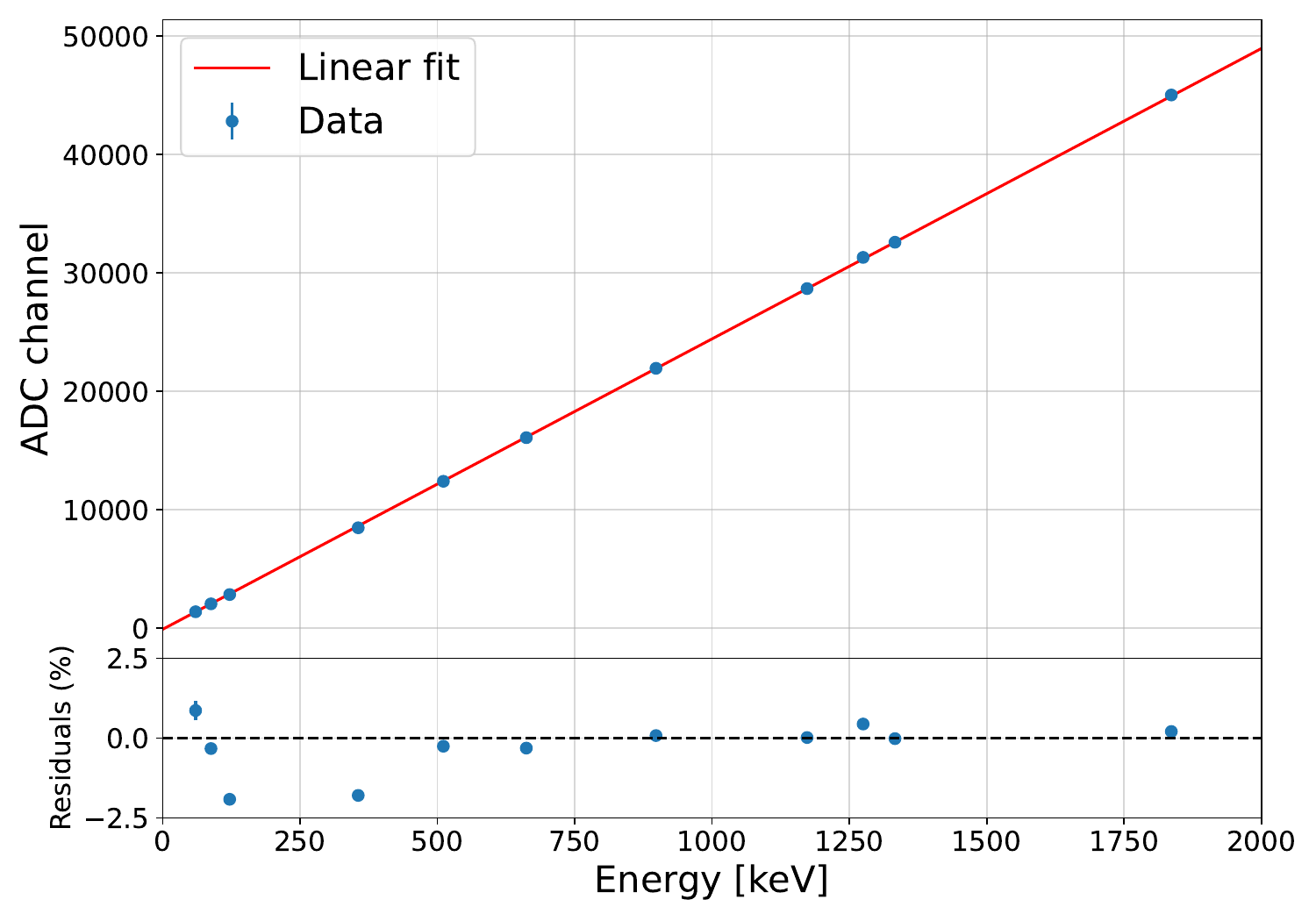}
\caption{Channel-energy relation for the NRL (left) and SSL (right) datasets and linear fit. The error bars are not visible because they are smaller than the size of the data points}\label{fig:channel-energy}
\end{figure}

We perform an energy calibration to convert the ADC channels into energies and obtain for each source the corresponding energy spectrum. The conversion from channel to energy is achieved as follows. We fit the photopeak with a Gaussian for each source and get the centroid. To account for the non-photopeak component, we use a linear model. For low-energy sources (like $^{241}$Am, $^{109}$Cd and $^{57}$Co), where the Compton tail has a larger overlap with the photopeak, instead of a linear model we found a Gaussian better describes the non-photopeak component. For $^{133}$Ba and $^{60}$Co, we use multiple Gaussian models to account for their partially overlapping photopeaks. In Fig. \ref{fig:fit_Am241_Ba133} we show the cases (from the SSL dataset) of $^{241}$Am, for which we use two Gaussian functions, and $^{133}$Ba, for which we use four Gaussian distributions to model the 276, 302, 356 and 383 keV overlapping photopeaks, plus a linear model to account for the continuum. Once we performed the fitting for each source, we associate to each centroid obtained from the fit the corresponding line energy of the source. Finally, we fit with a linear model ($C = m\cdot E+q$, with $C$ being the channel and $E$ the energy) the centroids and the corresponding energies to find the channel-energy relation, which is shown in Fig. \ref{fig:channel-energy} for both NRL and SSL datasets. The best-fit parameters are: $m = 0.904\pm 0.001$, $q=-10.1\pm0.8$ for NRL, $m = 24.524\pm 0.002$, $q=-86.2\pm1.3$ for SSL.

\subsubsection{Experimental energy resolution} \label{sec:ene_rel_exp}

From the Gaussian fitting of the photopeaks we obtain, for each source, the full width at half maximum (FWHM). The energy resolution $R$ at a particular energy $E$ is defined as
\begin{equation}
    R(E) = \frac{\text{FWHM}}{E}\,\,,\label{eq:resolution}
\end{equation}

In Table \ref{tab:resolutions} we list the energy resolution for each measured emission line and both calibration campaigns. We measure an energy resolution of approximately 55-60\% at 60 keV, $\sim$16\% at 511 keV and $\sim$8.5\% at 1836 keV. Besides the different readout system, the NRL and SSL resolutions are consistent within the 2$\sigma$ confidence level.

\begin{table}
\caption{Energy resolution ($R$) for each emission line}\label{tab:resolutions}%
\begin{tabular}{@{}lll@{}}
\toprule
\textbf{Emission line (keV)} & $\mathbf{R}$ \textbf{(\%) at NRL} & $\mathbf{R}$ \textbf{(\%) at SSL}\\
\midrule
60   & $54.5\pm 3.4$ & $60.1\pm1.8$  \\
88   & / & $44.9\pm 1.6$  \\
122   & / & $37.7\pm 1.1$  \\
356   & / & $19.7\pm 0.4$  \\
511   & $15.8\pm 0.3$ & $16.3\pm 0.1$
\\
662   & $14.0\pm 0.1$ & $14.3\pm 0.1$
\\
898   & / & $12.3\pm 0.04$  \\
1173   & / & $10.8\pm 0.04$  \\
1275   & $10.3\pm 0.3$ & $10.0\pm 0.1$  \\
1333   & / & $10.1\pm 0.1$  \\
1836   & / & $8.5\pm 0.02$  \\
\botrule
\end{tabular}
\end{table}

\begin{figure}
\centering
\includegraphics[width=0.7\textwidth]{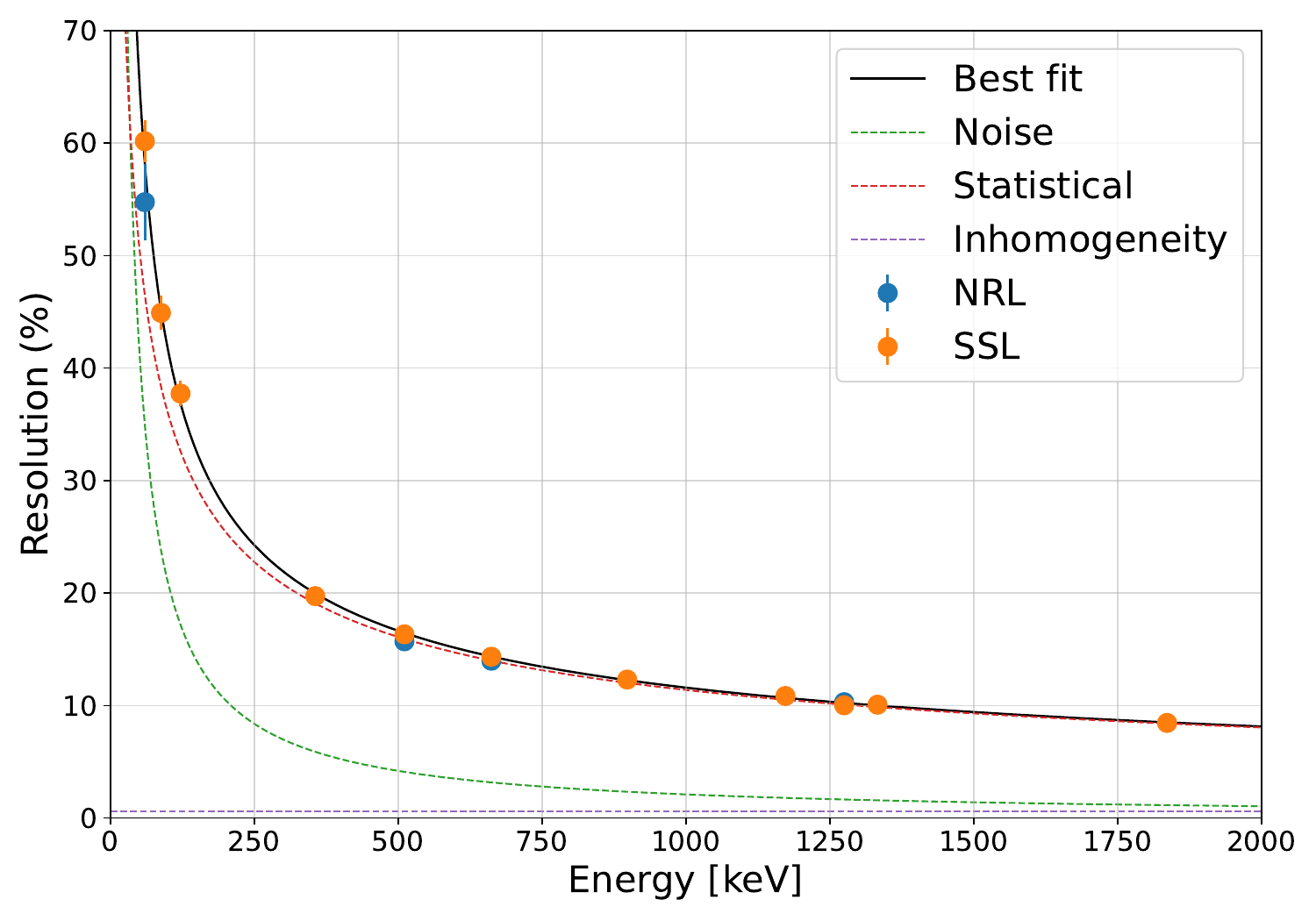}
\caption{Experimental BGO energy resolution as a function of the energy. The best-fit model (reported in Eq. \eqref{eq:fwhm}) along with the individual contributions to the resolution are also shown. The best-fit parameters are: $a=20.98\pm1.19$, $b=3.60\pm0.01$, $c=0.007\pm0.004$, with a reduced $\chi_\text{red}^2$ = 3.2}\label{fig:exp_resolution}
\end{figure}

The dependence of the resolution on the energy can be estimated by simple considerations about the photoelectron statistics. In fact, the energy corresponding to the peak pulse is proportional to the number of photoelectrons producing the signal. If the average number of photoelectrons is $N$, their statistical fluctuations would go as $\sim \sqrt{N}$. If we assume that the width of the photopeak is only due to this statistical uncertainty, we have $R = \text{FWHM}/E \propto 2.355\,\sqrt{N} / N = 2.355\,N^{-1/2}$. However, in a real detector there are other effects that can change the energy resolution, like the non-uniform optical light collection efficiency across the detector and the electronic noise of the SiPMs. To take into account all these effects, we follow the parametrization given in \citep{2009ExA....24...47B}:
\begin{equation}
    R=\frac{\text{FWHM}}{E} = \frac{\sqrt{a^2 + b^2\,E + c^2\,E^2}}{E}\,\,.
    \label{eq:fwhm}
\end{equation}
In the above equation, the parameter $a$ describes the limiting electronic resolution. The parameter $b$ accounts for the statistical nature of the light production, attenuation, photon-electron conversion, and electron amplification. The parameter $c$ represents the position dependence of the light transmission from the scintillator to the SiPM. The evaluated energy resolution for each emission line and calibration campaign, along with the best-fit model following Eq. \eqref{eq:fwhm}, is shown in Fig. \ref{fig:exp_resolution}. We also underline in the same plot the individual components contributing to the resolution. The best-fit parameters are: $a=20.98\pm1.19$, $b=3.60\pm0.01$, $c=0.007\pm0.004$, with a reduced $\chi_\text{red}^2 = \chi^2/dof=38.4/12=3.2$. The energy resolution is primarily determined by stochastic fluctuations in the generation and detection of optical photons across all energy levels. However, at lower energies, the electronic noise increases significantly and becomes comparable to the statistical fluctuations, thereby contributing notably to the overall resolution. The inhomogeneity component, which acts as an energy-independent factor, has a negligible contribution to the overall energy resolution.

\subsubsection{BGO response spatial distribution} \label{sec:rel_response}
\begin{figure}
\centering
\begin{minipage}[c]{0.45\textwidth}
    \centering
    \includegraphics[width=\textwidth]{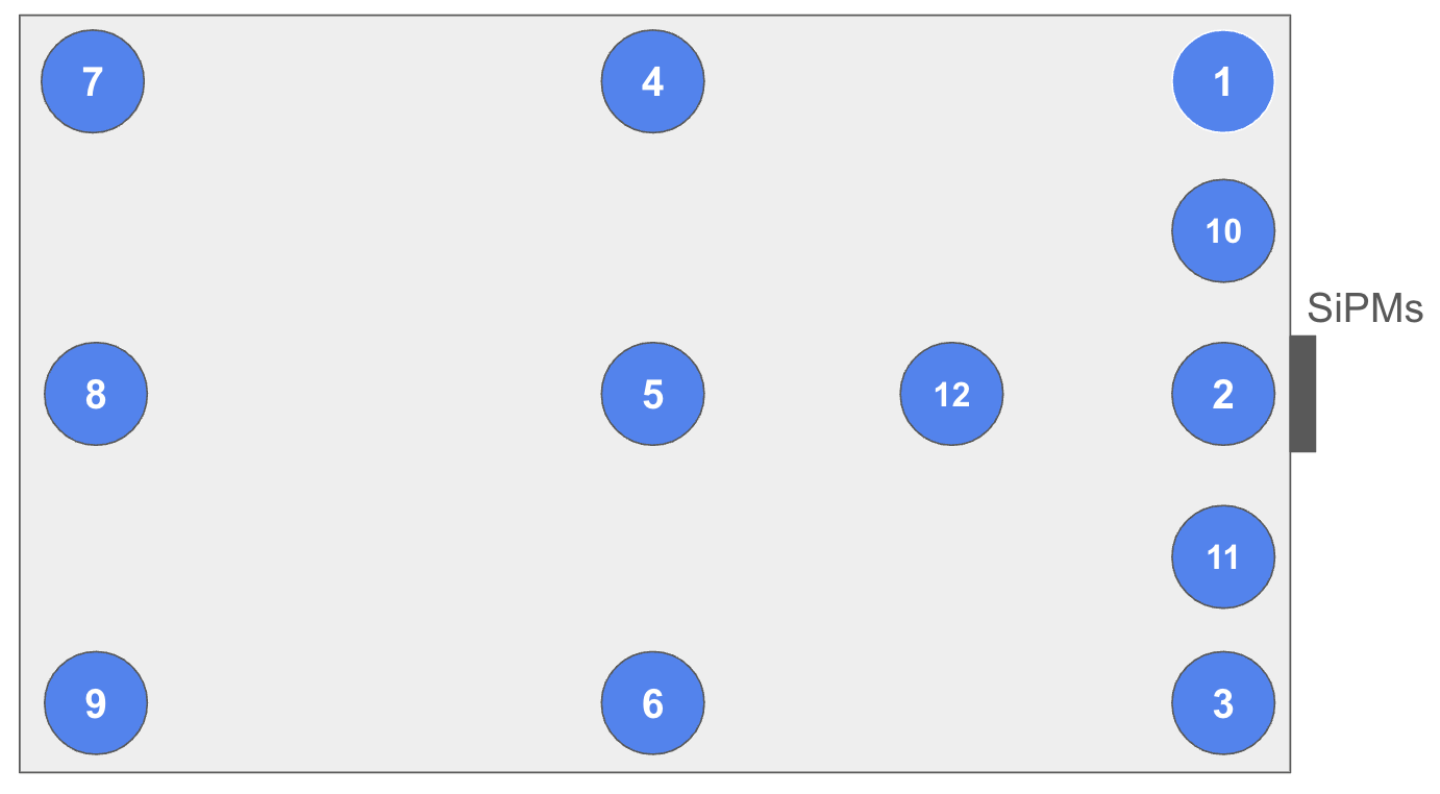}
\end{minipage}
\hspace{-0.5em}
\begin{minipage}[c]{0.45\textwidth}
    \centering
    \includegraphics[width=\textwidth]{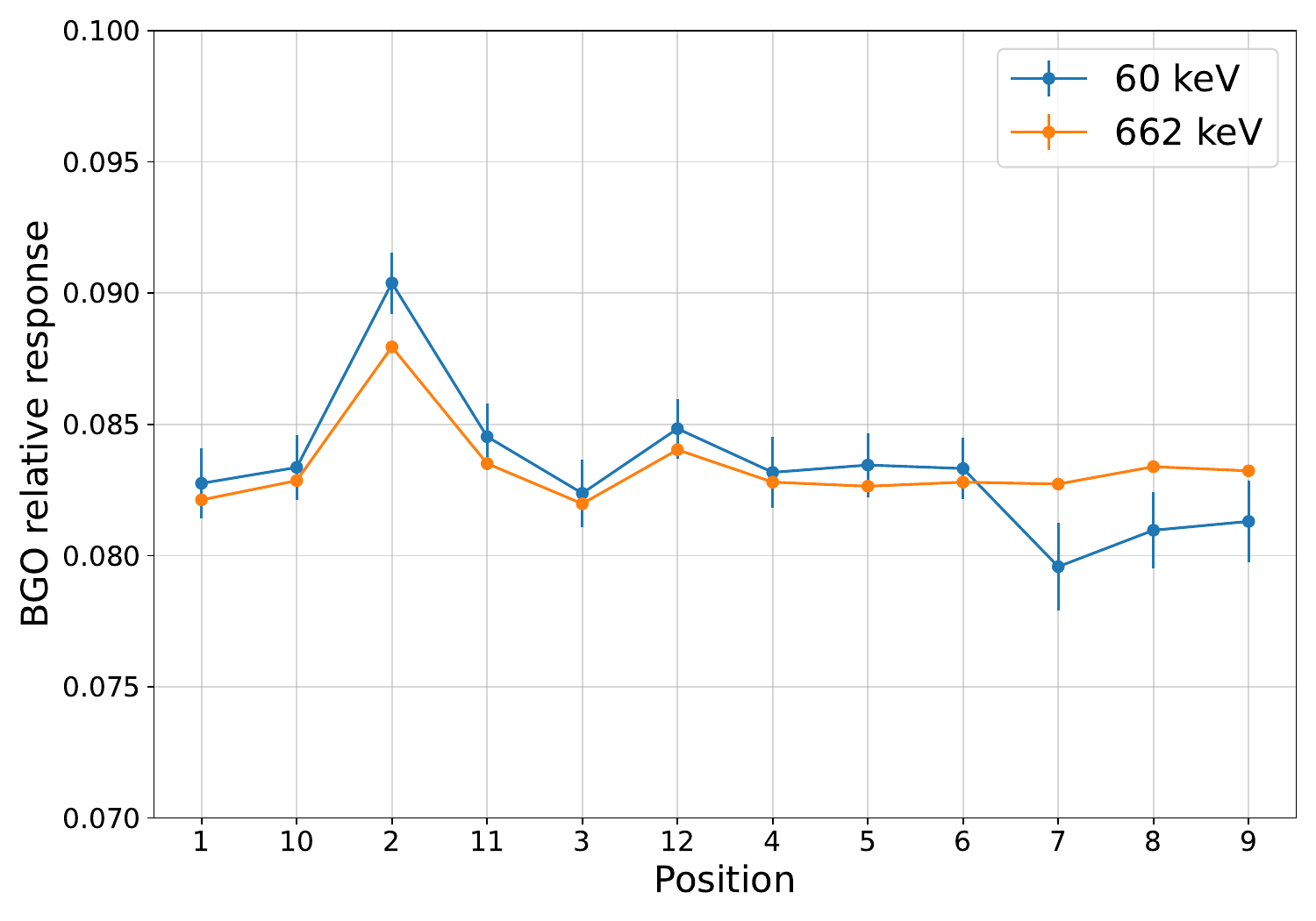}
\end{minipage}
\caption{Left: positions of the collimated sources on top of the $198 \times 118$ mm BGO surface. Right: experimental BGO relative response, measured at NRL, as a function of the collimated positions, for both $^{241}$Am and $^{137}$Cs sources}
\label{fig:rel_resp}
\end{figure}

The measurements with collimated sources conducted at NRL are used to scan the response across the BGO surface. In fact, different interaction positions of the high-energy photons within the crystal imply a different generation location for the optical photons, possibly affecting their chance to be detected. Having an enhanced/suppressed number of detected optical photons would produce a shift of the photopeak towards higher/lower channels (which indeed are proportional to the number of detected optical photons by the SiPMs), as shown in Fig. \ref{fig:spectra_pos_NRL2}. This shift introduces an additional uncertainty in the determination of the energy deposit in the BGO. To assert this effect, we fit the 60 keV (from $^{241}$Am) and 662 keV (from $^{137}$Cs) photopeaks for each collimated position and extract the centroid. Since we are interested in the relative response across the BGO surface, we normalize the ADC channels measured at each position by the sum of the channels from all positions. In Fig. \ref{fig:rel_resp} we show the positions of the collimated sources and the corresponding measured relative responses, for both 60 keV and 662 keV lines. The result shows an overall homogeneous response across the BGO surface for both energies. We observe an enhancement of $\sim$8\% with respect to the average response for the position in front of the SiPMs, and a slight suppression of $\sim$4\% far from the SiPMs for 60 keV.

\subsection{Geant4 simulation} \label{sec:Geant4_sim}
\subsubsection{The simulation framework: BoGEMMS-HPC}
We use the BoGEMMS-HPC simulation framework to reproduce the calibration measurements and produce the X-wall correction matrix. The same simulator has already been used for a preliminary verification against analytical models of the optical physics and experimental data with a photomultiplier tube (PMT) as readout device \citep{2024SPIE13093E..7YC}. BoGEMMS-HPC is based on the Bologna Geant4 Multi-Mission Simulator (BoGEMMS) \citep{2012SPIE.8453E..35B}, an astronomy-oriented Geant4-based application. The framework implements both Geant4 built-in multithreading and G4MPI library \citep{2016arXiv160501792D} along with dedicated I/O interfaces to run the simulation on HPC architectures. All results presented throughout the work are obtained with Geant4 v11.1.

\subsubsection{The Geant4 optical physics}
The scintillation process within the BGO and the photon propagation are simulated with the Geant4 optical physics library \citep{Geant4manual}. The generation and transport of the optical photons are regulated through a set of simulation parameters. The \texttt{SCINTILLATIONYIELD} parameter specifies the mean number of optical photons generated per unit of energy deposited in the material. We set the BGO light yield to 8.2/keV \citep{12684}. The decay time of the BGO pulse shape is defined with the \texttt{SCINTILLATIONTIMECONSTANT1} parameter and set to 300 ns \citep{1981NIMPR.188..403M}. The \texttt{SCINTILLATIONCOMPONENT1} is used to define the spectral shape of the generated optical photons, and we set it according to the emission spectrum shown in \citep{2011NIMPA.640...91A}. 

The transport of optical photons through matter is affected by two processes: absorption and boundary effects. The absorption capability of a medium is defined through the \texttt{ABSLEN} parameter and represents the mean free path of an optical photon in that medium. The absorption length of the BGO crystals used in the experiments was not specified, and several values are found in literature \citep{Roncali_2019, 2011NIMPA.640...91A}. We will therefore use the benchmarking to constrain its value. The boundary effects are regulated through a set of optical properties of the surface between two materials and the reflection and refraction probabilities. Such surface treatment is modelled in Geant4 in the so-called \texttt{UNIFIED} model \citep{591410}, that applies to dielectric-dielectric interfaces and allows the simulation of the surface finish and reflector coating. The BGO crystal has a polished surface and is coated with a reflective material, where optical photons are either reflected or absorbed. The distribution of the reflected photons, from specular to diffusive Lambertian, depends on the BGO and coating surface finish. In our simulations, the BGO-VM2000 reflective surface is set to \texttt{polishbackpainted}, i.e. a polished coating with the presence of air gap, with 100\% specular reflection and reflectivity of $\sim$99\% @ 480 nm \citep{2012ITNS...59..490J}. The BGO-Tetratex interface in the SiPM face is set to \texttt{groundbackpainted}, i.e. also with air gap but a ground reflecting surface, with 100\% diffusive reflection and reflectivity of $\sim$94\% @ 480 nm \citep{2012ITNS...59..490J}. Finally, we assign a wavelength-dependent refractive index to the BGO (2.15 @ 480 nm \citep{1996ApOpt..35.3562W}) and to the optical pad (1.4118 @ 589 nm, info from the manufacturer).

\begin{figure}
\centering
\includegraphics[width=0.7\textwidth]{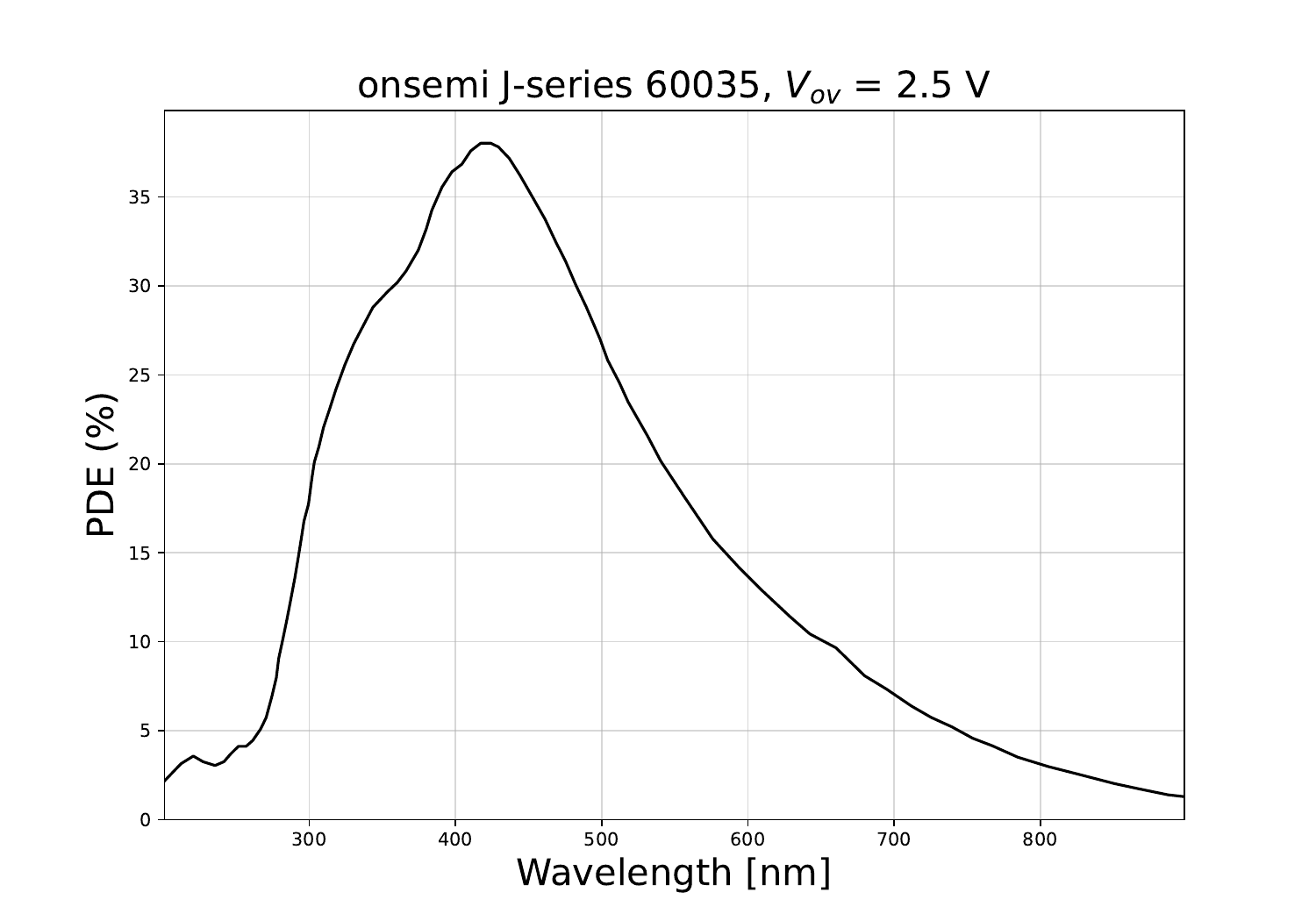}
\caption{Wavelength-dependent photon detection efficiency for onsemi J-series 60035 SiPM with an overvoltage of 2.5 V (reproduced from \href{https://www.onsemi.com/download/data-sheet/pdf/microj-series-d.pdf}{onsemi datasheet})}\label{fig:pde}
\end{figure}

The SiPMs are simulated as empty, fully absorbing boxes, so that every optical photon that reaches the SiPMs is automatically absorbed. The actual number of optical photons detected is then obtained by applying the SiPM energy-dependent PDE\footnote{The PDE also depends on the incident angle of the optical photons \citep{2020JInst..15P1019N}. In our simulation, we applied the manufacturer-provided PDE uniformly, due to the lack of angular dependence data.} for an overvoltage of 2.5 V (see Fig. \ref{fig:pde}). The final output of the simulation is a histogram of the numbers of detected optical photons, for each radioactive source. 

\subsubsection{Simulated mass model of the experimental setups}
We reproduce in BoGEMMS-HPC the calibration campaigns at NRL and SSL modeling the corresponding experimental setups as accurately as possible. We built the BGO mass model as a $198 \times 118 \times 23$ mm block coated with two layers of VM2000, one layer of Tetratex and one layer of Teflon on the faces without SiPMs, and with two layers of Tetratex on the face with the SiPMs attached. The $18 \times 18 \times 1$ mm optical pad is placed at the center of the $118 \times 23$ mm face, with the Tetratex layers having a cavity of the same size to allow the pad to be in direct contact with the BGO. The $3 \times 3$ array of SiPMs is coupled with the optical pad.

We simulate the mass model of the aluminum housing using its CAD model, whose file was provided by NRL. For the NRL experiment, we reproduce the metal table and the plastic piece below the X-wall. The aluminum frame and the foam block are included for the uncollimated $^{241}$Am and $^{137}$Cs. For the collimated measurements, we reproduce the collimator as a lead cylinder with a central hole of 16 mm diameter. In the SSL simulation we included the wooden table, the anti-static mat, the stand holding the sources and a lead block that was in proximity to the X-wall. 

For all the experiments, we simulate the radioactive sources using the Geant4 \texttt{G4RadioactiveDecay} physics list. The sources were modelled as 5 mm disks, with a thickness of 1 mm for NRL and 3.18 mm for SSL, and enclosed in a $25.4 \times 6.35$ mm plastic disk. Moreover, $^{22}$Na for NRL and $^{133}$Ba, $^{60}$Co, $^{88}$Y for SSL were further encased in plastic and placed on top of a 1 mm thick foam piece, and we included these materials in the mass model.

\subsection{Simulation benchmarking} 
\label{sec:results} 
The results presented in this section are based on simulations using a BGO absorption length of 5 m, selected as the optimal trade-off between agreement with collimated and uncollimated source measurements. Increasing the absorption length enhances the spatial homogeneity of the BGO response; however, it also decreases the FWHM of the photopeaks, hence worsening the comparison with the experimental energy resolution. This decrease occurs because a larger number of detected optical photons, thanks to the increased absorption length, reduces statistical fluctuations, thereby improving the simulated energy resolution. The chosen value of 5 m provides a balance that maintains acceptable the benchmarking using both collimated and uncollimated sources, with an uncertainty of the order of $\pm1$ m. We carried out this trade-off with the goal of limiting the discrepancy with the experiment to approximately 20\%. In App. \ref{secA2}, we present the simulated and experimental relative response and energy resolution for BGO absorption lengths of 4, 5 and 6 m. Our results show that while absorption lengths greater than 6 m would lead to a discrepancy larger than 20\% in the energy resolution, absorption lengths below 4 m would result in a discrepancy exceeding 20\% in the relative response. 

Another key parameter of the benchmarking is the electronic noise, which contributes to the experimental energy resolution, as discussed in Sec. \ref{sec:ene_rel_exp}. While the Geant4 simulation models the statistical nature of optical photon production and detection, as well as light transport inhomogeneities within the crystal, it does not account for the electronics of the SiPMs. We therefore incorporate the electronic noise in the simulation by adding a Gaussian smearing to the simulated energies, with a FWHM corresponding to the experimental value of $a \simeq$ 21 keV.

Table \ref{tab:parameters} summarizes the key parameters relevant to the benchmarking, including their values or ranges, whether they were treated as fixed or free in the analysis, and the corresponding references.

\begin{table}
\caption{List of simulation parameters used for the benchmarking}\label{tab:parameters}%
\begin{tabular}{@{}llll@{}}
\toprule
\textbf{Parameter} & \textbf{Value/Range} & \textbf{Fixed/Free} & \textbf{References}\\
\midrule
BGO light yield & 8.2/keV & Fixed & \citep{12684}  \\
BGO decay time & 300 ns & Fixed & \citep{1981NIMPR.188..403M} \\
BGO emission range & 330-620 nm (peak @ 480 nm) & Fixed & \citep{2011NIMPA.640...91A} \\
BGO refractive index & 2.15 @ 480 nm & Fixed & \citep{1996ApOpt..35.3562W} \\
Optical pad refractive index & 1.4118 @ 589 nm & Fixed & Manufacturer \\
VM2000 reflectivity & 99\% @ 480 nm & Fixed & \citep{2012ITNS...59..490J} \\
VM2000 reflection type & Specular & Fixed & \citep{2012ITNS...59..490J} \\
Tetratex reflectivity & 94\% @ 480 nm & Fixed & \citep{2012ITNS...59..490J} \\
Tetratex reflection type & Diffusive & Fixed & \citep{2012ITNS...59..490J} \\
PDE & 38\% @ 420 nm & Fixed & \href{https://www.onsemi.com/download/data-sheet/pdf/microj-series-d.pdf}{onsemi datasheet} \\
BGO absorption length & 5 m & Free & / \\
Electronic noise & 21 keV & Free & / \\
\botrule
\end{tabular}
\end{table}

\subsubsection{Relative response} \label{sec:results_relative_response}
\begin{figure}
\centering
\includegraphics[width=0.5\textwidth]{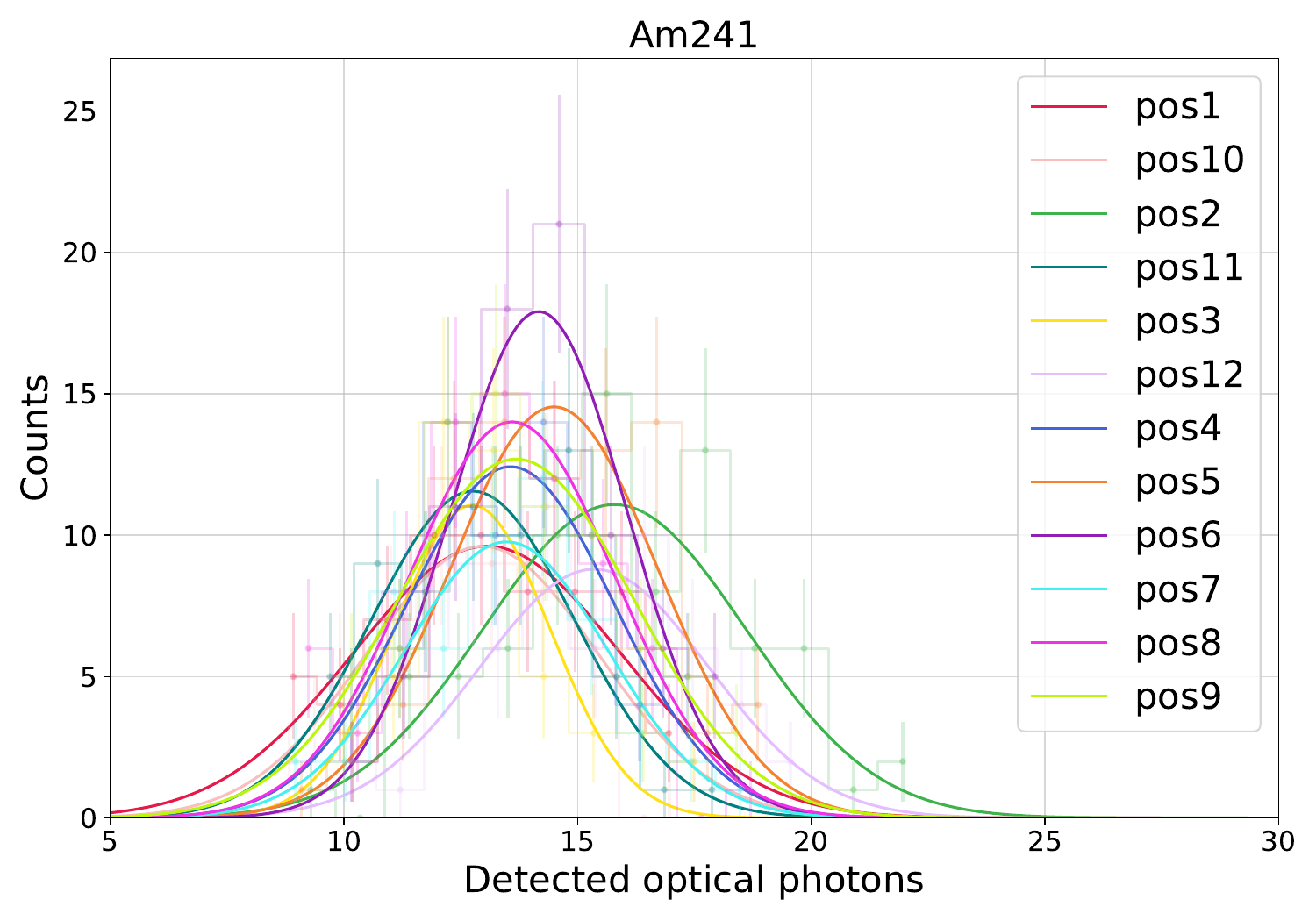}\includegraphics[width=0.5\textwidth]{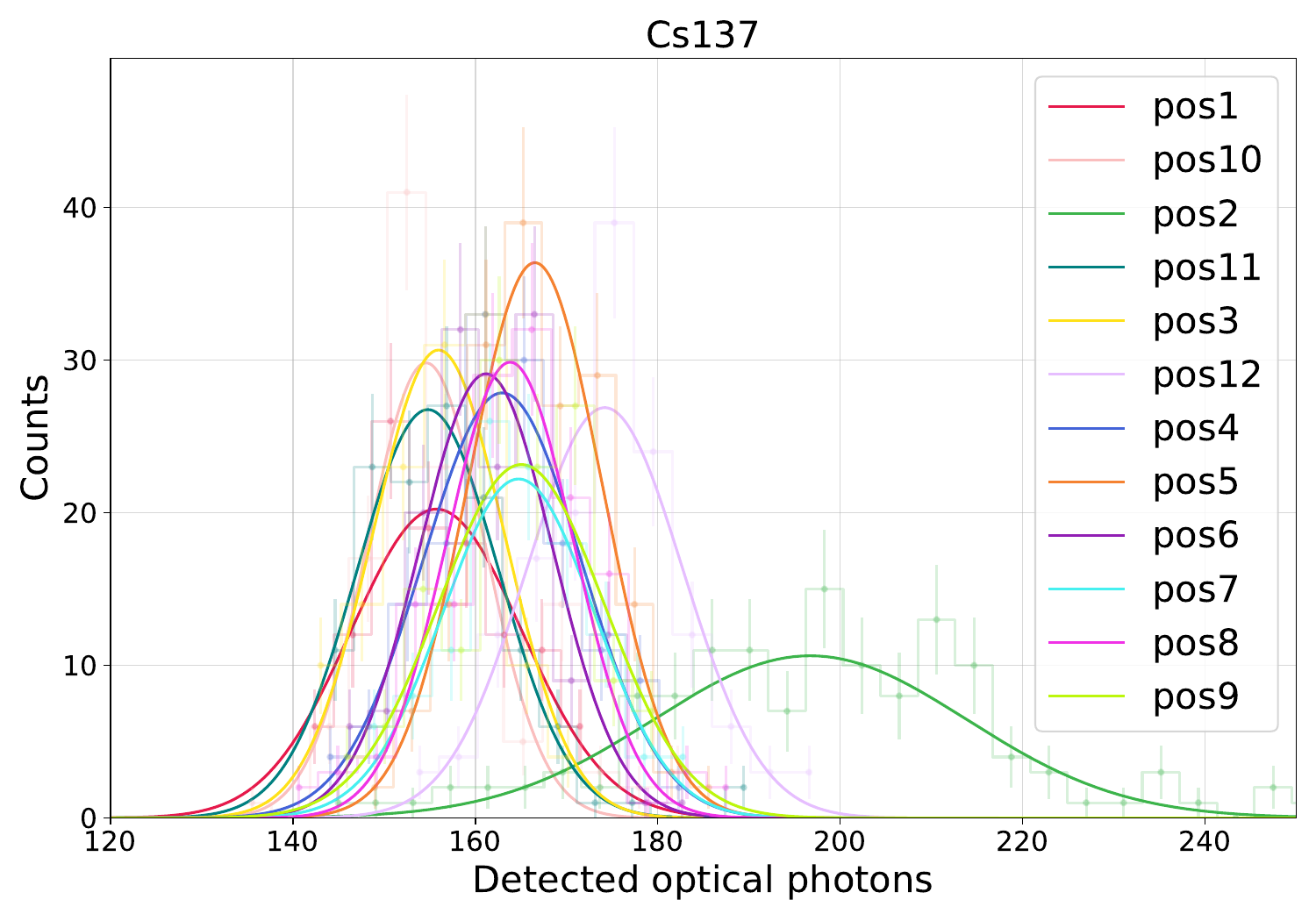}
\caption{Simulated photopeaks for each position and corresponding Gaussian fit for 60 keV (left) and 662 keV (right)}\label{fig:sim_collimated}
\end{figure}

We use the collimated simulations to compare the response spatial distribution with the experimental one. Following the same procedure conducted for the experimental data (Sec. \ref{sec:rel_response}), we quantify the response of the ACS for each position of the collimated source. We fit with a Gaussian the simulated photopeaks for both 60 and 662 keV (Fig. \ref{fig:sim_collimated}) and assign to each position the number of detected optical photons, that defines the ACS response. We then normalize such value by the sum of the response of all the positions to get the relative response. This normalization also allows for a direct comparison between the experimental and simulated relative responses, as ADC channels and $N_\text{det}$ are linearly proportional.

The uncertainty in the positioning of the collimated sources, estimated to be $\sim$1 mm, represents a source of systematic error in the response estimation. We quantify its impact by repeating the simulations with the source positions shifted by 1 mm in four orthogonal directions and evaluating the change in response for each configuration. The maximum change in the response due to the positioning error is taken into account as systematic error and summed to the statistical one. The same systematic error is also included in the measured response of Fig. \ref{fig:rel_resp}.

\begin{figure}
\centering
\includegraphics[width=0.5\textwidth]{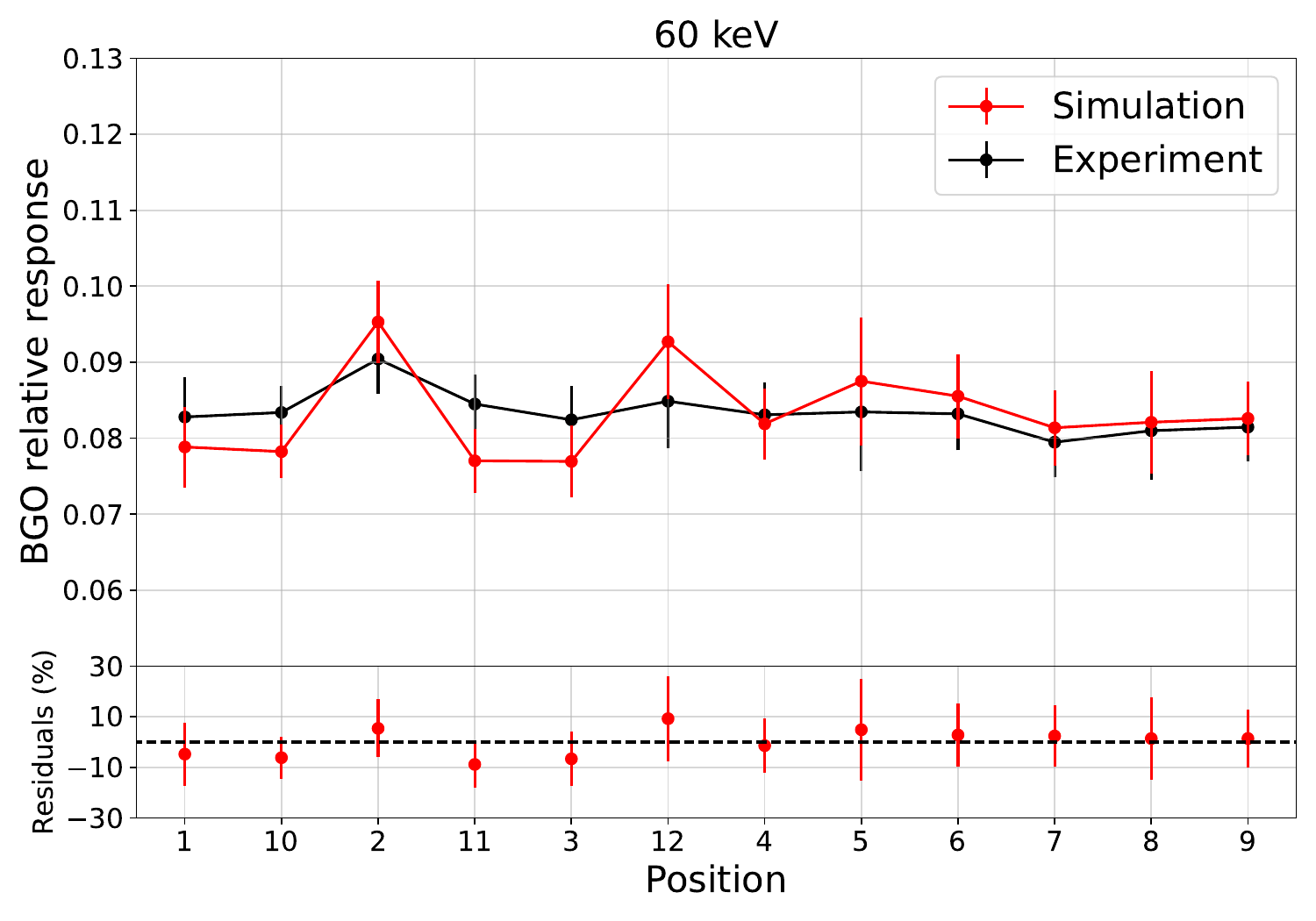}\includegraphics[width=0.5\textwidth]{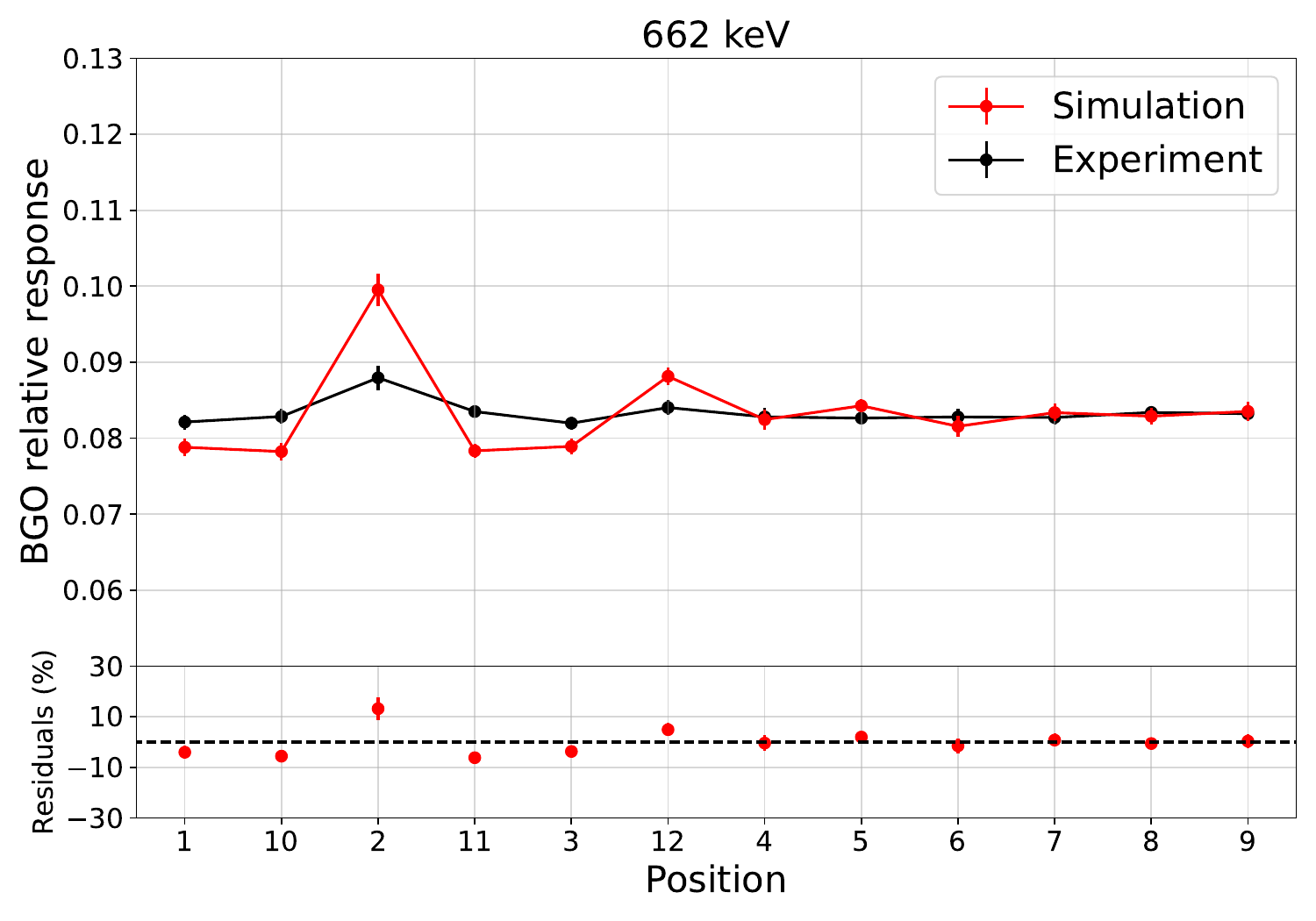}
\caption{Comparison between experimental and simulated relative response for each collimated position, for $^{241}$Am (left) and $^{137}$Cs (right)}\label{fig:relative_response}
\end{figure}

We show in Fig. \ref{fig:relative_response} the simulated relative response for each position, for both $^{241}$Am and $^{137}$Cs, compared with the collimated measurements obtained at NRL. The error bars represent the total (statistical and systematic) error for both simulated and experimental data points. All the residuals shown in the following plots are defined as $(S-L)/L$ (\%), where $S$ and $L$ are simulated and laboratory data, respectively. Overall, the simulation reproduces the homogeneity of the experimental response with a maximum discrepancy of $\sim $10\%. 
The largest discrepancy is observed near the SiPMs, where the simulation is predicting a larger inhomogeneity compared with the laboratory measurements. Many factors may contribute to this discrepancy. Increasing the BGO absorption length to 6 m reduces the discrepancy between simulation and experiment by approximately $\sim$15\%. Another potential source of discrepancy is the uncertainty in the positioning of the collimated sources. Although a systematic uncertainty of 1 mm was already included, this may underestimate the true positional uncertainty. To investigate this, we considered a systematic offset of 8 mm, corresponding to the radius of the collimator hole (see Fig. \ref{fig:test_different_sys}). The response at the position directly in front of the SiPMs is particularly sensitive to small positional shifts. Assuming a systematic uncertainty of 8 mm, the simulated results become consistent with the experimental measurements within the associated uncertainties. Nevertheless, with the current setup, the maximum discrepancy between the experiment and the simulated response spatial distribution remains well below 20\%, consistently meeting our target.

\subsubsection{Spectral comparison}
\begin{figure}
\centering
\includegraphics[width=0.5\textwidth]{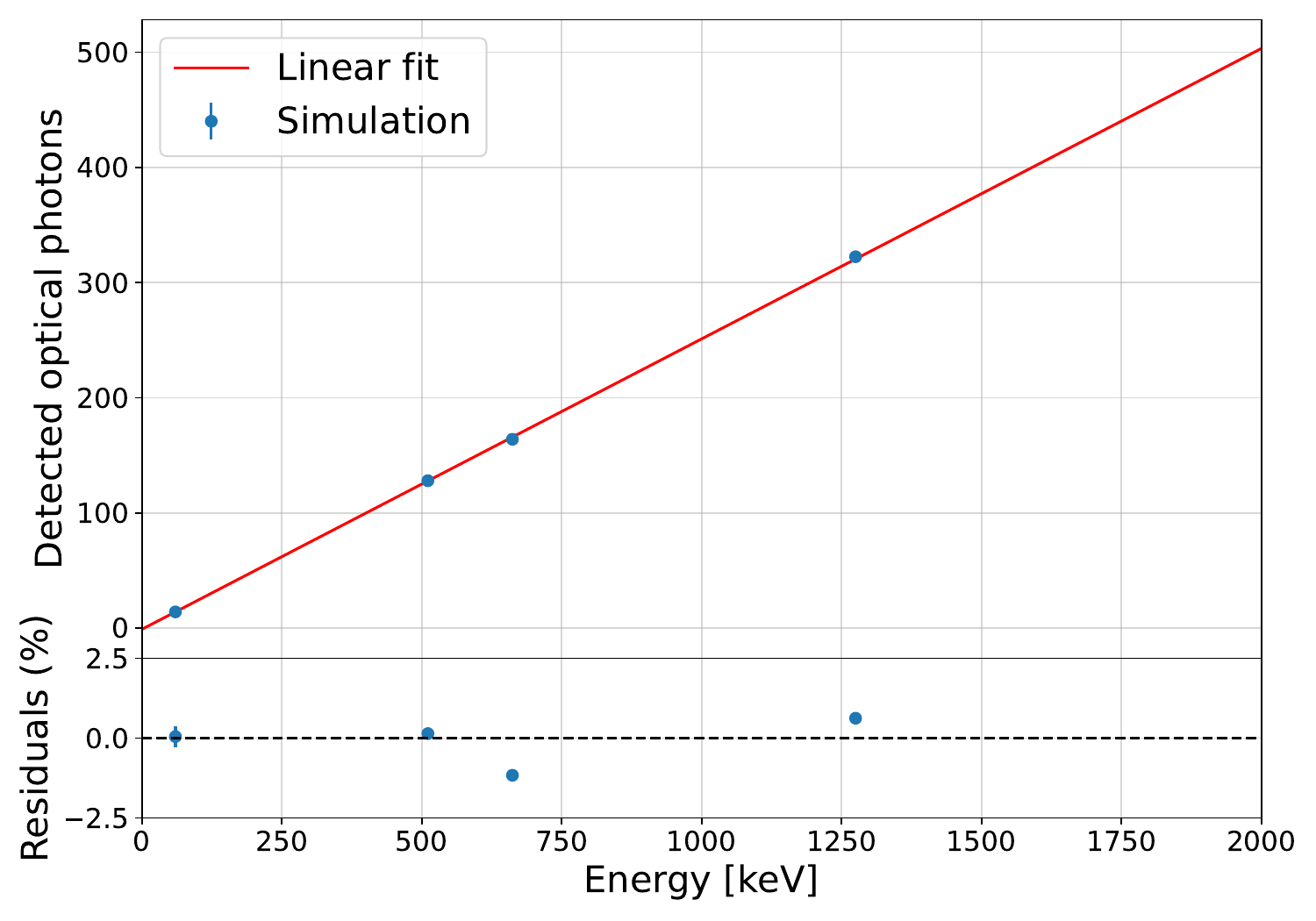}\includegraphics[width=0.5\textwidth]{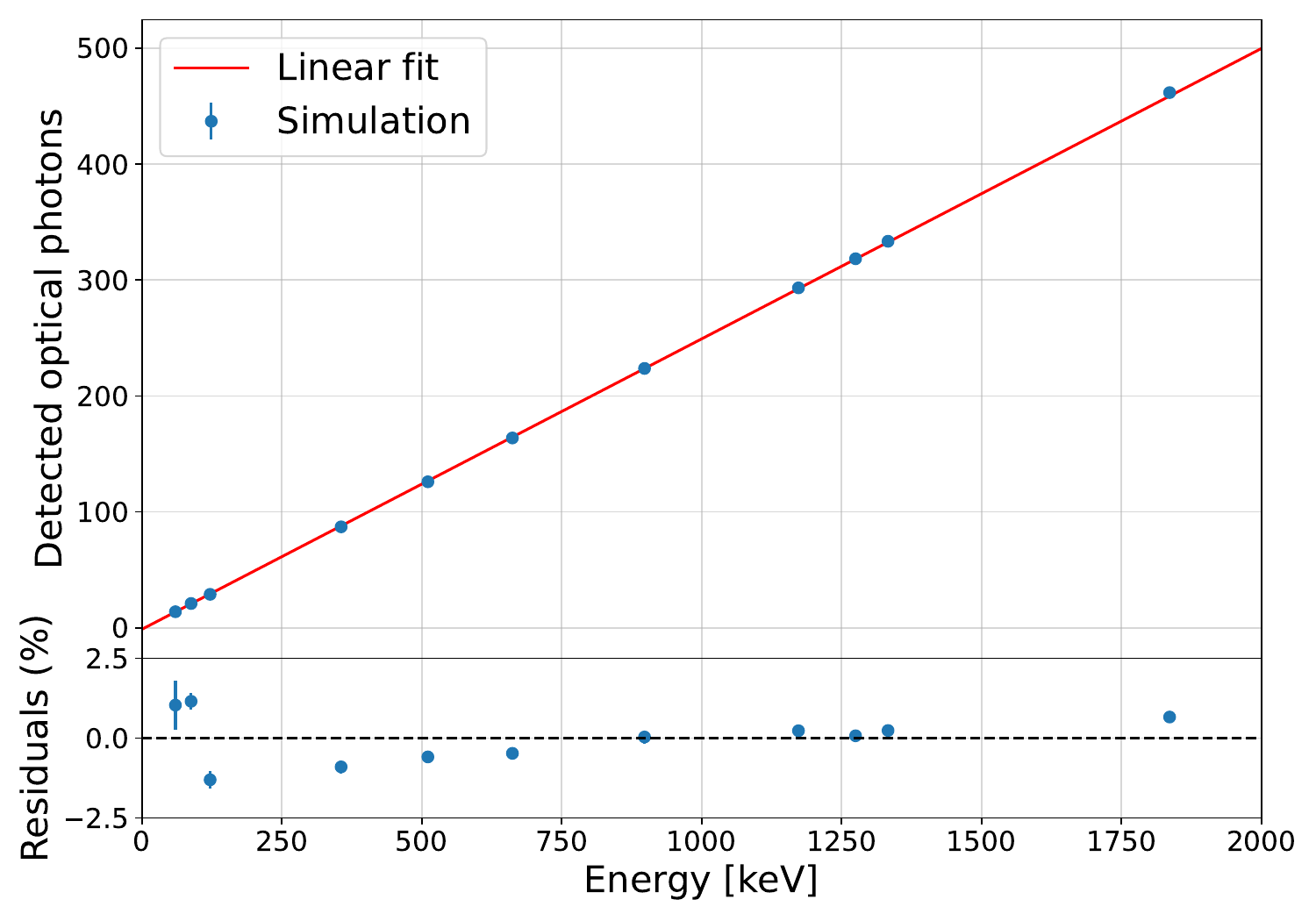}
\caption{Photon-energy relation for the simulated NRL (left) and SSL (right) datasets and linear fit}\label{fig:energy_calibration}
\end{figure}

The simulated spectra are first converted to energy and then compared with the calibrated spectra obtained in the laboratory with uncollimated sources. The conversion from optical photons to energy is obtained following the same procedure as for the experimental energy calibration, described in Sec. \ref{sec:ene_cal_exp}. The photon-energy relation obtained is shown in Fig. \ref{fig:energy_calibration}. The best-fit parameters are: $m = 0.2522\pm 0.0002$, $q=-1.13\pm0.05$ for NRL, $m = 0.2506\pm 0.0002$, $q=-1.36\pm0.05$ for SSL.

After applying the energy calibration for the simulations, we can compute the simulated energy resolution for each energy line, fit the data with the model described in Eq. \eqref{eq:resolution} and compare it to the experimental ones. 

\begin{figure}
\centering
\includegraphics[width=0.5\textwidth]{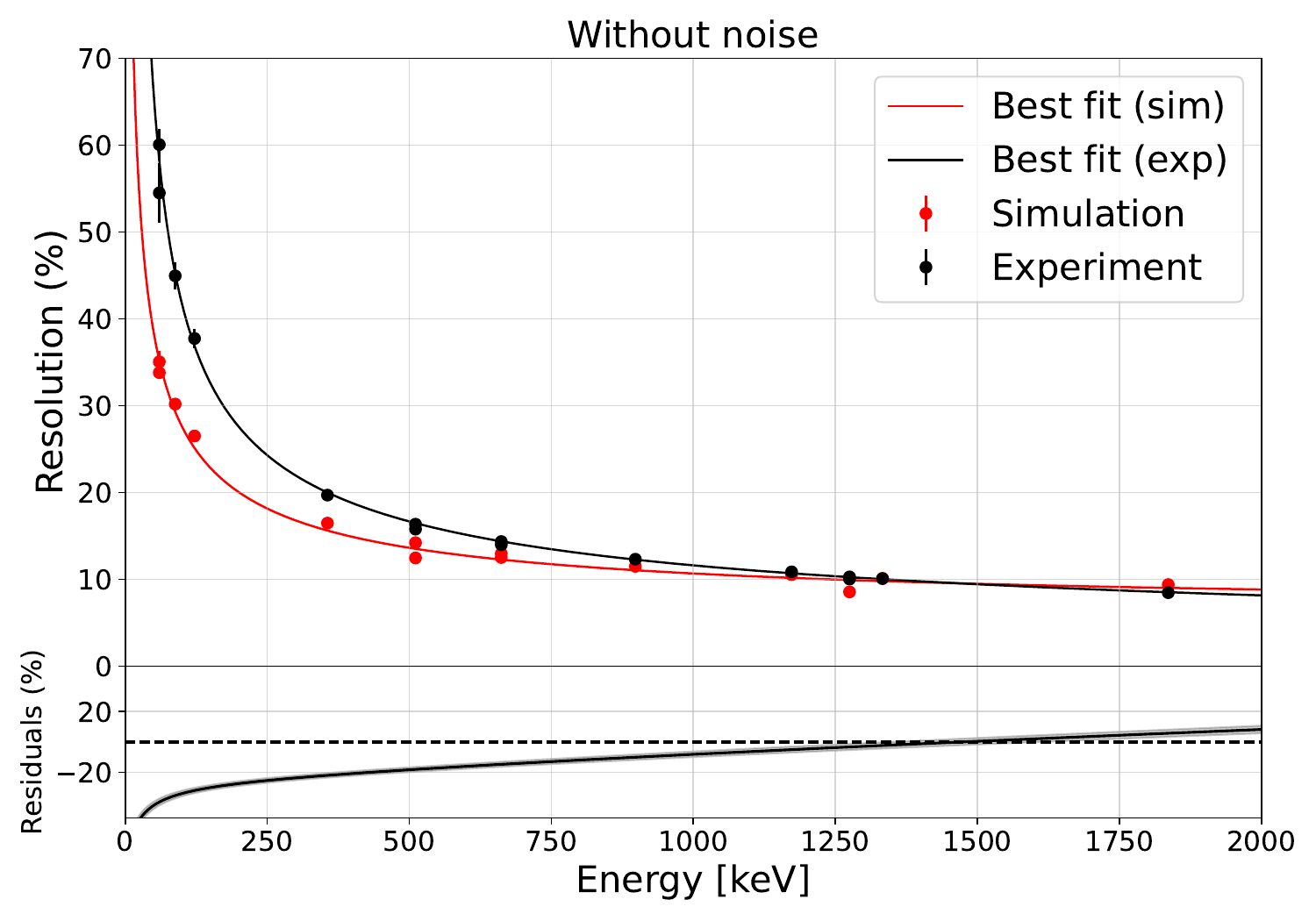}\includegraphics[width=0.5\textwidth]{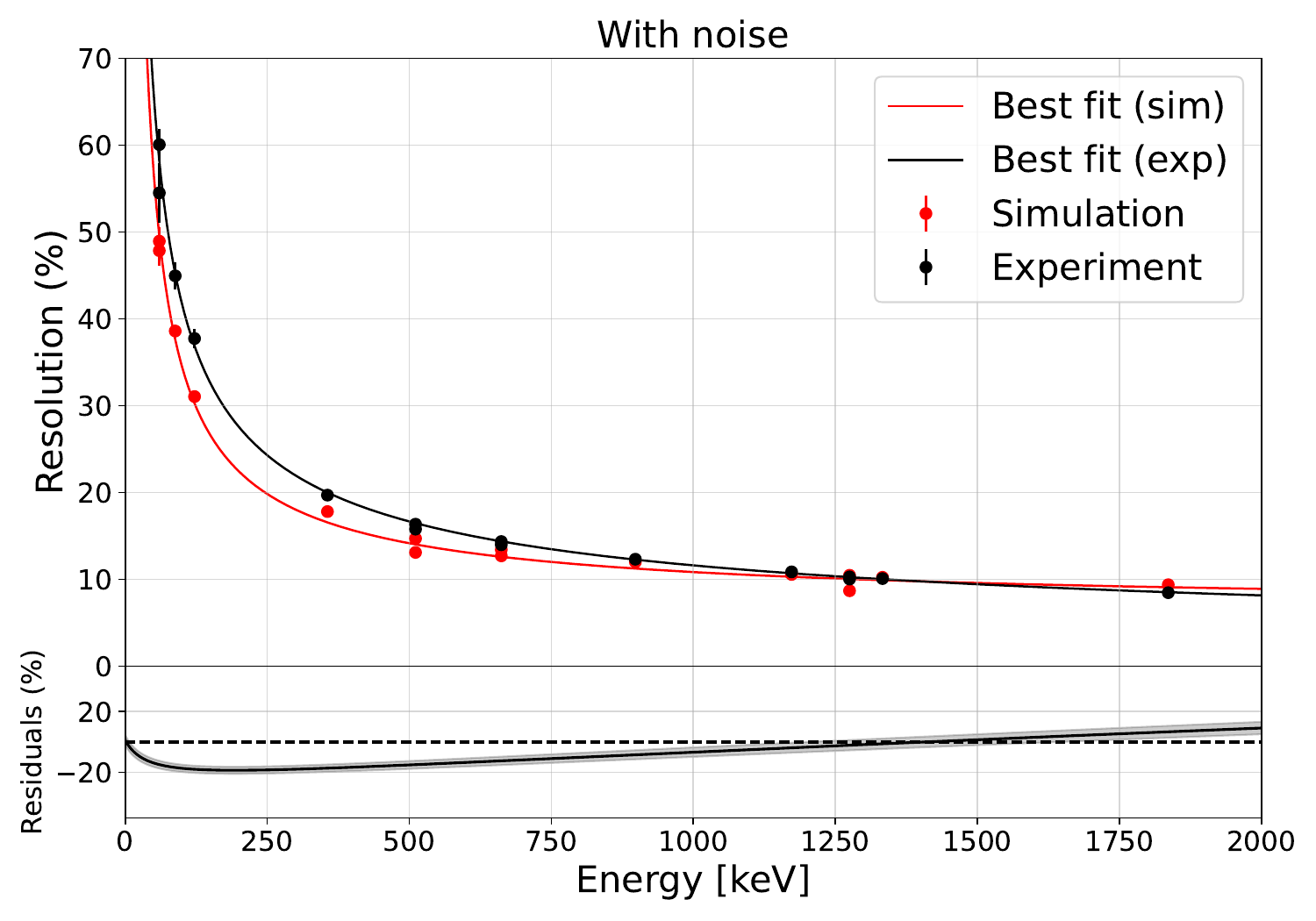}
\caption{Experimental and simulated energy resolution for each source energy and the corresponding best-fit model, without (left) and with (right) the application of the electronic noise to the simulation}\label{fig:resolution}
\end{figure}

In Fig. \ref{fig:resolution} we show the comparison between the experimental energy resolution with the simulated results, with and without the inclusion of electronic noise. In the absence of electronic noise, the best-fit parameters are: $a = 0.63 \pm 0.58$, $b = 2.67 \pm 0.02$, and $c = 0.06 \pm 0.01$. The value of $a$ is consistent with zero, confirming the lack of electronic noise in the simulation, as expected. With the inclusion of the electronic noise, the new best-fit parameters are: $a = 20.77 \pm 0.69$, $b = 2.63 \pm 0.06$ and $c = 0.06 \pm 0.01$. While $b$ and $c$ remain largely unchanged, the value of $a$ increases significantly, becoming consistent with its experimental counterpart. Incorporating the experimental electronic noise into the simulation notably improves the agreement, particularly at low energies, resulting in a maximum discrepancy of 20\% across the investigated energy range. The discrepancy arises in the range $100$ keV $\lesssim E \lesssim 500$ keV, where the simulation predicts a better energy resolution compared with the experimental results. Although we accounted for electronic noise in the simulation, some components contributing to the statistical term may still be missing, such as gain fluctuations arising from statistical variations in breakdown voltage and temperature \citep{DOROSZ2013202}. This effect would introduce additional broadening in the measured photopeaks, leading to a degradation of the experimental energy resolution. Unlike electronic noise, this additional source of statistical uncertainty cannot be isolated and extracted through the fitting of the experimental energy resolution.

\begin{figure}
\centering
\includegraphics[width=0.7\textwidth]{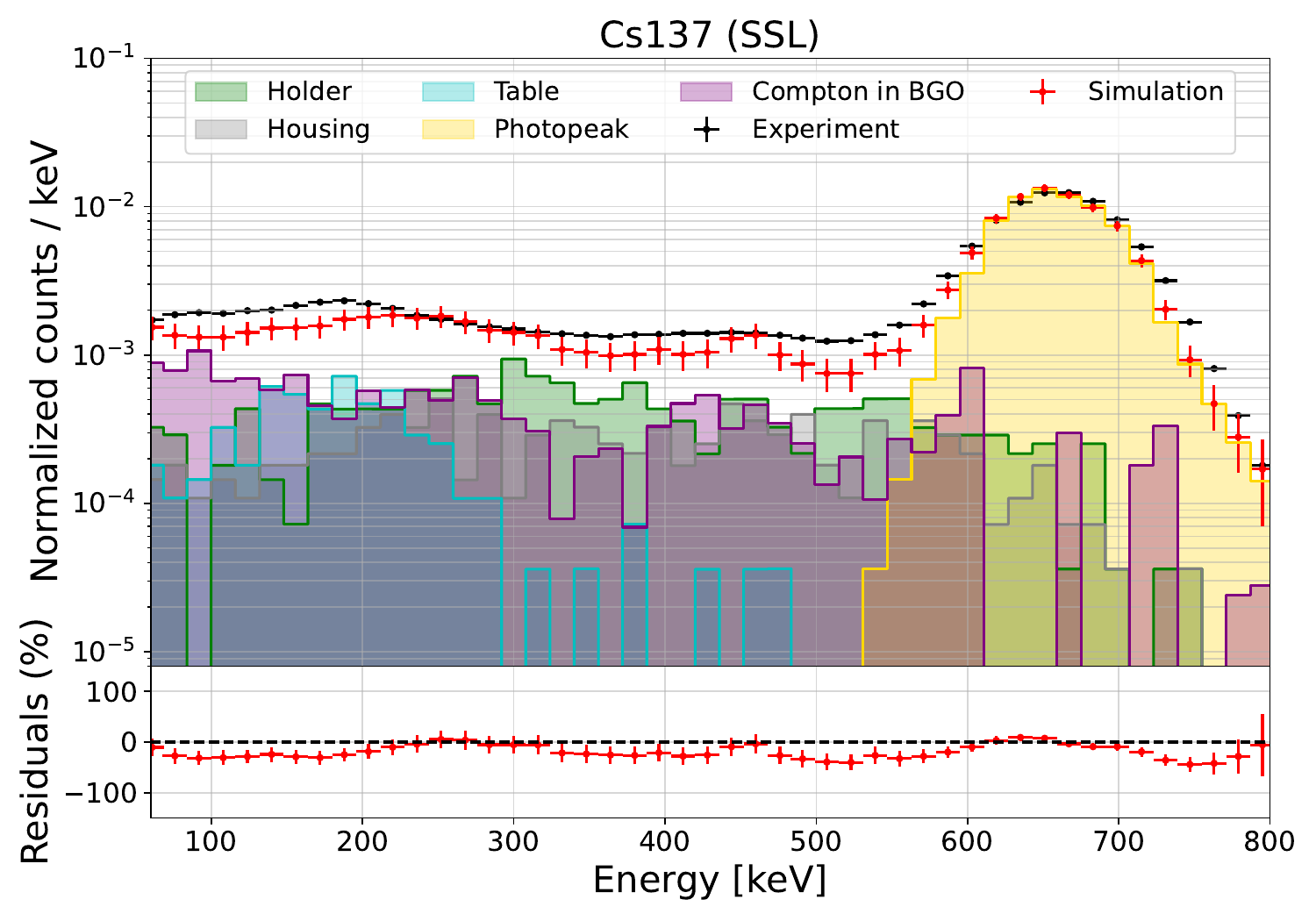}
\caption{SSL experimental and simulated $^{137}$Cs spectrum, along with the scattering components (i.e. those photons which scattered first on that component before entering the BGO) contributing to the simulated spectrum}\label{fig:Cs137_components}
\end{figure}

Once the electronic noise is applied to the energy resolution, we compare the experimental and simulated spectra. We normalized both spectra by the total number of counts in the photopeak, within one FWHM from the peak. The error in the simulated count rate is the sum of the statistical fluctuations propagated with the uncertainty in the photon-to-energy conversion and the electronic noise. For the experimental counts, we propagated the uncertainty in the channel-to-energy conversion and the systematic in the background subtraction. On average, the systematic uncertainty contributes to the total error by $\sim$25\% for the simulation, and by $\sim$45\%  for the experiment. In Fig. \ref{fig:Cs137_components} we show the comparison between the SSL experimental and simulated $^{137}$Cs spectrum. 

\begin{figure}
\centering
\includegraphics[width=0.5\textwidth]{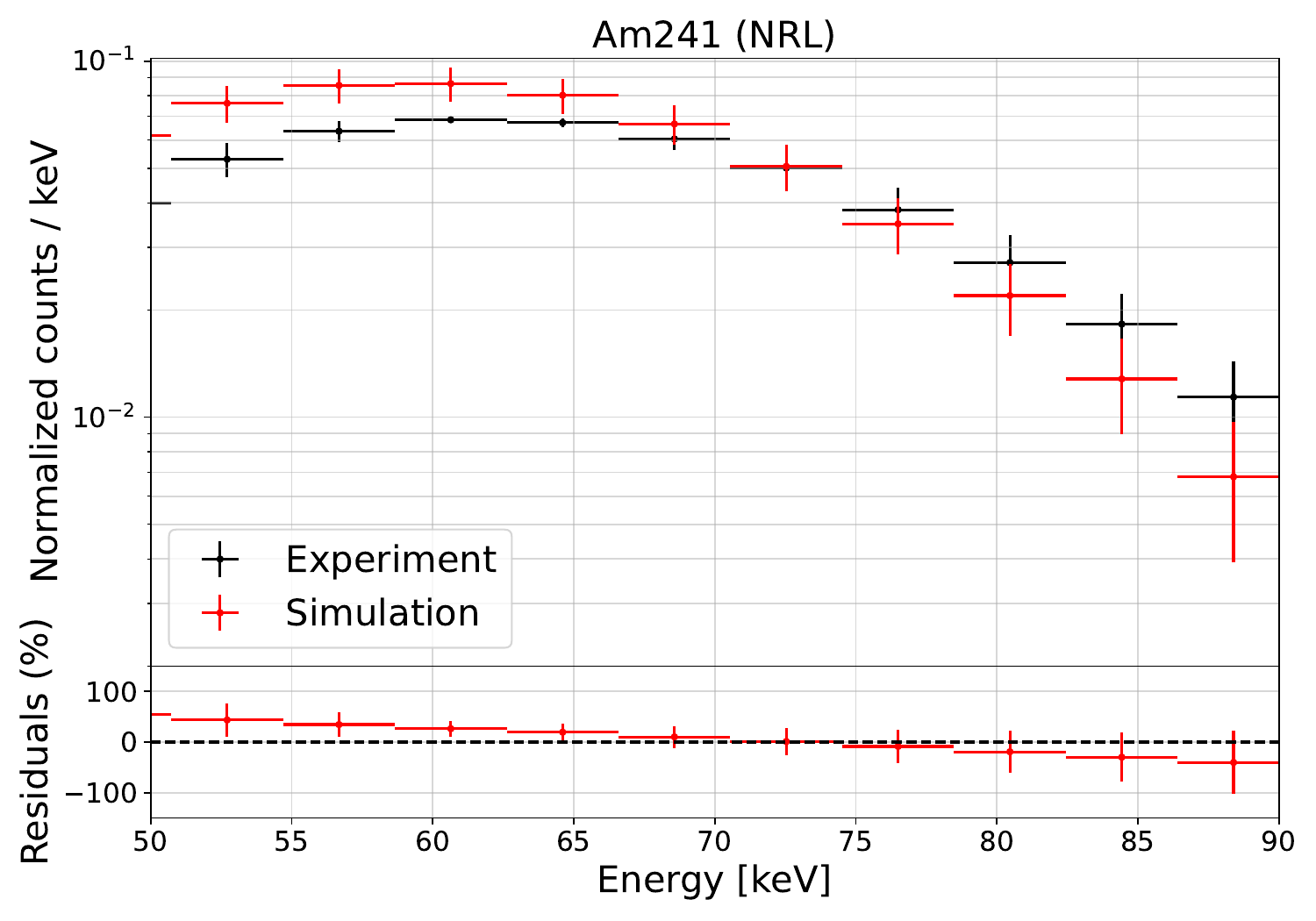}\includegraphics[width=0.5\textwidth]{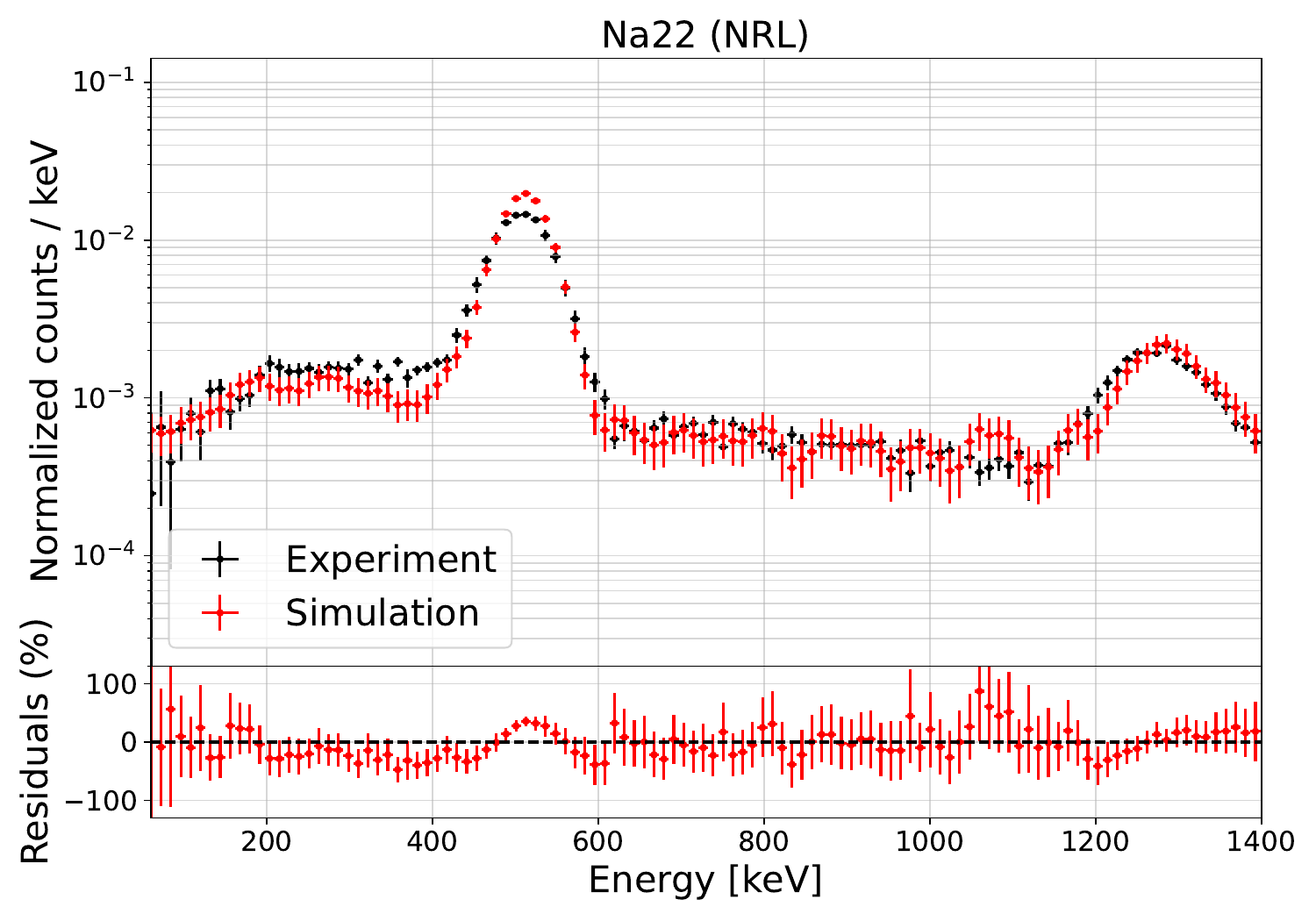}\\\includegraphics[width=0.5\textwidth]{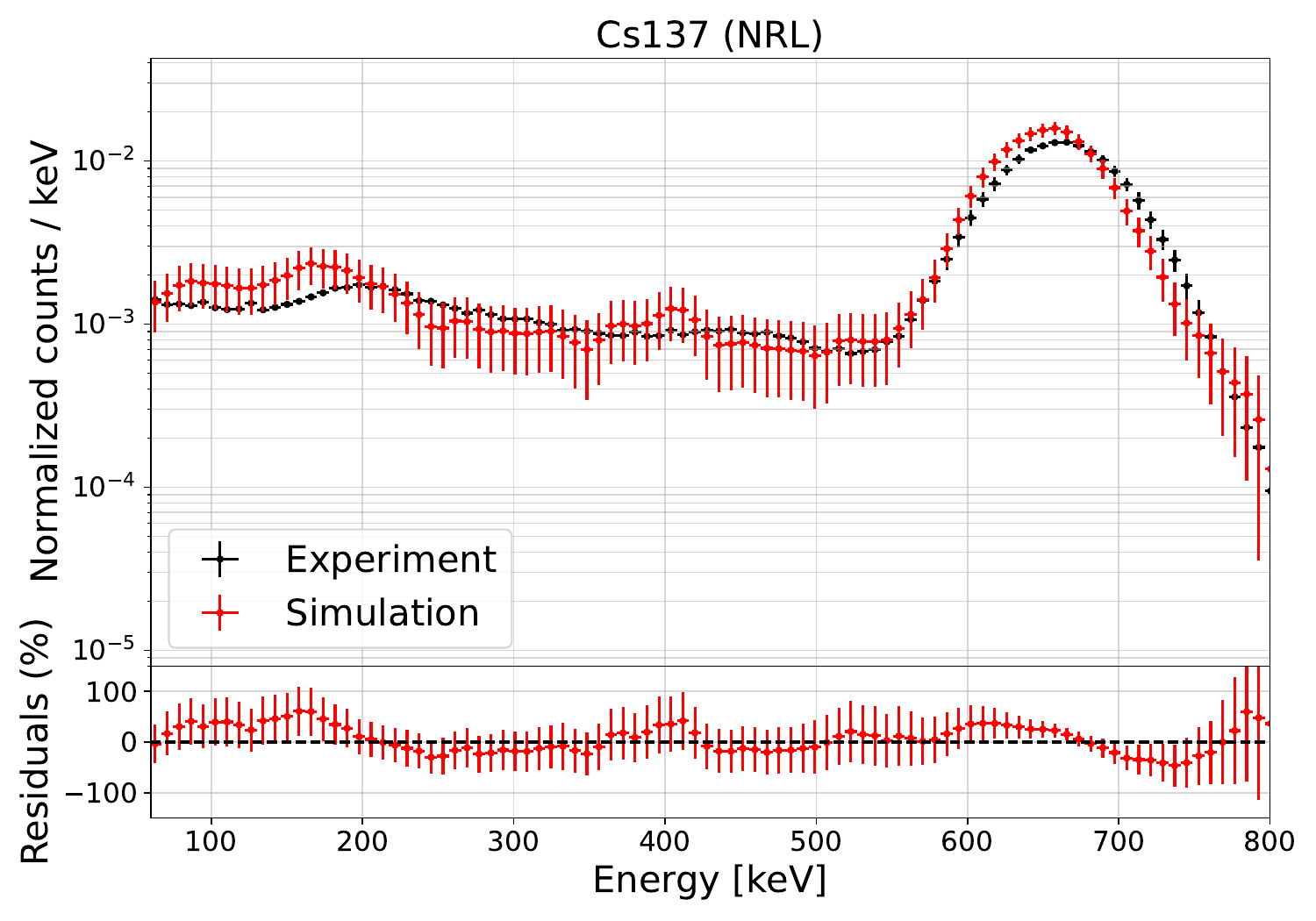}
\caption{Comparison between NRL experimental (black) and simulated (red) energy spectra for each radioactive source}\label{fig:comp_NRL}
\end{figure}

\begin{figure}
\centering
\includegraphics[width=0.42\textwidth]{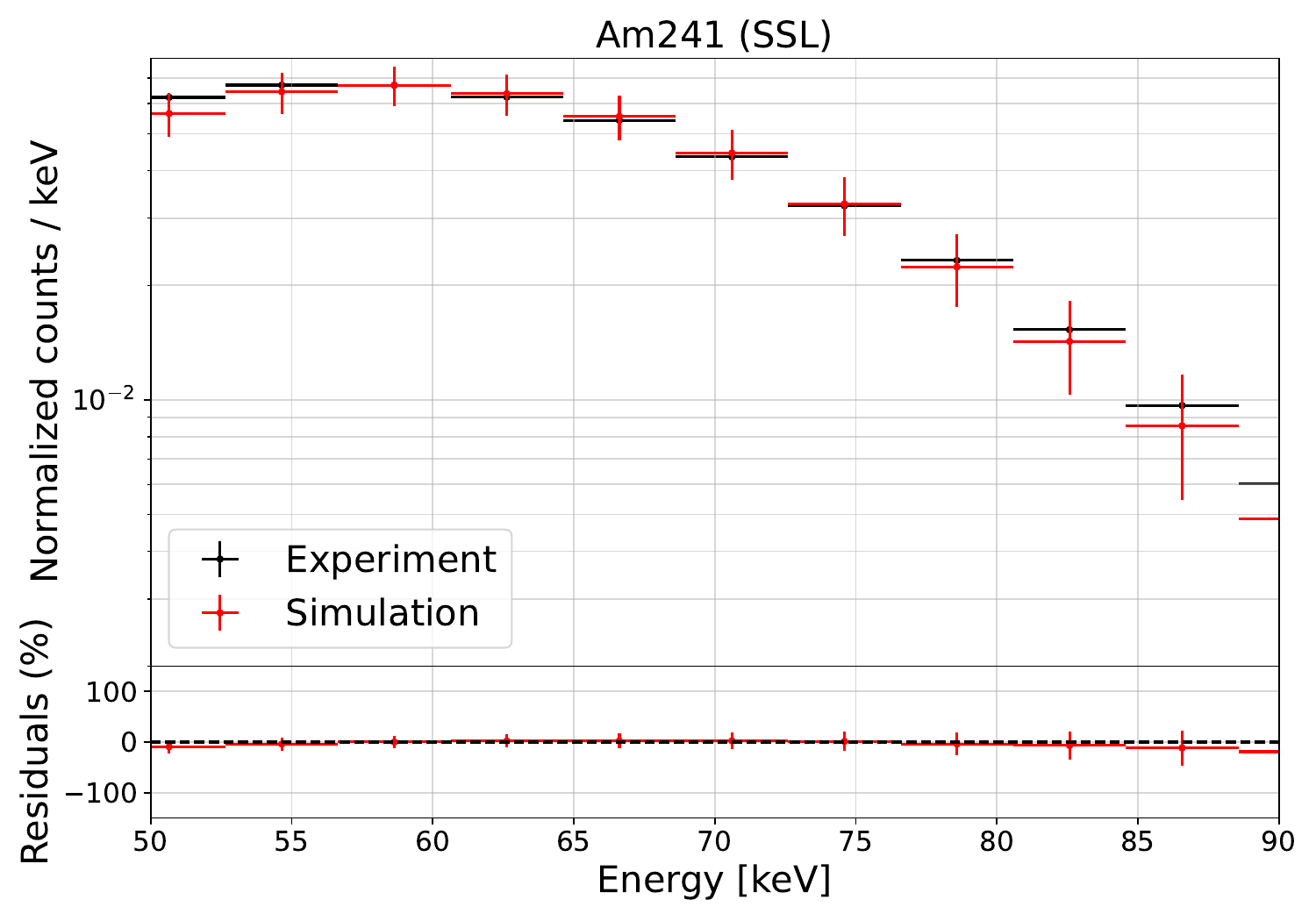}\includegraphics[width=0.42\textwidth]{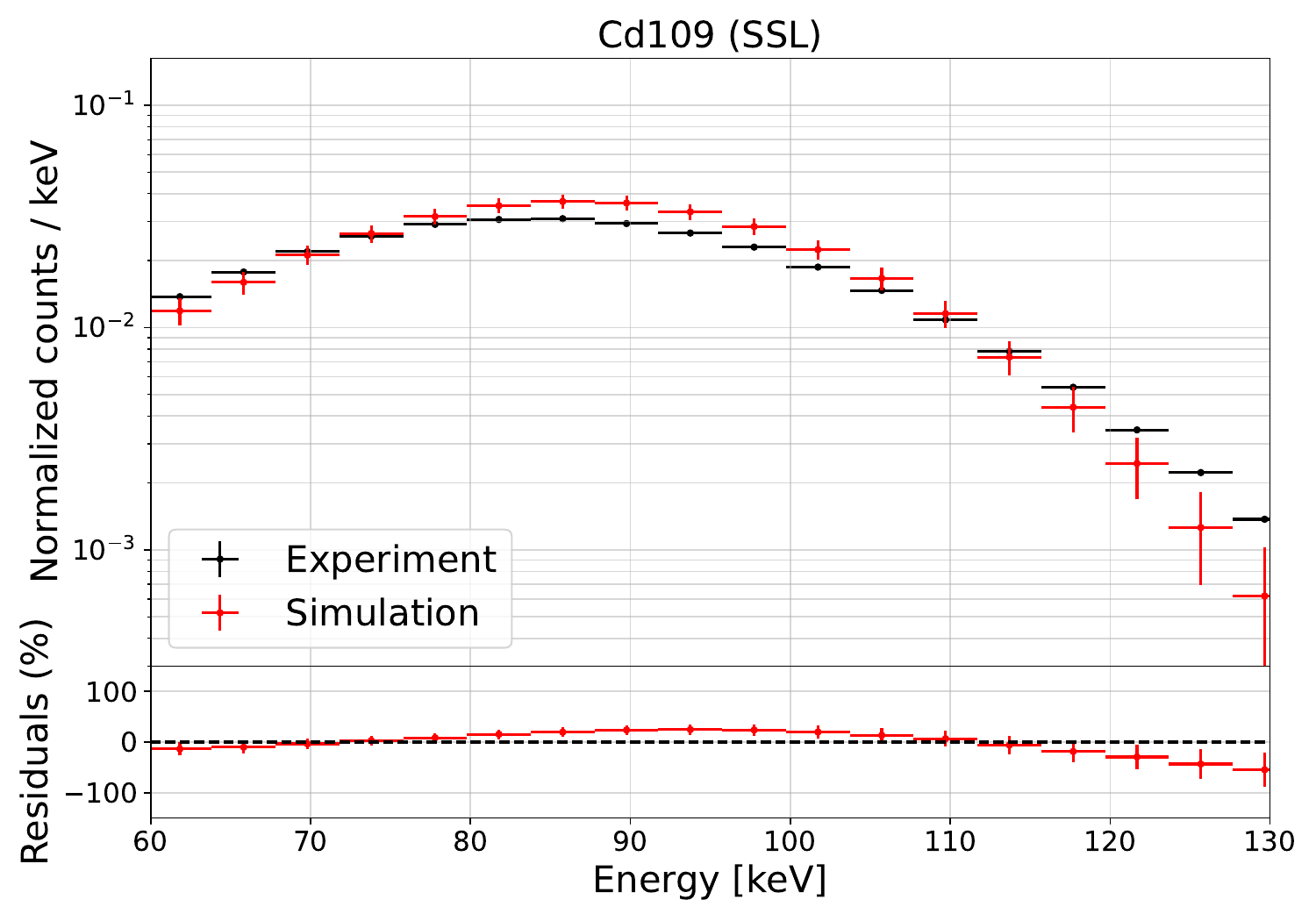}\\\includegraphics[width=0.42\textwidth]{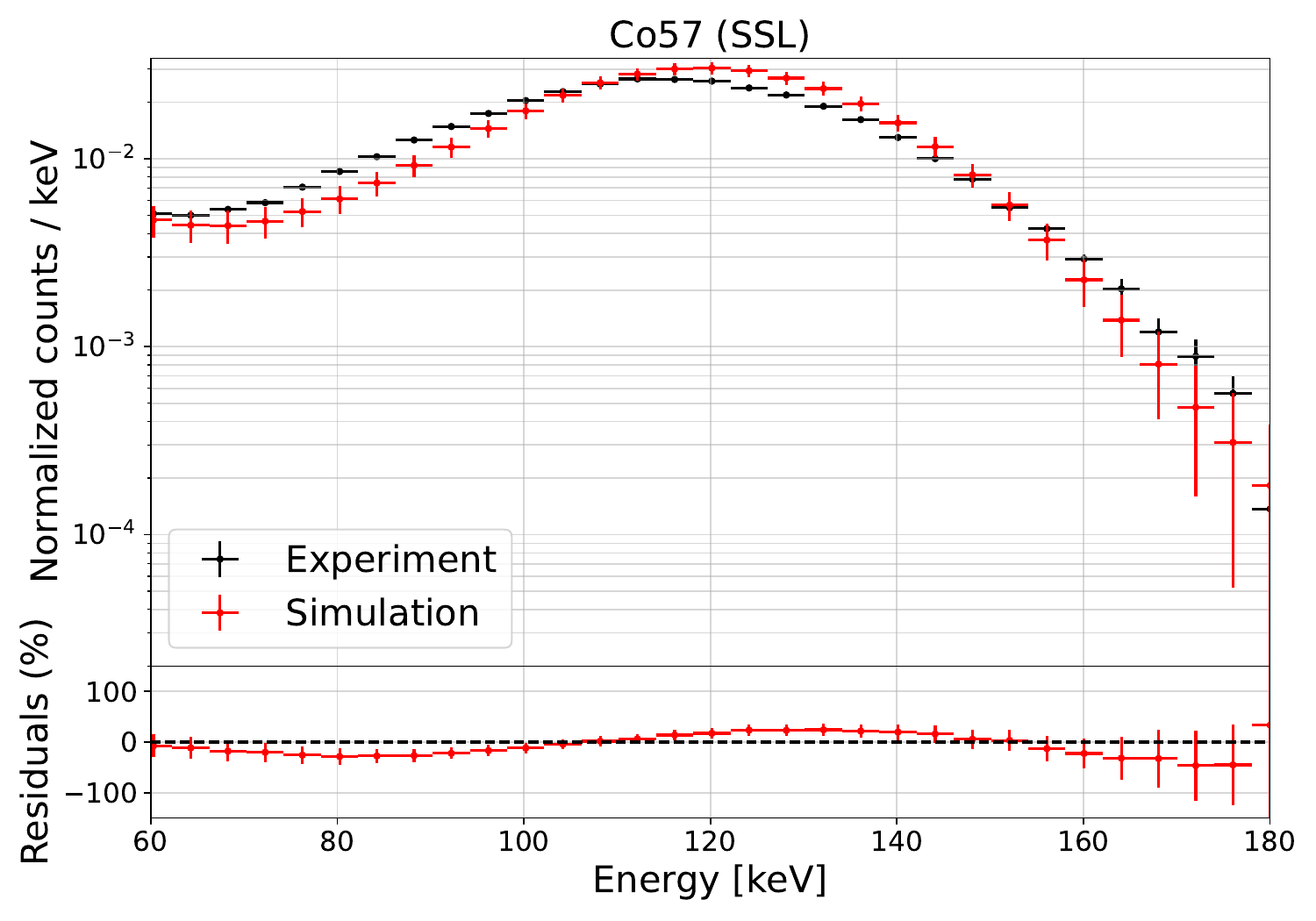}\includegraphics[width=0.42\textwidth]{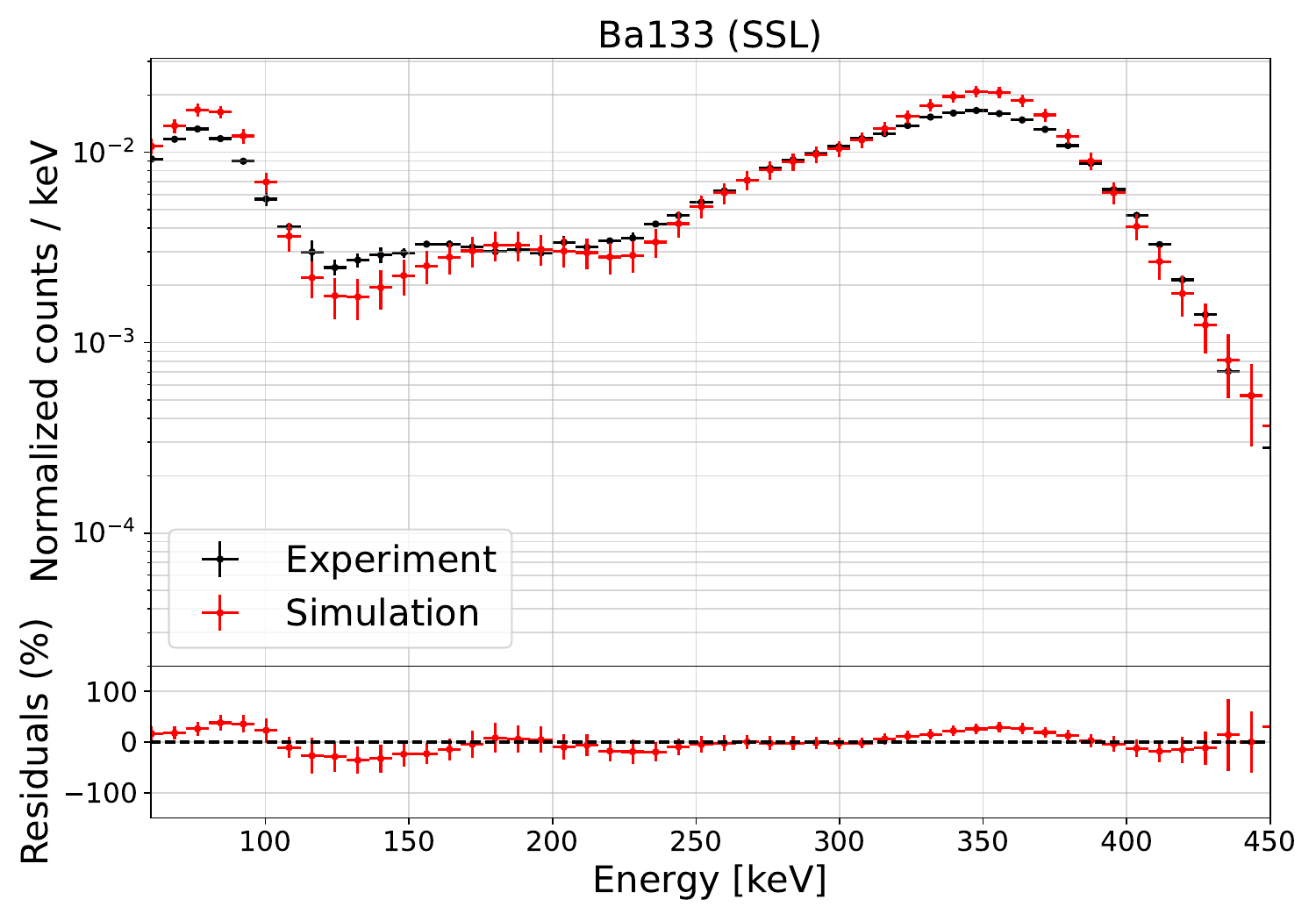}\\\includegraphics[width=0.42\textwidth]{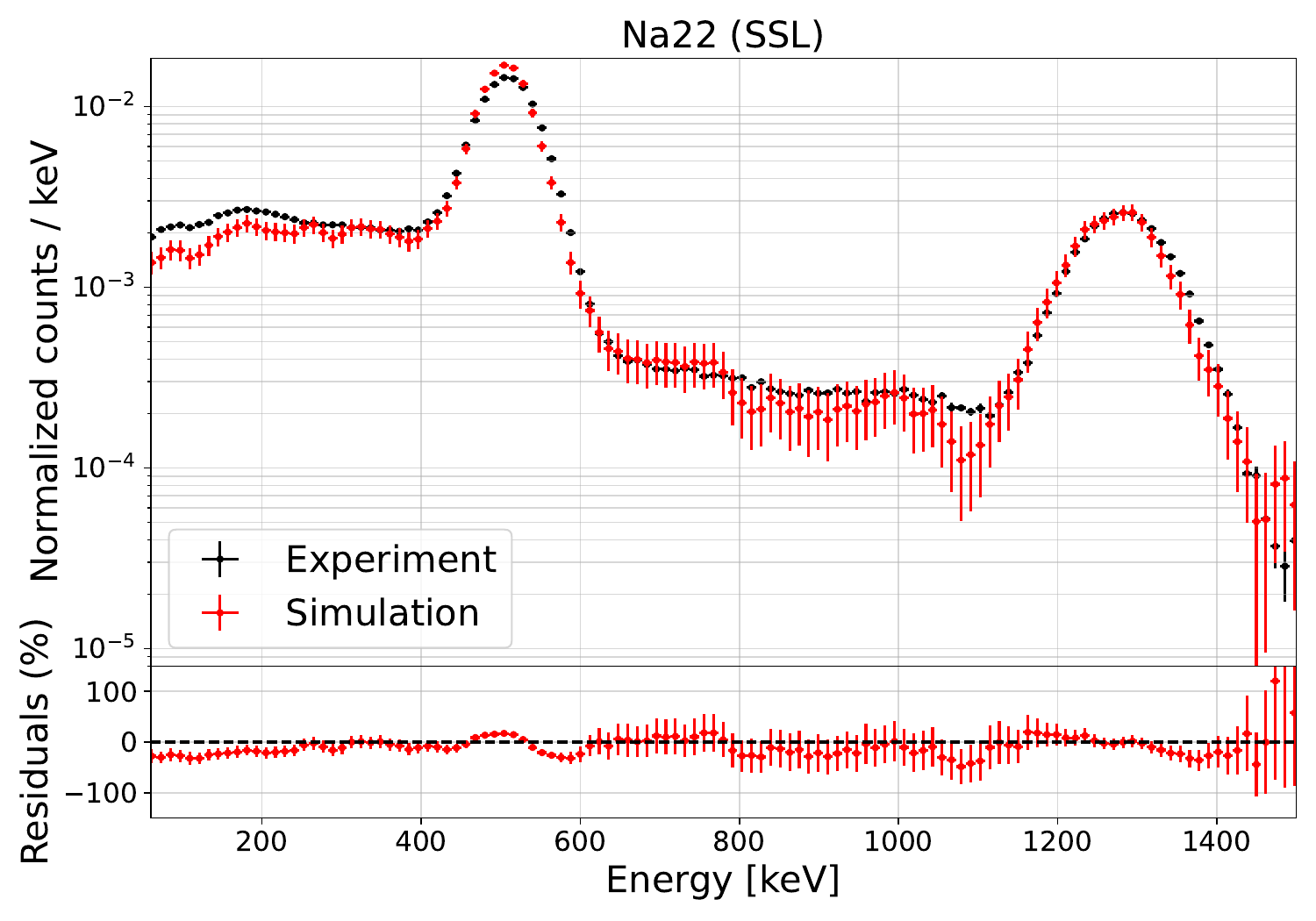}\includegraphics[width=0.42\textwidth]{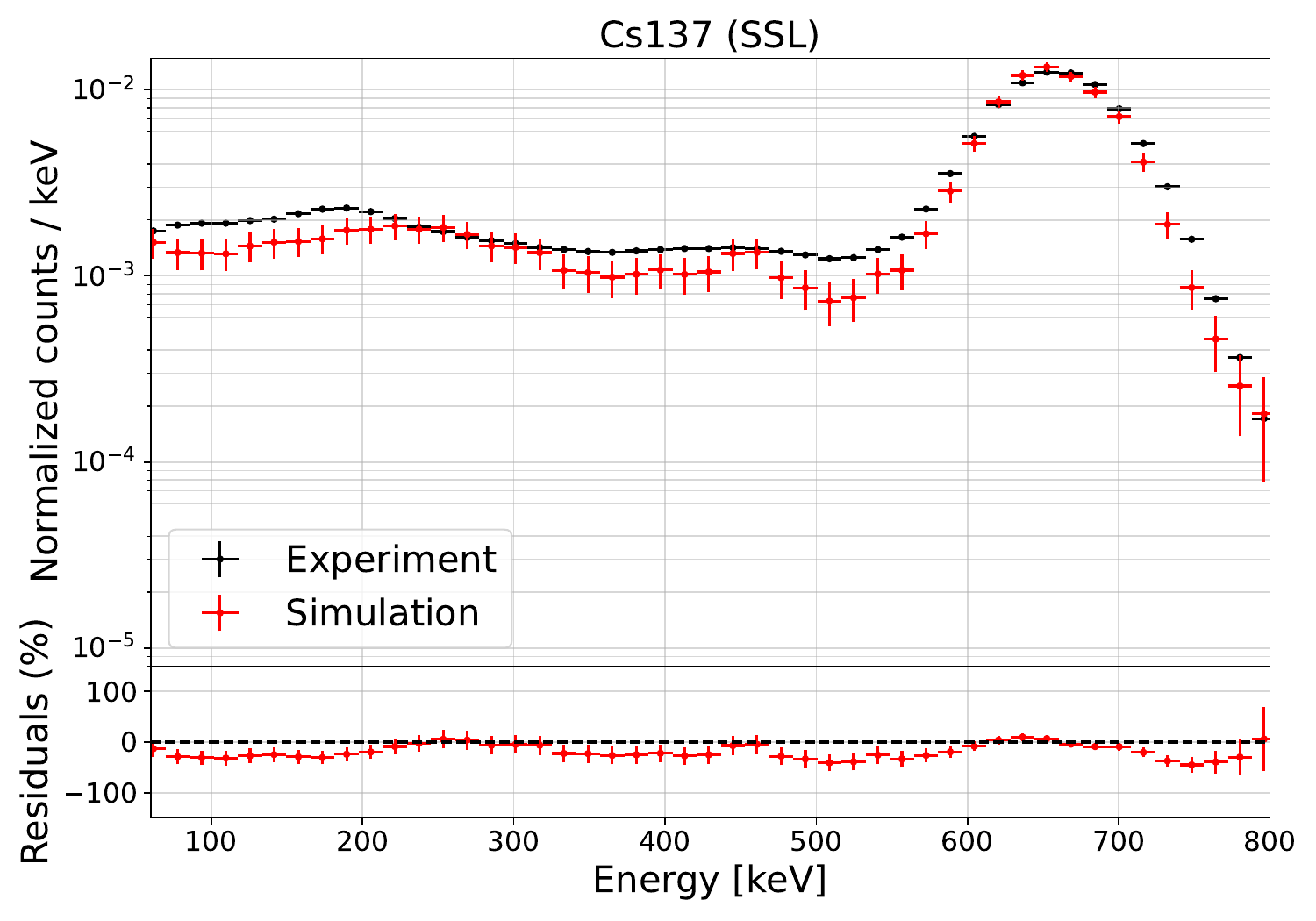}\\\includegraphics[width=0.42\textwidth]{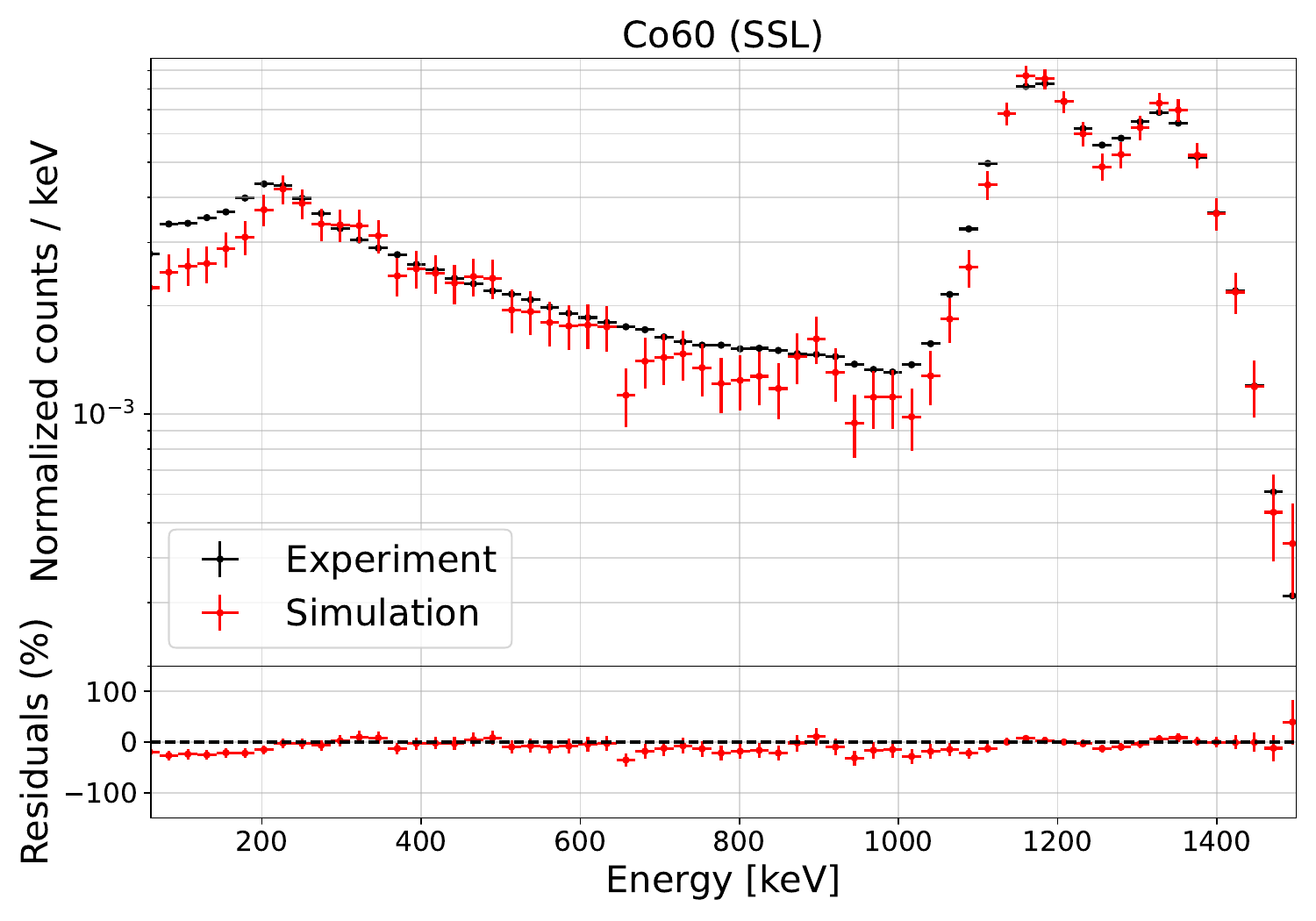}\includegraphics[width=0.42\textwidth]{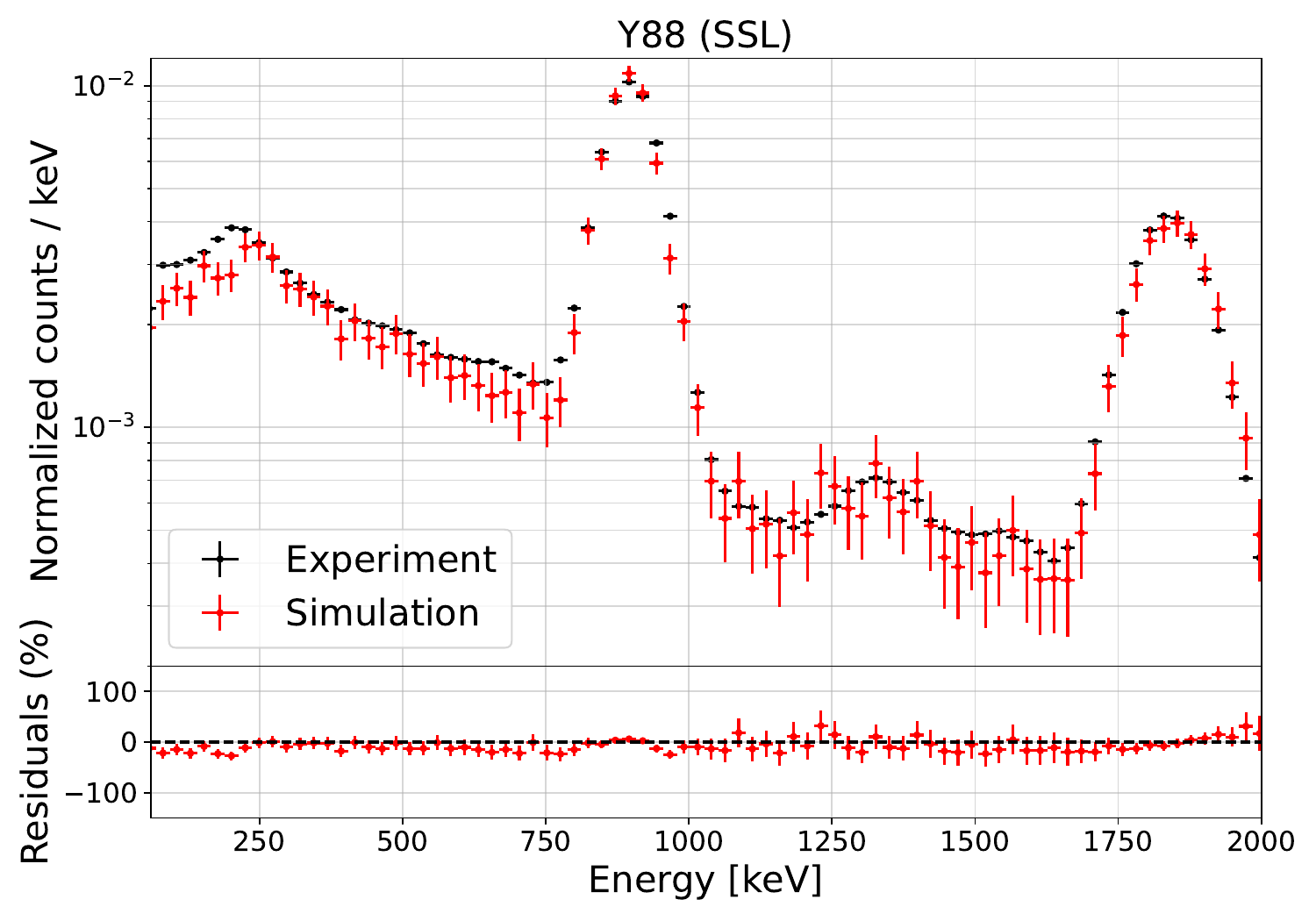}
\caption{Comparison between SSL experimental (black) and simulated (red) energy spectra for each radioactive source}\label{fig:comp_SSL}
\end{figure}

The simulation provides an opportunity to investigate the individual components contributing to the spectrum: the photopeak arises from photons depositing their entire energy in the BGO, while the low-energy tail results from Compton scattering of the primary photons within the BGO or the surrounding environment. At $E \gtrsim 300$ keV, the source holder is the dominant contribution, followed by the Compton in the BGO and the aluminum housing. At $E \sim 200$ keV, an important contribution arises from back-scattered photons from the table, while the lowest energy region is dominated by Compton scattering in the BGO. This continuum region of the spectrum is therefore highly dependent on the specific experimental setup, which can vary between calibration campaigns and with unknown density and composition of the surrounding materials, increasing the discrepancy with the simulation. For example, the table used in the SSL calibration is made of wood, a material with a wide range of densities \citep{wood}, introducing an inherent source of uncertainty. Additionally, the simulation does not account for all objects present in the laboratory, focusing instead on those immediately surrounding the BGO detectors. In Figs. \ref{fig:comp_NRL} and \ref{fig:comp_SSL} we present the comparison for all sources and calibration campaigns. The error bars in the simulated spectra are generally larger than those in the experimental ones, as the laboratory measurements have significantly higher counts per bin, resulting in lower statistical fluctuations. Also, the statistics vary among the sources due to differences in source activity and their distances from the BGO. The average discrepancy between simulation and experiment is $\sim$12\% in reproducing the photopeaks and $\sim$18\% in reproducing the Compton tails. In particular, all photopeaks are reproduced within the $2\sigma$ confidence level, except for the 511 keV line from the $^{22}$Na source. In this case, the experimental photopeaks from the NRL and SSL measurements are broader than the simulated ones, with the discrepancy reaching up to $3\sigma$. A likely explanation for this behavior is the Doppler broadening of 511 keV photons, an intrinsic effect arising from electron-positron annihilation which additionally degrades the experimental energy resolution \citep{DONASCIMENTO2005723}. Although this broadening is more visible in high-resolution detectors such as germanium detectors, it might still be contributing to the experimental BGO resolution and increasing the observed discrepancy with the simulation. Regarding the continuum regions, a larger discrepancy is observed in the low-energy portion of the spectra for SSL high-energy sources. In particular, the simulated counts are systematically lower than the laboratory ones. As previously discussed, this portion of the spectrum is strongly influenced by the experimental environment, suggesting that some components may be absent in the simulation. In contrast, the low-energy comparison shows better agreement with the NRL measurements.

\section{ACS correction matrix} \label{sec:dee}
Benchmarking the ACS simulations allows us to integrate realistic effects (energy resolution, electronic noise etc.) and reproduce the real data. This corresponds to tuning and developing a DEE module that applies all these corrections to simulated data. Currently, a dedicated DEE for the ACS is not available. Standard MEGAlib simulations consider only the true energy deposit from gamma-ray interactions in the BGO. In reality, the relationship between the energy deposit in the BGO and the energy actually measured is not straightforward. The energy deposit is first converted into
scintillation optical photons, a fraction of which is eventually detected by the SiPMs to produce the signal. Using the channel-energy relation obtained from the energy calibration, the signal is finally converted back into the measured energy. This entire process is intrinsically affected by the details of the scintillation process in the BGO, i.e. the generation of optical photons and their transport through the crystal until they are either absorbed or detected by the SiPMs. 

Two key parameters define the ACS performance: the optical light collection efficiency and the energy resolution. The former is related to the fraction of optical photons that reach the SiPMs and are therefore detected. This probability depends on several factors, including the BGO absorption length, the reflectivity of the layers wrapping the BGO, and the SiPM PDE. Moreover, the scintillation light collection efficiency varies throughout the crystal, as photons generated at different locations have different probabilities of being detected, depending on their proximity to the SiPMs. This non-uniformity has also an impact on the energy threshold, which tends to be higher where the light collection efficiency is lower, and vice versa. The energy resolution, as described in Eq. \eqref{eq:fwhm}, is determined by both the statistical fluctuations in the number of optical photons and the spatial inhomogeneity of the BGO light collection efficiency. The explicit simulation of the optical physics requires a large computational effort\footnote{Simulating 100000 photons at 511 keV requires approximately 250 core-hour with the inclusion of the optical physics. In contrast, the same simulation without scintillation is completed in less than a minute on a single core.} due to the massive number of optical photons, making most of the simulations unfeasible. 

To address this, we perform dedicated Geant4 simulations with the inclusion of the optical physics to encode the optical light collection efficiency, energy resolution and their spatial distribution into a correction matrix, to be finally applied to ACS data in standard MEGAlib simulations. This provides more realistic ACS simulations that better reflect the real behavior, without explicitly simulating the scintillation process in the shields. We present here the characterization of the optical light collection efficiency, energy resolution and their positional dependence for a BGO X-crystal with dimension $198\times 118\times 23$ mm. We encode this information in a correction matrix to be applied to ACS data in standard MEGAlib simulations. We used the same simulation parameters selected from the benchmarking (Sec. \ref{sec:results}), as well as a BGO absorption length of 5 m and an electronic noise of 21 keV. We conducted the simulations using the Leonardo supercomputer at CINECA, Bologna, granted through ICSC resources.

\subsection{Optical light collection and energy resolution map}
\begin{figure}
\centering
\includegraphics[width=0.7\textwidth]{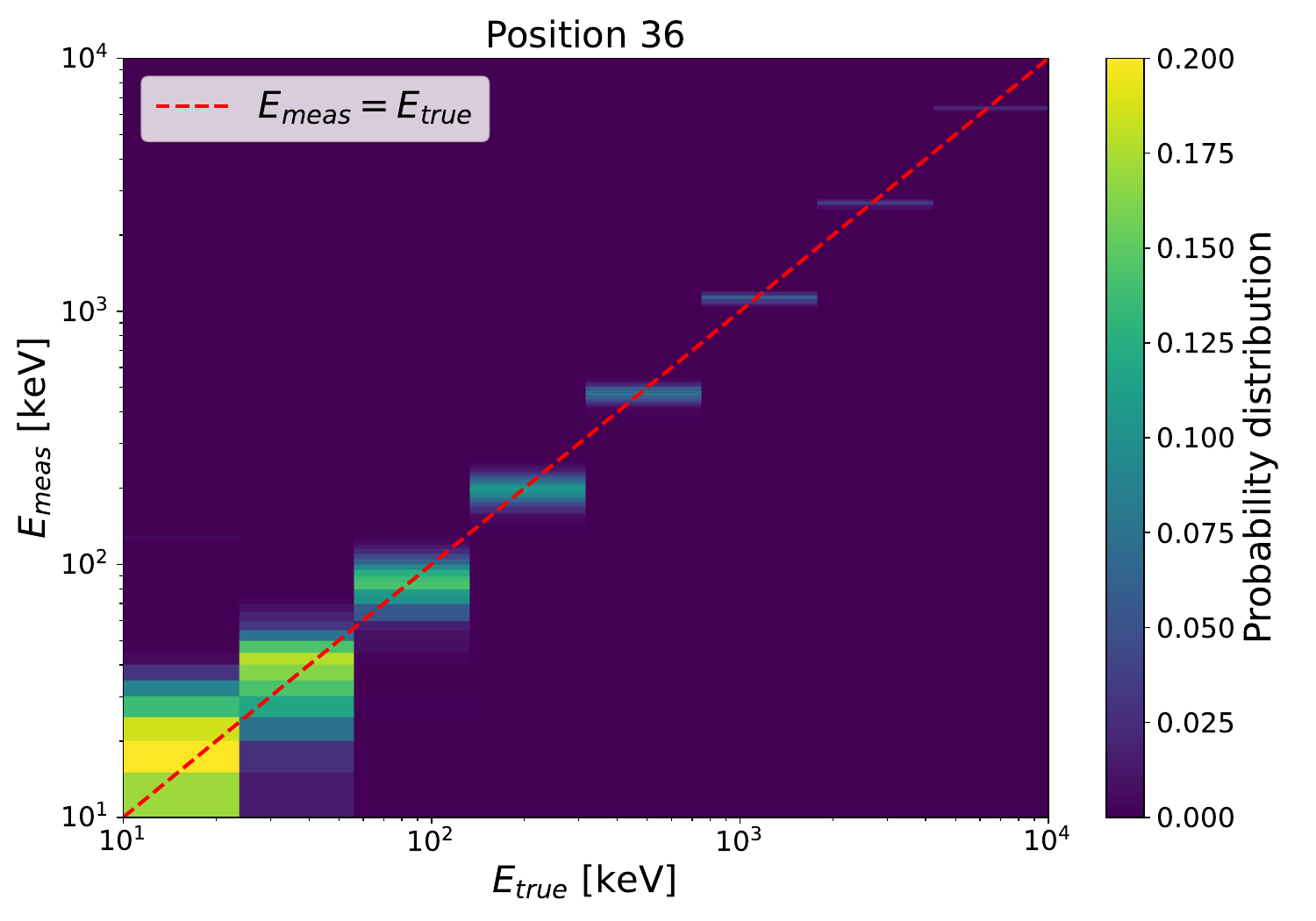}
\caption{Response matrix ($E_\text{meas}$ against $E_\text{true}$) for the BGO X-crystal for a particular spatial bin}\label{fig:response_matrix}
\end{figure}

\begin{figure}
\centering
\includegraphics[width=\textwidth]{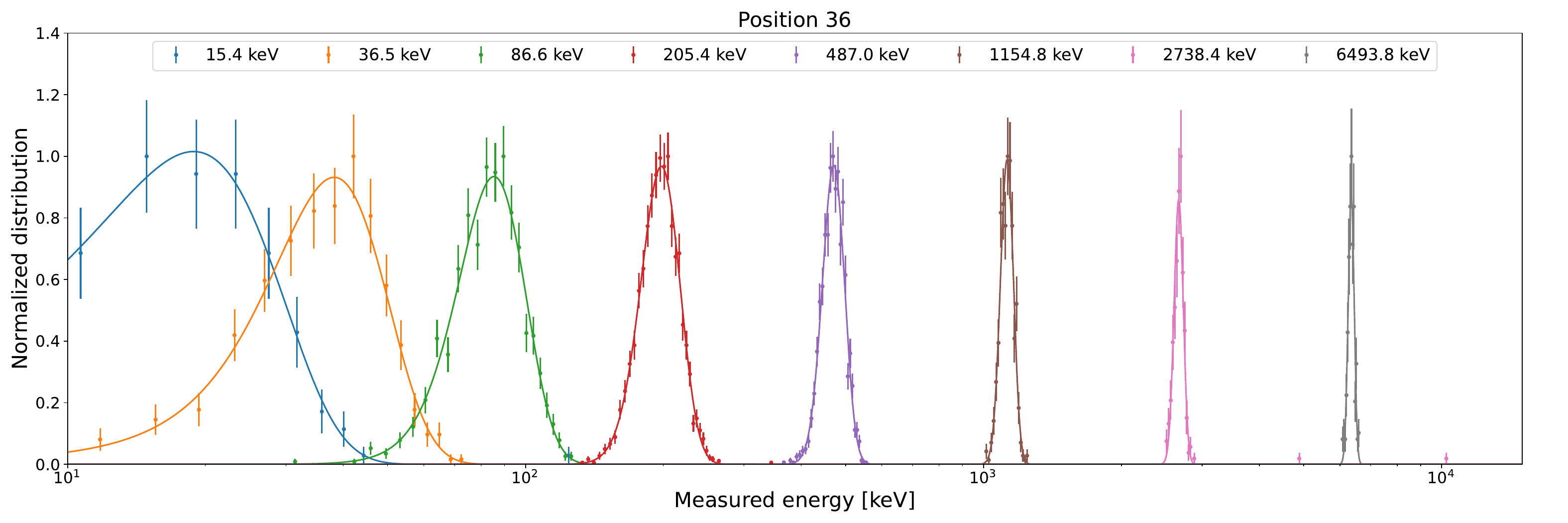}
\caption{Simulated normalized distributions of the deviations of $E_\text{meas}$ from $E_\text{true}$ for each energy bin, for a particular spatial bin taken as example}\label{fig:gaussians_DEE}
\end{figure}

We simulate photons from 10 keV to 10 MeV hitting perpendicularly the $198\times 118$ mm face of the BGO and, for each photon, we record the energy deposit ($E_\text{true}$), the 2-dimensional position of the deposit ($x$ and $y$) and the corresponding number of optical photons detected by the SiPMs\footnote{For each high-energy photon interacting with the crystal, multiple energy deposits can occur at different locations within the crystal, and each energy deposit generates its own set of optical photons. In this case, we treat each energy deposit independently, each of which is associated with its own $x$-$y$ position and number of detected optical photons.}. We then use the energy calibration obtained from the SSL simulations (left plot of Fig. \ref{fig:energy_calibration}) to convert the number of detected optical photons into measured energies ($E_\text{meas}$). At the end of this process, we obtain \{$E_\text{true}$, $x$, $y$, $E_\text{meas}$\}. We then perform a spatial $x$-$y$ binning of the $198\times 118$ mm BGO surface into $20\times 10$ bins. Within each spatial bin, $E_\text{true}$ is further divided into 8 logarithmic energy bins spanning from 10 keV to 10 MeV. For each bin, we analyze the distribution of the differences between $E_\text{meas}$ and $E_\text{true}$, shifted to the corresponding energy bin center, and fit them with a Gaussian model. In Fig. \ref{fig:response_matrix} we show the measured energy distribution against the true energy deposit for a particular spatial bin. In Fig. \ref{fig:gaussians_DEE} we show, for the same spatial bin, the same distributions with the corresponding Gaussian fitting.

\begin{figure}
\centering
\includegraphics[width=0.5\textwidth]{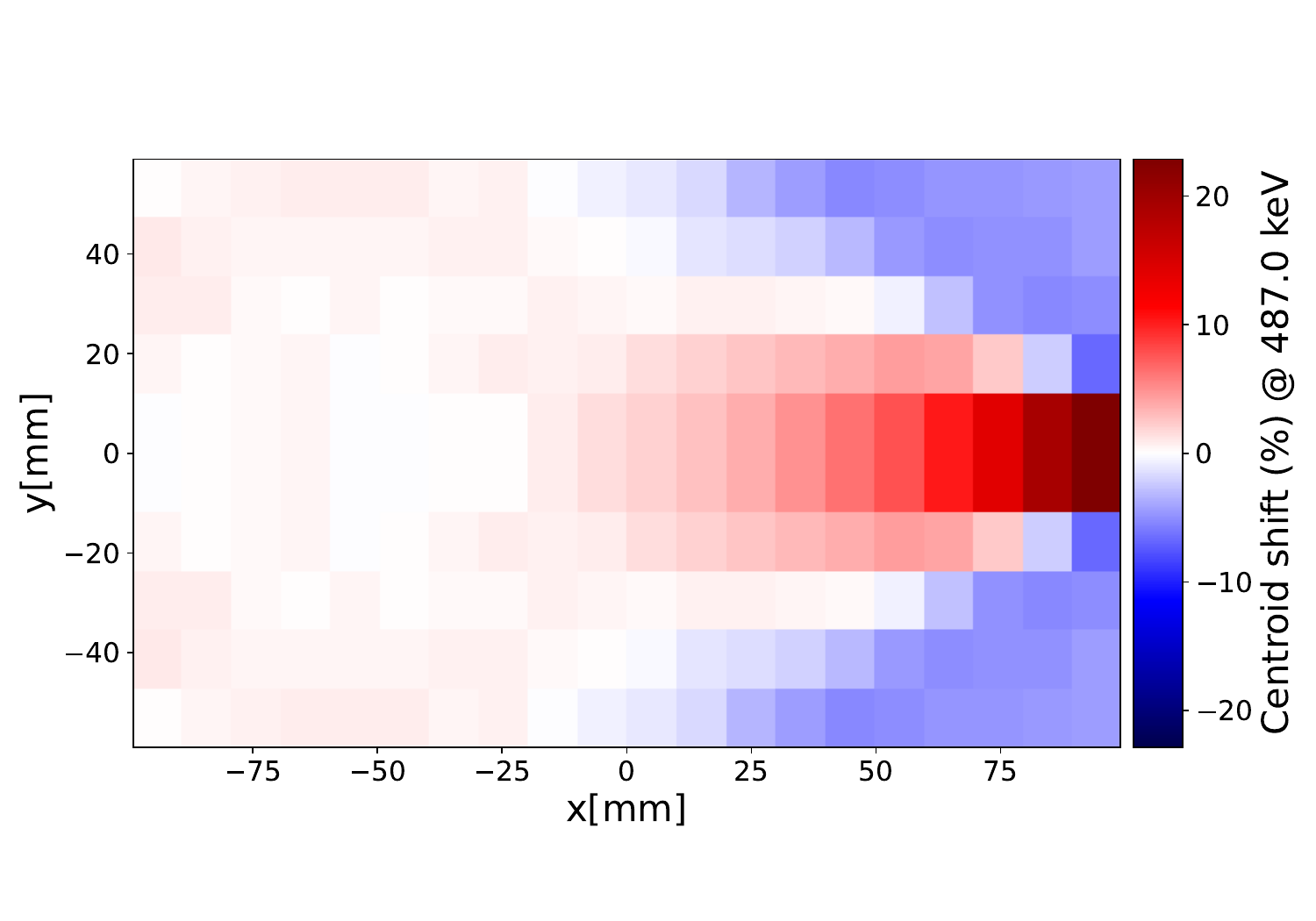}\includegraphics[width=0.5\textwidth]{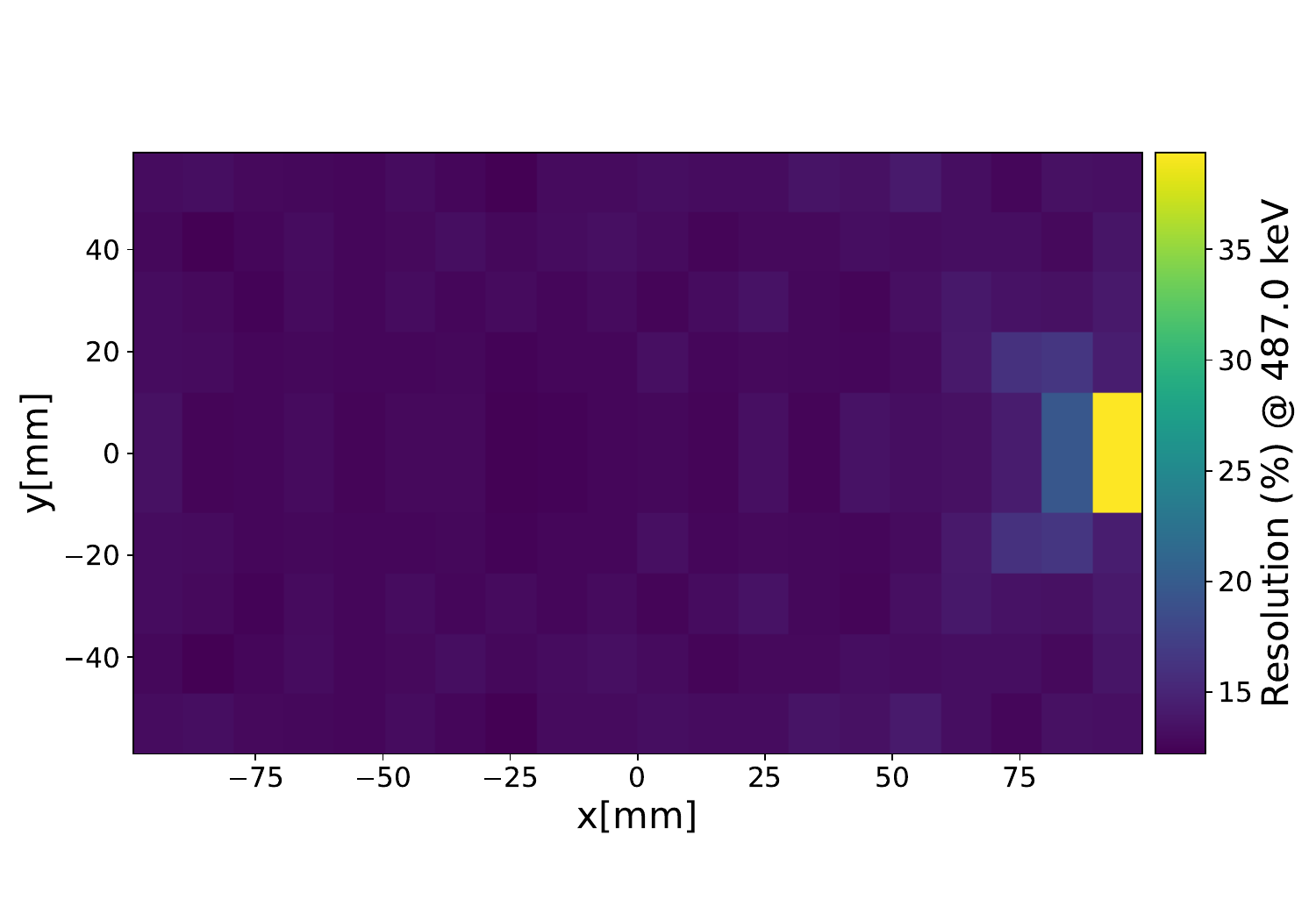}
\caption{Centroid shift map (left) and energy resolution map (right) at 487 keV}\label{fig:centroid_resolution_maps}
\end{figure}

The fit of the distributions provides, for each spatial and energy bin, two key parameters: the centroids ($\bar E_\text{meas}$) and the FWHMs. These two quantities are related to two fundamental properties of the BGO performance: the optical light collection efficiency and the energy resolution, respectively. The relationship between FWHM and energy resolution is straightforward: the FWHM quantifies the precision of the true energy deposit estimation. Regarding the centroids, if for a particular spatial and energy bin we have an enhanced light collection efficiency, the mean measured energy $E_\text{meas}$ would systematically be larger than the true energy $E_\text{true}$ and the centroid of the distribution would be shifted towards higher energies. This arises from using a single average calibration to convert the number of optical photons into energy, which fails to account for the position-dependent variation in the relationship between detected optical photons and energy. This introduces a bias in the energy determination that must be accounted for in simulations to reflect the real ACS performance. In Fig. \ref{fig:centroid_resolution_maps} we show the relative centroid shift (with respect to the bin center) and the energy resolution map for the energy bin at 487 keV (the closest one to 511 keV). The SiPMs are centered on the right side of the map. The centroid shift distribution is affected by the SiPMs position: it is higher for the central bins in front of the SiPMs, reaching a maximum enhancement of $\sim$22\%, suppressed by $\sim$5\% in the corners adjacent to the SiPMs, and nearly uniform in the region distant from the SiPMs. The energy resolution is instead highly uniform, except for the two bins right in front of the SiPMs. In these bins, the detector response exhibits significant variability depending on the exact location of the energy deposition. Due to the close proximity between the interaction point and the SiPMs, even small displacements within a single bin can lead to relatively large variations in the light transmission.

\begin{figure}
\centering
\includegraphics[width=0.5\textwidth]{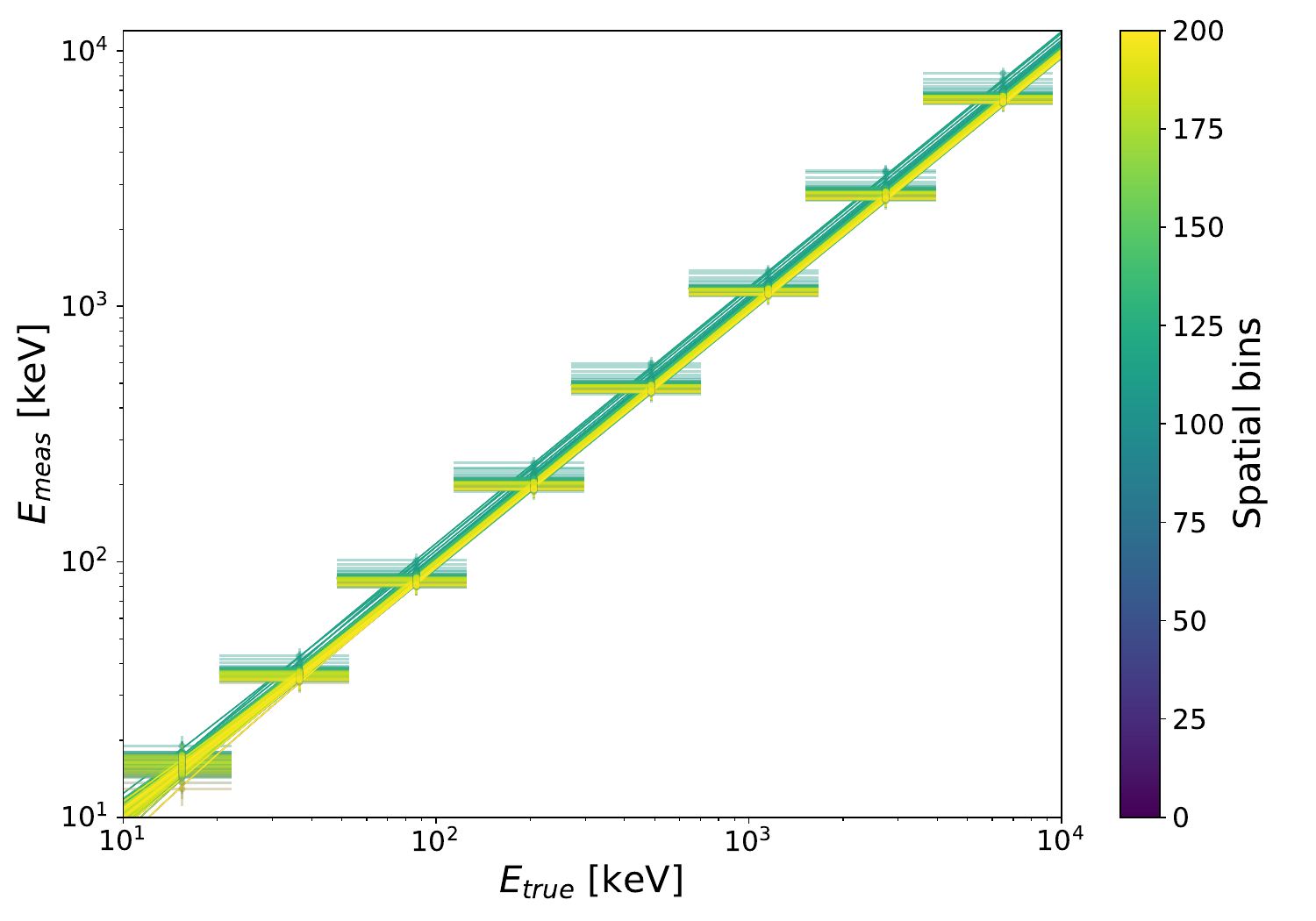}\includegraphics[width=0.5\textwidth]{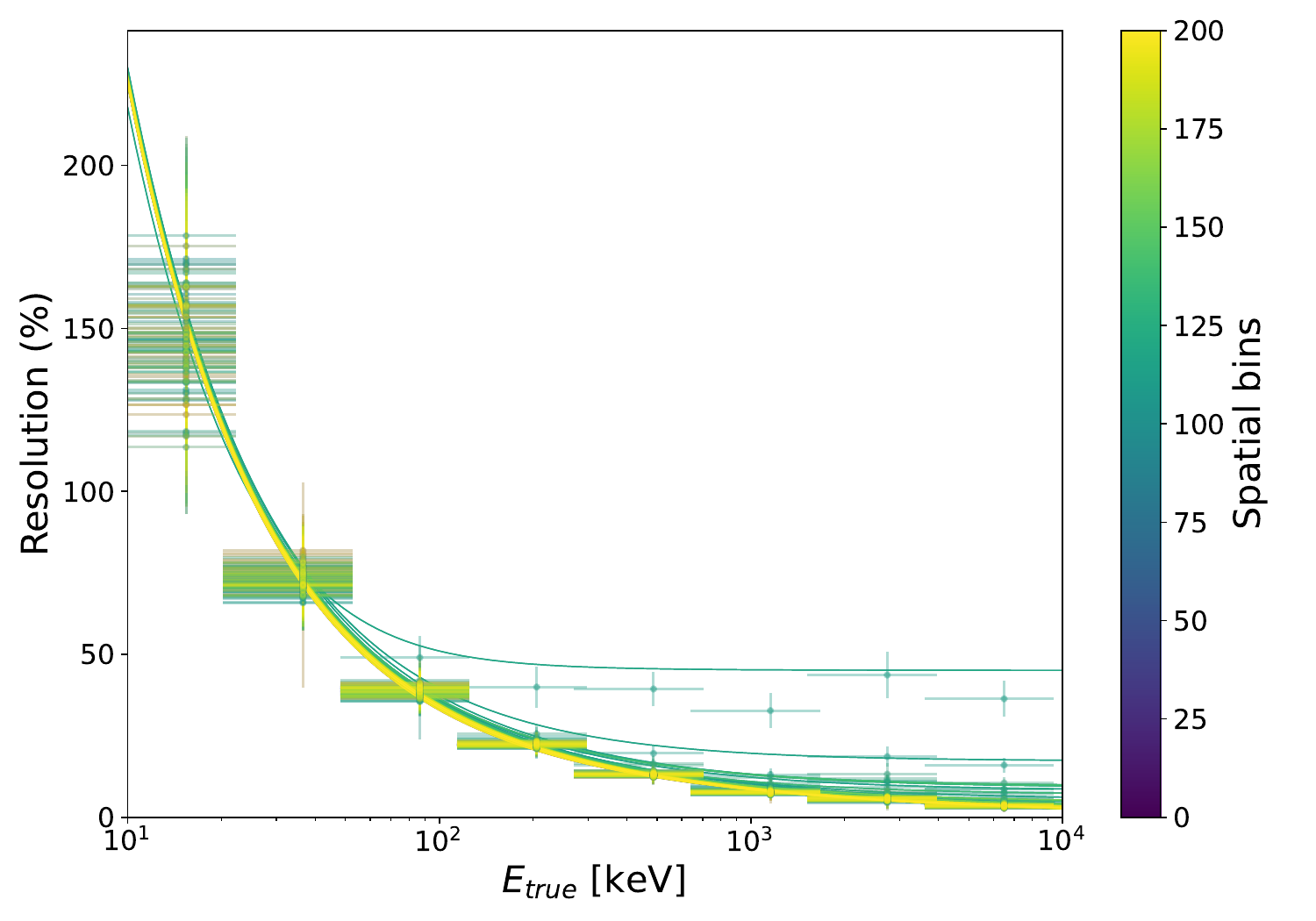}
\caption{Centroids (left) and energy resolutions (right) as a function of the energy, color coded with the corresponding spatial bin, and best-fit models for each distribution. The outlier curve for the energy resolution, visible in the right panel, corresponds to the spatial bins in front of the SiPMs, which have a large value of $c_{xy}$ due to the strong variability of the detection probability within the bin}\label{fig:centroid_resolution_fits}
\end{figure}

In Fig. \ref{fig:centroid_resolution_fits} we show the centroids and energy resolutions as a function of the energy bins, color coded with the $x$-$y$ bin. We fit the two relations with the following models:
\begin{equation}
    \bar E_\text{meas} = m_{xy} * E_\text{true} + q_{xy}\,\,, \label{eq:centroid_DEE}
\end{equation}
\begin{equation}
    \text{FWHM} = \sqrt{a_{xy}^2 + b_{xy}^2\,E_\text{true} + c_{xy}^2\,E_\text{true}^2}\,\,, \label{eq:fwhm_DEE}
\end{equation}
with the second equation having the same description as Eq. \eqref{eq:fwhm}. The best-fit provides, for each $x$-$y$ bin, a set of five parameters \{$m_{xy}$, $q_{xy}$, $a_{xy}$, $b_{xy}$, $c_{xy}$\} that encodes the whole BGO performance and the impact of the optical physics. These parameters allow us to take as input the true energy deposit $E_\text{true}$ and its $x$-$y$ position, and give as output a new measured energy $E_\text{meas}$ drawn from a Gaussian distribution with centroid $\bar E_\text{meas}$ and FWHM as given by Eqs. \ref{eq:centroid_DEE} and \ref{eq:fwhm_DEE}.

\begin{figure}
\centering
\includegraphics[width=0.5\textwidth]{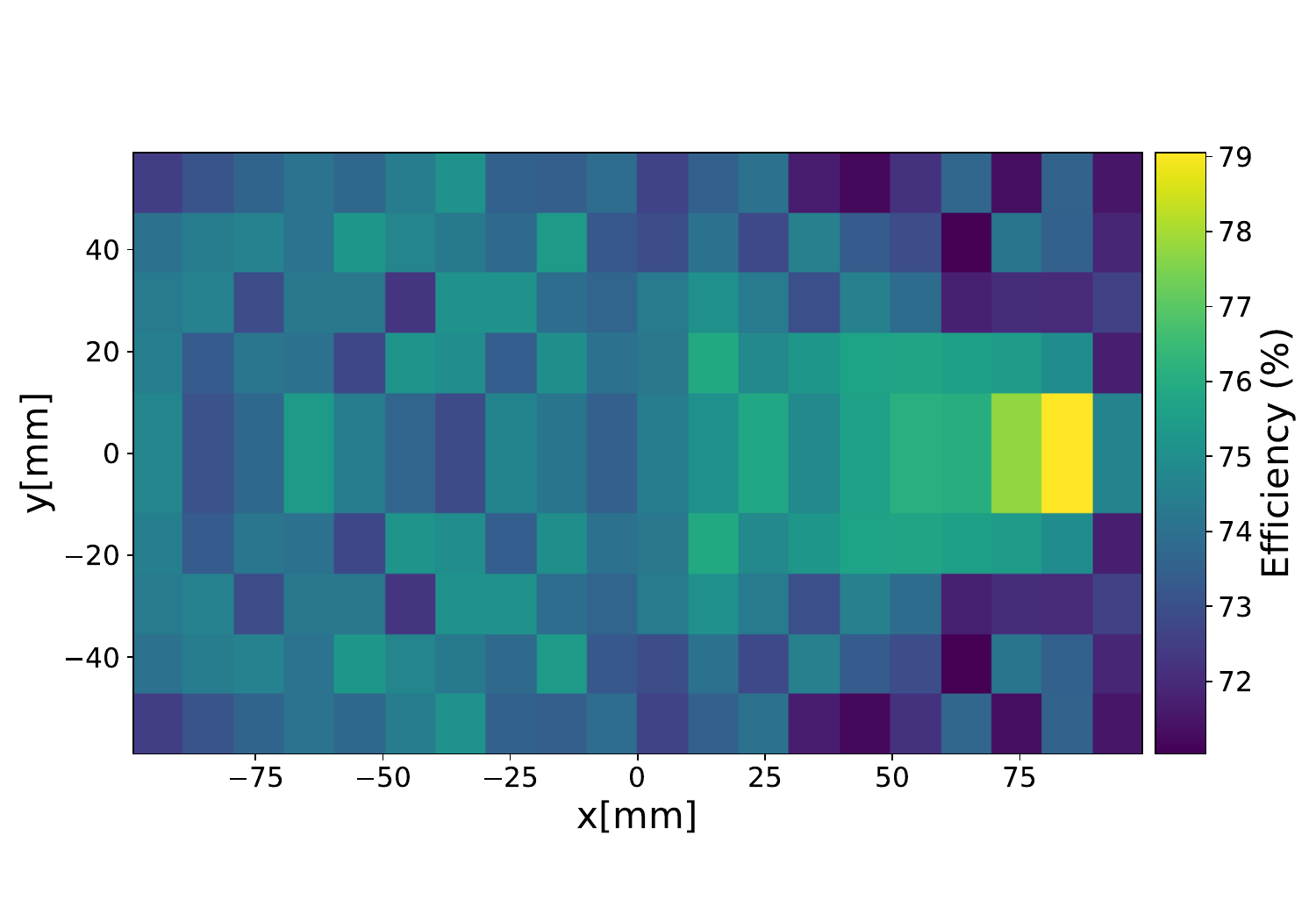}\includegraphics[width=0.5\textwidth]{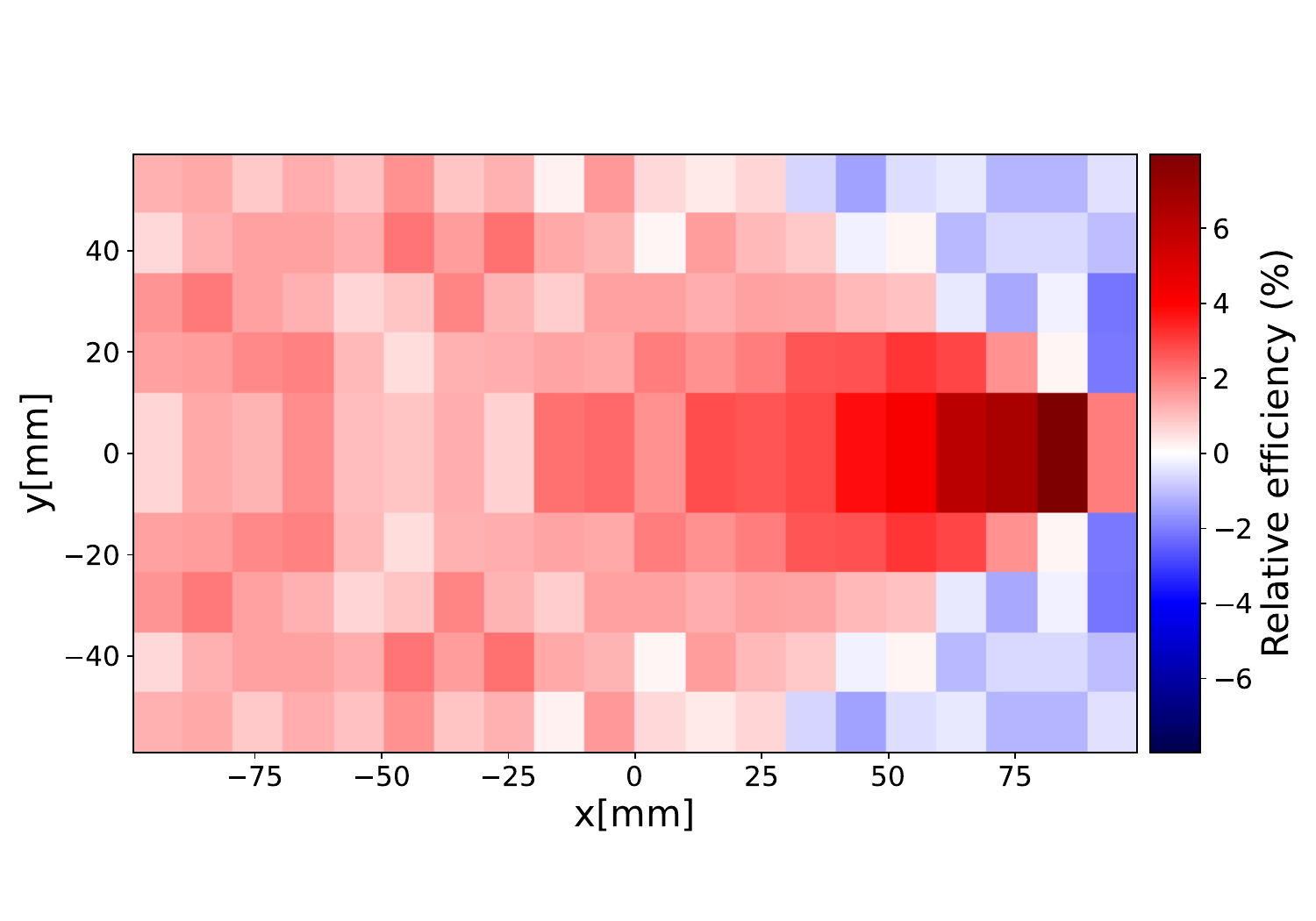}
\caption{Quantum detection efficiency map (left) and relative quantum detection efficiency map (right)}\label{fig:efficiency_maps}
\end{figure}

This correction also has an impact on the overall gamma-ray quantum detection efficiency of the BGO. The quantum detection efficiency is defined as the fraction of detected high energy photons (i.e. with $E_\text{meas} > 80$ keV) out of the total incident photons. We simulate a flat energetic spectrum of photons with an initial energy $< 300$ keV hitting the BGO and calculate the detection efficiency for each $x$-$y$ bin. The detection efficiency map is shown in the left plot of Fig. \ref{fig:efficiency_maps}, while on the right plot we show the relative difference between the evaluated efficiency and the same efficiency evaluated without the correction, to underline the impact of the optical physics. We observe a similar distribution obtained for the centroid shift in Fig. \ref{fig:centroid_resolution_maps}: the efficiency is nearly uniform far from the SiPMs, larger when approaching the SiPMs, and suppressed in the corner adjacent to the SiPMs. The effect of implementing the optical physics correction is to enhance the efficiency by at most $\sim$8\% close to the SiPMs, and suppress it by at most $\sim$2\% in the adjacent corners.

All this analysis requires a large number of simulated photons in order to populate every energy and spatial bin and obtain sufficient statistics to adequately perform the fit of the centroids and FWHM. We plan to execute an additional simulation campaign to refine the results presented in this paper, to extend the characterization of the BGO performance to higher energies and repeat the same study to the bottom crystals. A global characterization of the ACS performance will pave the way for the development of a dedicated DEE module for the BGO.

\section{Bottom ACS simulation} \label{sec:bottom}
Benchmarked simulations represent a powerful and flexible tool to predict the performance of the ACS or investigate different design solutions, without necessarily performing explicit measurements in the laboratory. We used the simulations validated against the calibration data presented in the previous sections to investigate the performance of the BGO bottom crystals (Z-crystals), in particular to estimate their energy threshold and resolution. The main difference of the bottom BGO crystals with respect to the side panels is their smaller dimensions. The bottom ACS contains 10 BGO crystals of varying dimensions: four measuring $198 \times 62.5 \times 24$ mm (Z-crystal A), four measuring $171 \times 64.5 \times 24$ mm (Z-crystal B) and two measuring $198 \times 80.5 \times 24$ mm (Z-crystal C).

\begin{figure}[h]
\centering
\includegraphics[width=\textwidth]{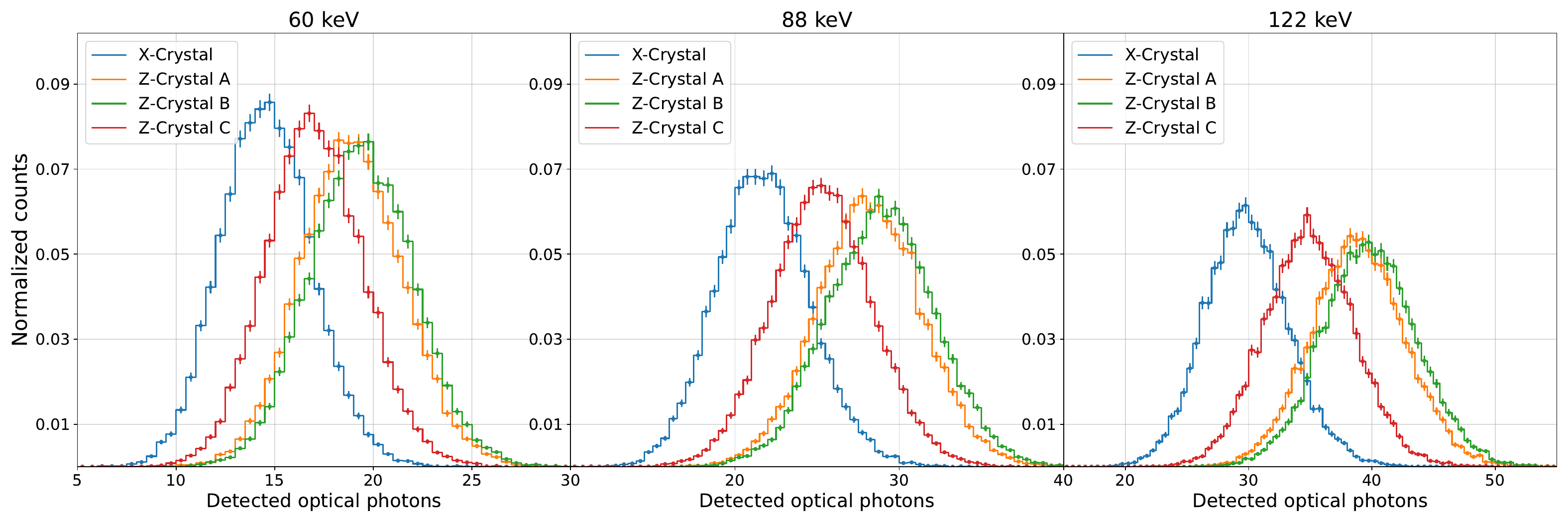}
\caption{Simulated photopeak distributions detected by the BGO X-crystal and Z-crystals, for 60 keV (left), 88 keV (middle) and 122 keV (right) source photons}\label{fig:photopeaks_bottom}
\end{figure}

The crystal size has a potential impact on the overall optical light collection efficiency of the SiPMs and, consequently, on the energy threshold. In fact, in a smaller crystal the active detection area corresponding to the SiPMs occupies a larger fraction of the total BGO surface, enhancing the probability for the optical photons to reach the SiPMs before being absorbed. Hence, if for the same energy deposit the average number of detected optical photons is larger, the energy threshold would be accordingly shifted to lower energy. To quantify this effect, we simulate 3 parallel monochromatic beams at 60, 88 and 122 keV, hitting orthogonally the largest surface of each type of Z-crystal, and one of the X-wall crystals, that have a dimension of $194 \times 118 \times 23$ mm. Then, for each energy and crystal, we collect the number of detected optical photons in the SiPMs by fitting with a Gaussian the photopeaks. The photopeaks, shown in Fig. \ref{fig:photopeaks_bottom}, show indeed that the smaller crystals have a larger number of detected optical photons, caused by an enhanced optical light collection efficiency of the SiPMs. 

\begin{figure}
\centering
\includegraphics[width=0.5\textwidth]{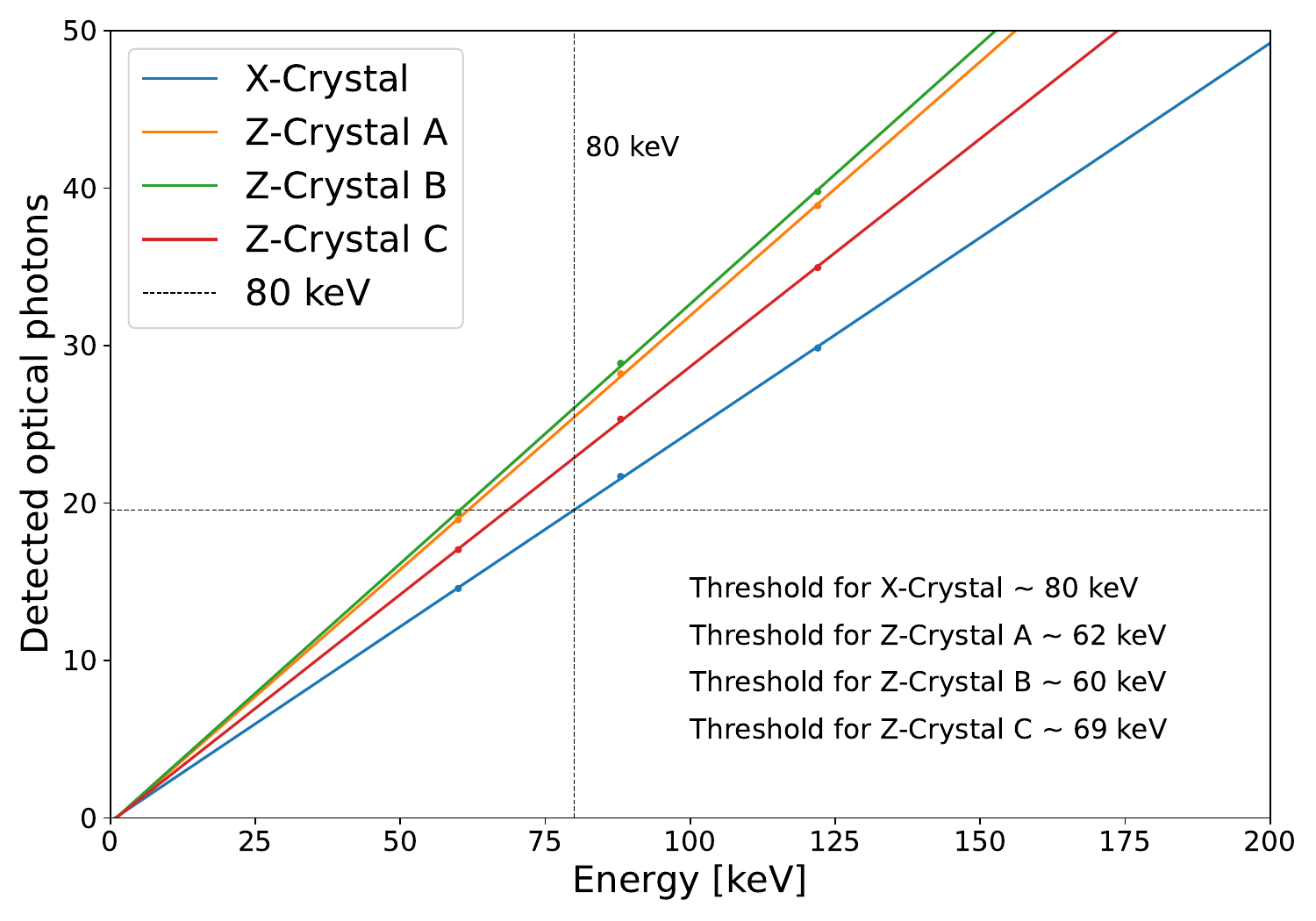}\includegraphics[width=0.5\textwidth]{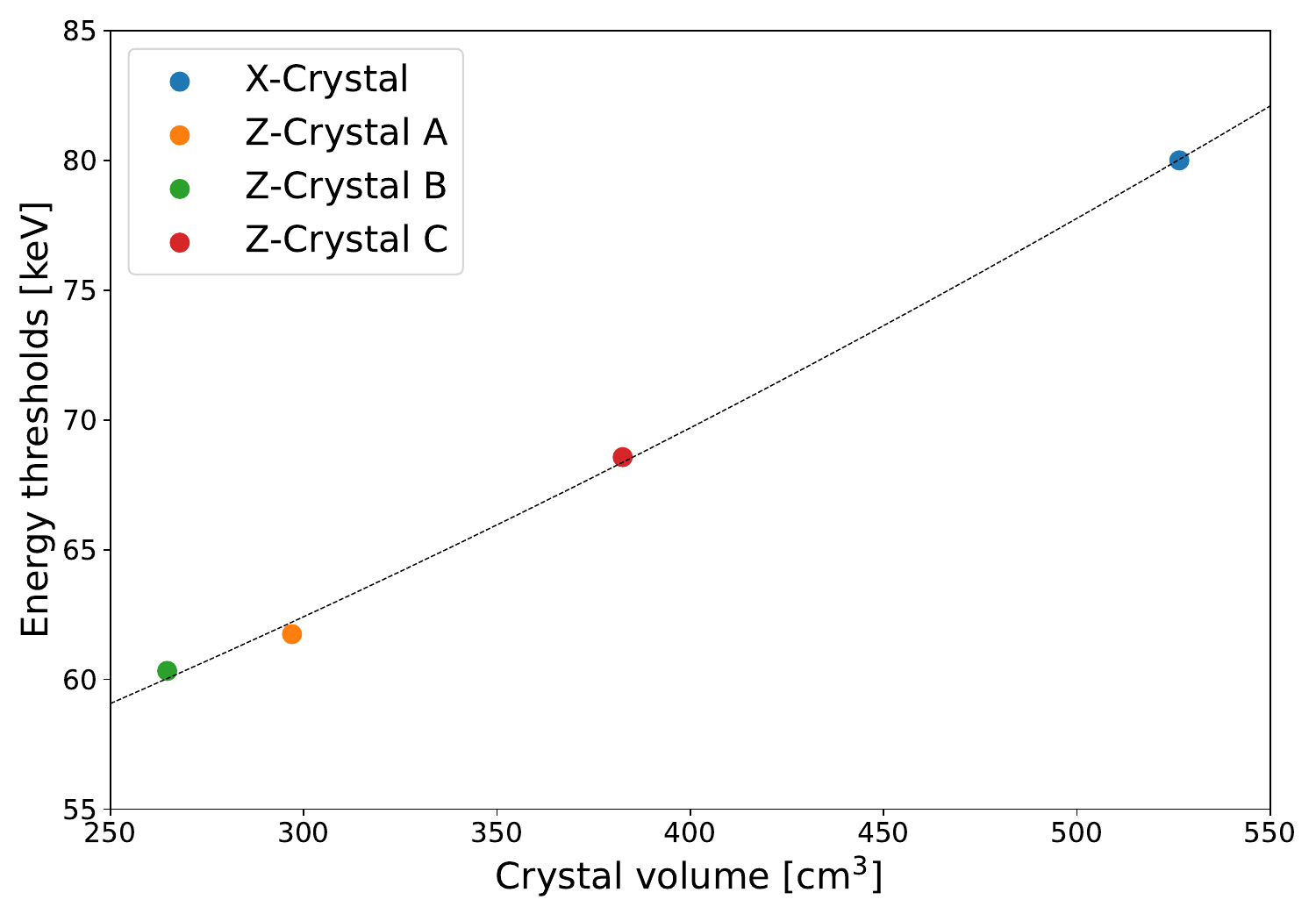}
\caption{On the left, energy calibrations for the lateral and bottom crystals, with highlighted the optical photons emitted at the 80 keV threshold in the lateral crystal. On the right, energy threshold as a function of the crystal volume}\label{fig:bottom_threshold}
\end{figure}

To estimate the impact on the energy threshold, we perform an energy calibration by taking the peak values from the Gaussian fitting of the photopeaks and associating to each of them the corresponding source energy, in a similar fashion as described in Sec. \ref{sec:ene_cal_exp}. From the linear fit of the data points, we obtain the photon-energy relation for each crystal, as shown in the left plot of Fig. \ref{fig:bottom_threshold}. Smaller crystals exhibit steeper slopes, as a result of an enhanced optical light collection efficiency. This means that, for smaller crystals, the same number of detected optical photons corresponds to a lower energy or, equivalently, the same amount of energy deposit produces a larger signal. From the photon-energy relation of the X-crystal, we derive the number of optical photons that corresponds to 80 keV, which is the goal for the ACS threshold. We then use the other photon-energy relations to convert back this number into energy for each Z-crystal. We find an energy of $\sim$62 keV for the Z-crystal A, $\sim$60 keV for the Z-crystal B and $\sim$69 keV for the Z-crystal C. These are the energy thresholds expected for the Z-crystals, if we assume an energy threshold of 80 keV for the X-crystal. In terms of relative variation, we have a reduction of $\sim$22\%, $\sim$25\% and $\sim$14\% for the Z-crystal A, Z-crystal B and Z-crystal C, respectively. In the right plot of Fig. \ref{fig:bottom_threshold} we show the energy thresholds as a function of the crystal volume. The trend demonstrates that the energy threshold progressively decreases with the crystal volume, as we indeed expected. This is also a useful relation to quickly estimate the energy threshold of a BGO crystal with arbitrary dimensions, assuming the same readout system.

Having a larger number of detected optical photons for smaller crystals has an impact also on the energy resolution. In fact, the FWHM increases with the energy, as described in Eq. \eqref{eq:resolution}, and is affected by the electronic noise (constant term), statistical fluctuations of the detected optical photons ($\propto \sqrt{E})$ and response inhomogeneity ($\propto E)$. We can assume the electronic noise to be the same for all the crystals, as it does not depend on the BGO geometry. However, for the same energy we have more detected optical photons for smaller crystals, resulting in lower statistical fluctuations. Moreover, for a smaller BGO we expect a smaller contribution in the resolution from the response inhomogeneity with respect to a larger crystal, where the increased size introduces larger variation in the optical photon collection. After converting the photopeaks in energy using the corresponding photon-energy relation for each crystal, we fit the obtained distributions with a Gaussian to compute the FWHM and, consequently, the energy resolution. Among the three investigated energies, we obtain an average improvement in the energy resolution of $\sim$8\% for Z-crystal A and Z-crystal B, and of $\sim$5\% for Z-crystal C.

\section{Conclusions}
In this paper, we present the characterization of the ACS energy resolution and optical light collection efficiency as a foundation for the development of an ACS DEE. For this purpose, we use calibration data of the ACS EM X-wall to benchmark the simulations and fine-tune the optical parameters for the scintillation physics. The calibration campaigns, one conducted at NRL and the other at SSL, use a set of radioactive sources to measure the corresponding energy spectra detected by the SiPMs and, with collimated measurements, the response at several positions on the BGO surface. The spatial response distribution, measured at 60 keV and 662 keV, demonstrates a high degree of homogeneity. An enhancement of approximately 8\% relative to the average response is observed directly in front of the SiPMs, while a suppression of about 4\% is noted at 60 keV in regions far from the SiPMs. Additionally, the experimental spectra exhibit an energy resolution of approximately 55-60\% at 60 keV, $\sim$16\% at 511 keV and $\sim$8.5\% at 1836 keV. We use Geant4 simulations to model the experimental setup of the calibration campaigns and replicate the corresponding measurements. We reproduce the spatial distribution of the response achieved with collimated sources, with a maximum discrepancy of 10\% for the position in front of the SiPMs. Using uncollimated sources, we reproduce the energy resolution with a maximum discrepancy of 20\% and the energy spectra for each source, achieving an average discrepancy of 12\% in the photopeaks and 18\% in the Compton tails.

We present the characterization of the optical light collection efficiency and energy resolution across the surface of a $198 \times 118 \times 23$ mm BGO crystal up to 10 MeV, obtained after extensive simulations with the inclusion of the optical physics. The integration of the optical physics influences the BGO detection efficiency, with a maximum enhancement of 8\% and a suppression up to 2\%, depending on the interaction position of the gamma photons. 
We further utilize these benchmarked simulations to investigate the impact of the smaller dimensions of the bottom BGO crystals on their energy threshold and resolution. Our results indicate an energy threshold reduced by $\sim$22\% for the Z-crystal A, $\sim$25\% for the Z-crystal B and $\sim$14\% for the Z-crystal C, compared with the X-crystal. We also obtain an average improvement in the energy resolution of $\sim$8\% for Z-crystal A and Z-crystal B and  $\sim$5\% for Z-crystal C. Future work will build on these results as we continue benchmarking simulations against new calibration data, in case additional measurements become available. Additionally, it will be essential to accurately model the temperature dependence of the BGO response (particularly the BGO light yield and scintillation timescale) and SiPM gain, and to update the simulation benchmarking accordingly. For the characterization of the optical light collection efficiency and energy resolution, we plan to launch further simulation campaigns to gather more statistics and extend the analysis to other BGO crystals, including the bottom crystals. This ongoing effort represents a fundamental step toward the development of a dedicated DEE for the ACS, ensuring accurate modeling of its real-world behavior and improving the simulations.

\bmhead{Acknowledgements} The Compton Spectrometer and Imager is a NASA Explorer project led by the University of California, Berkeley with funding from NASA under contract 80GSFC21C0059. This work is also supported in part by the Centre National d’Etudes Spatiales (CNES).

\bmhead{Funding} A.C. and V.F. are funded by Italian Research Center on High Performance Computing Big Data and Quantum Computing (ICSC), project funded by European Union - NextGenerationEU - and National Recovery and Resilience Plan (NRRP) - Mission 4 Component 2 within the activities of Spoke 2 (Fundamental Research and Space Economy), (CN 00000013 - CUP C53C22000350006). V.F. and A.B. are funded by ASI under contract 2024-11-HH.0. J.A.T., A.Z., P.P., L.M. and E.W. are funded by NASA under contract 80GSFC21C0059.

\begin{appendices}

\section{Experimental source spectra}\label{secA1}
\begin{figure}[h!]
\centering
\includegraphics[width=0.33\textwidth]{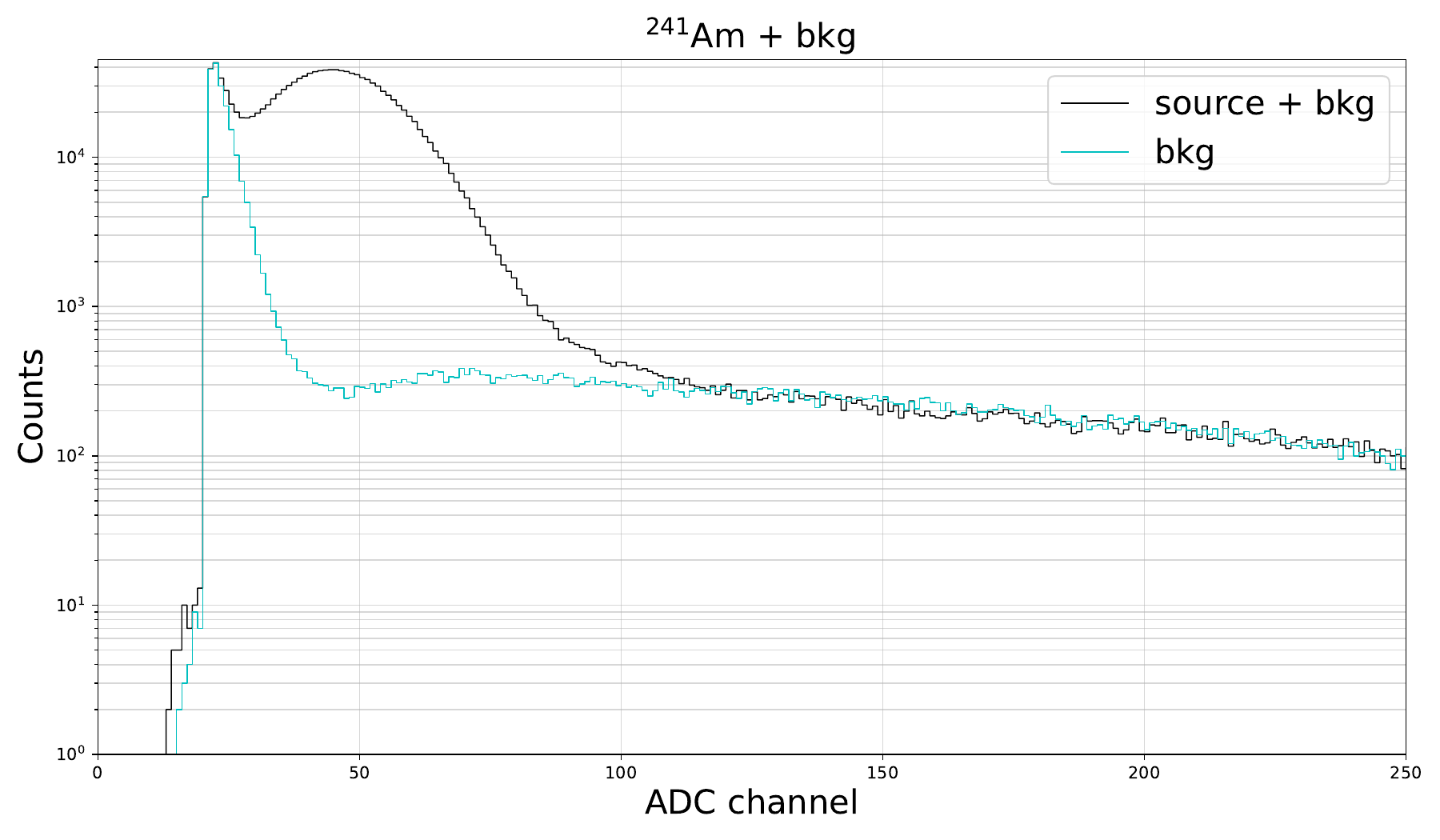}\includegraphics[width=0.33\textwidth]{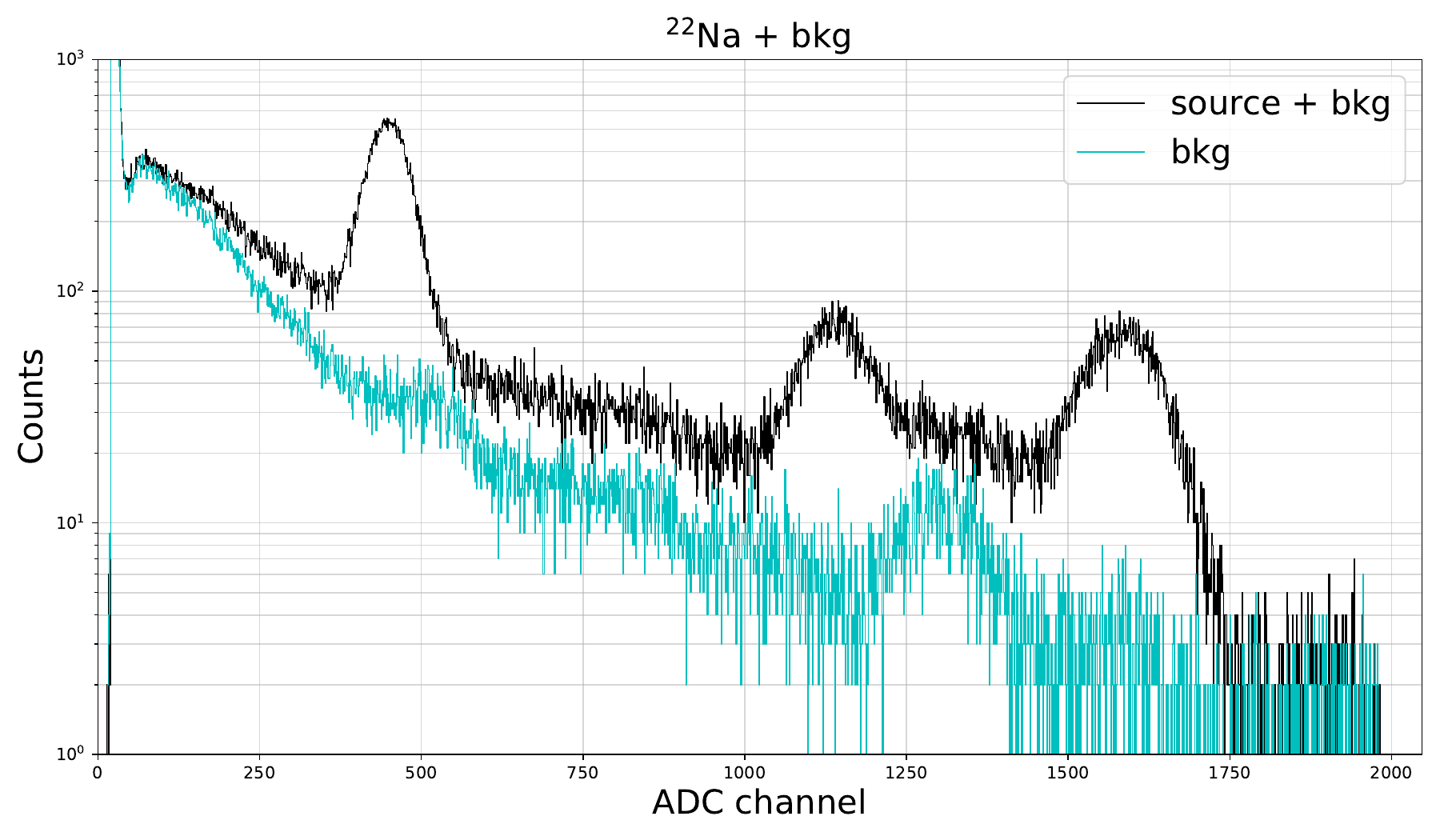}\includegraphics[width=0.33\textwidth]{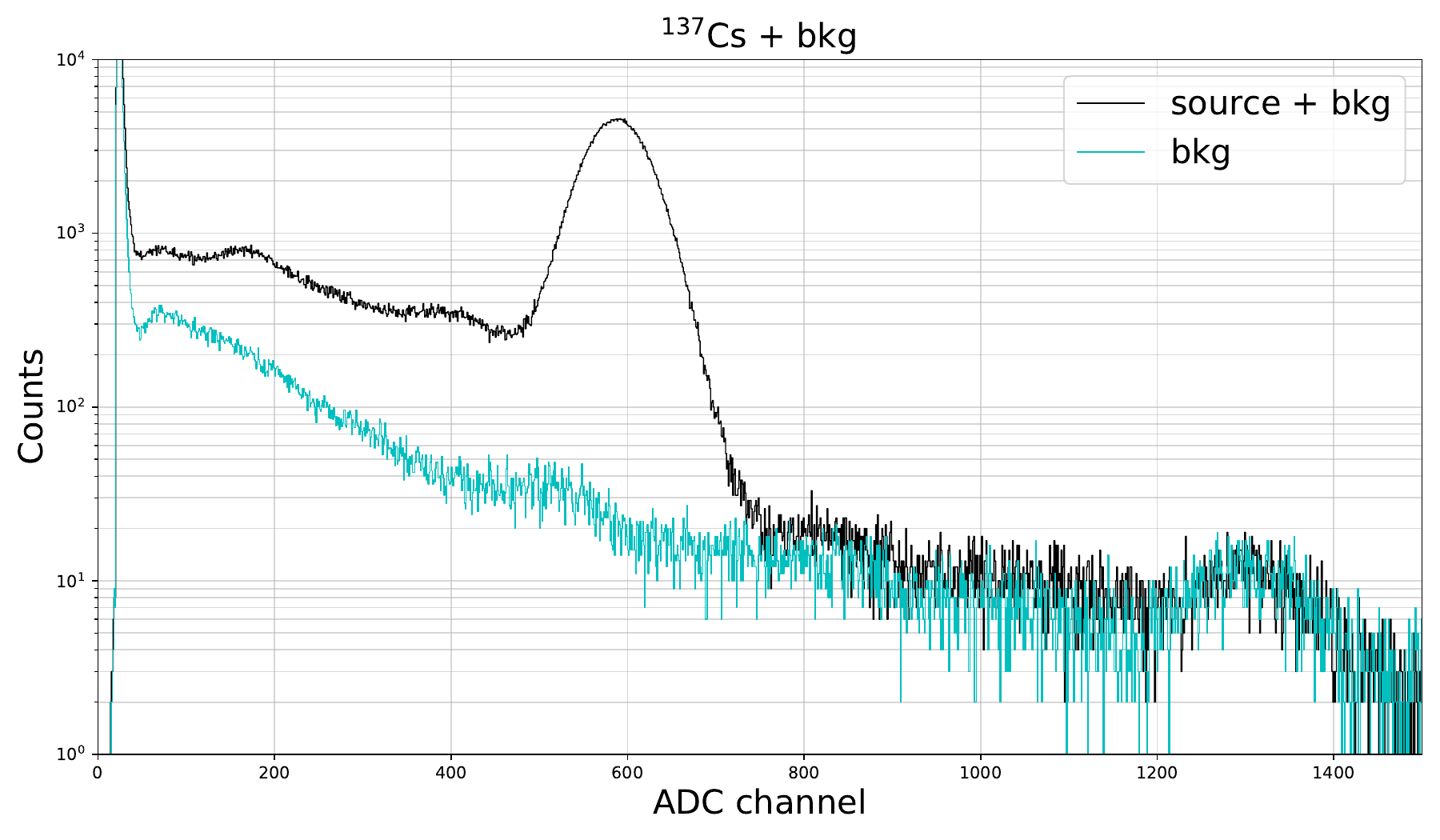}
\caption{Source + background spectra for the ACS X-wall measured at NRL, for each radioactive source}\label{fig:spectra_NRL2}
\end{figure}

\begin{figure}[h!]
\centering
\includegraphics[width=0.33\textwidth]{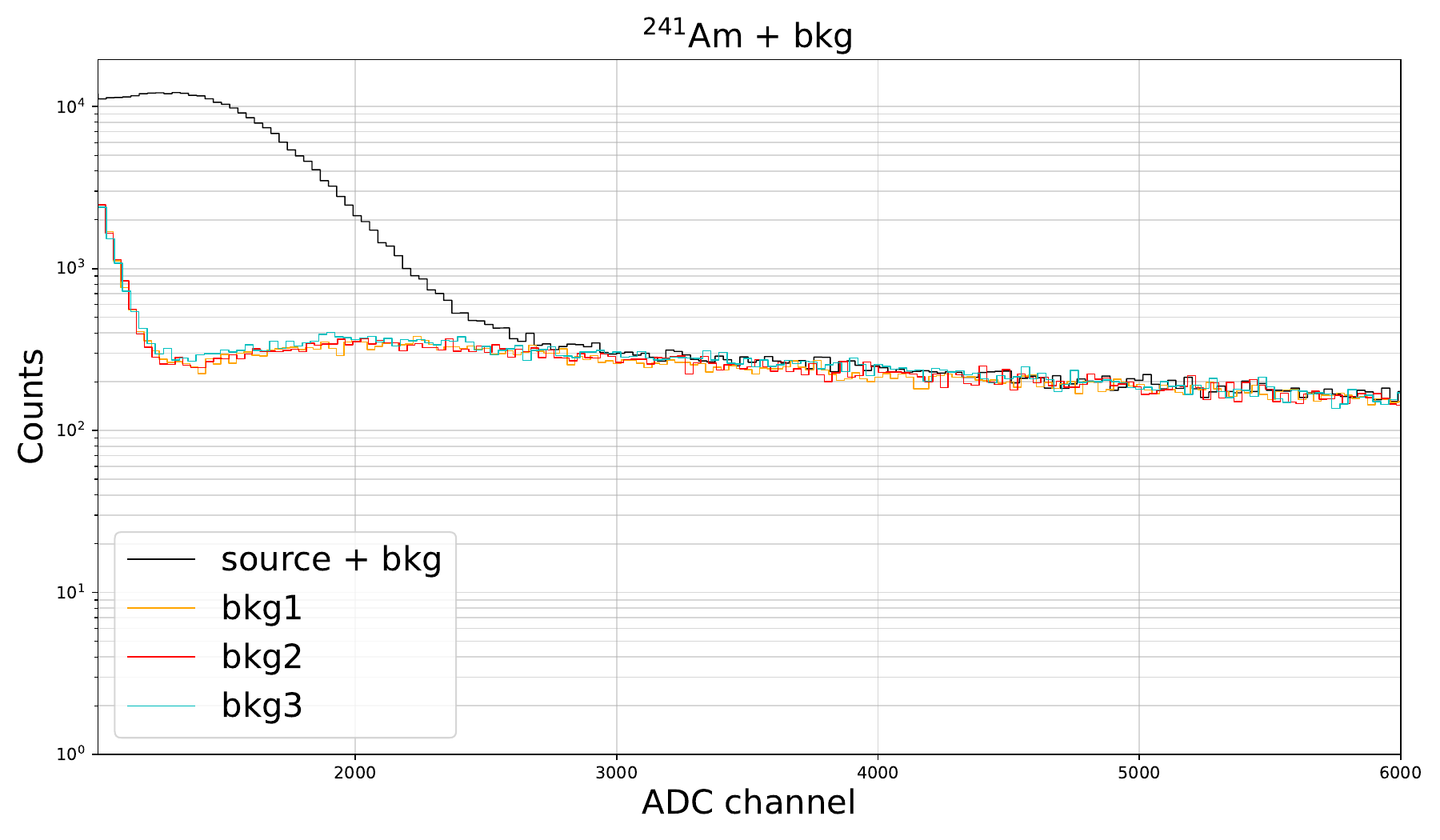}\includegraphics[width=0.33\textwidth]{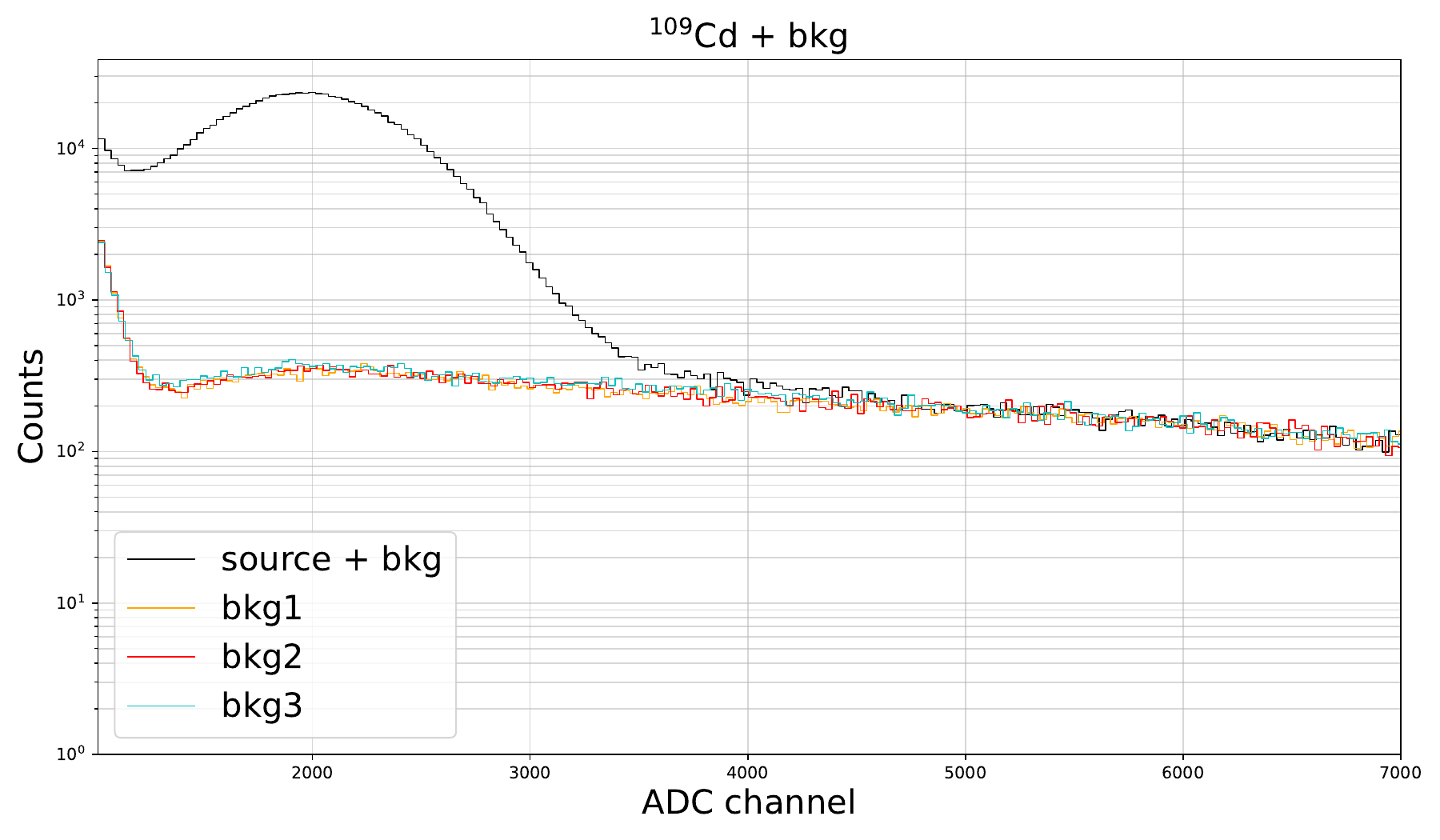}\includegraphics[width=0.33\textwidth]{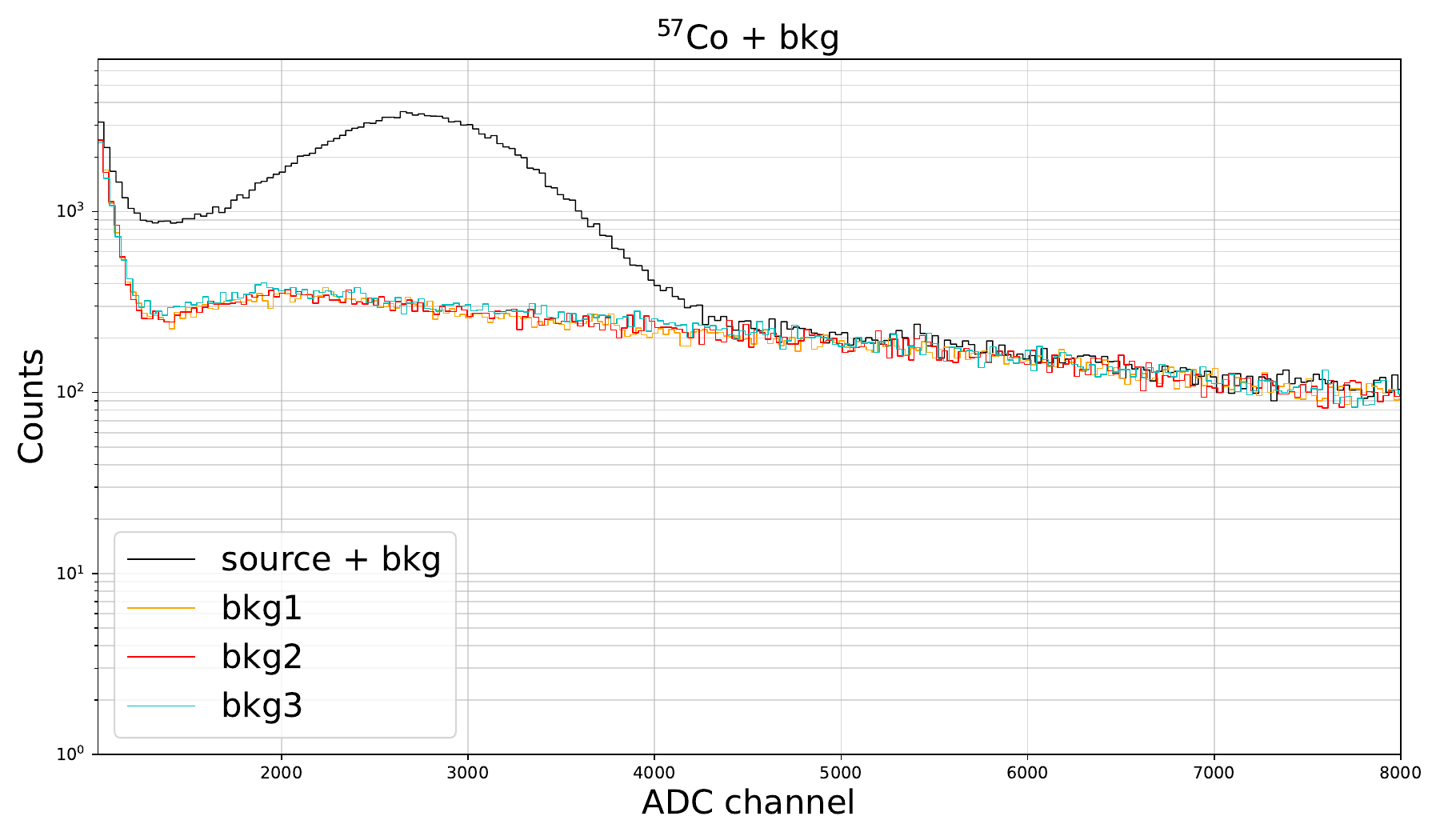}\\\includegraphics[width=0.33\textwidth]{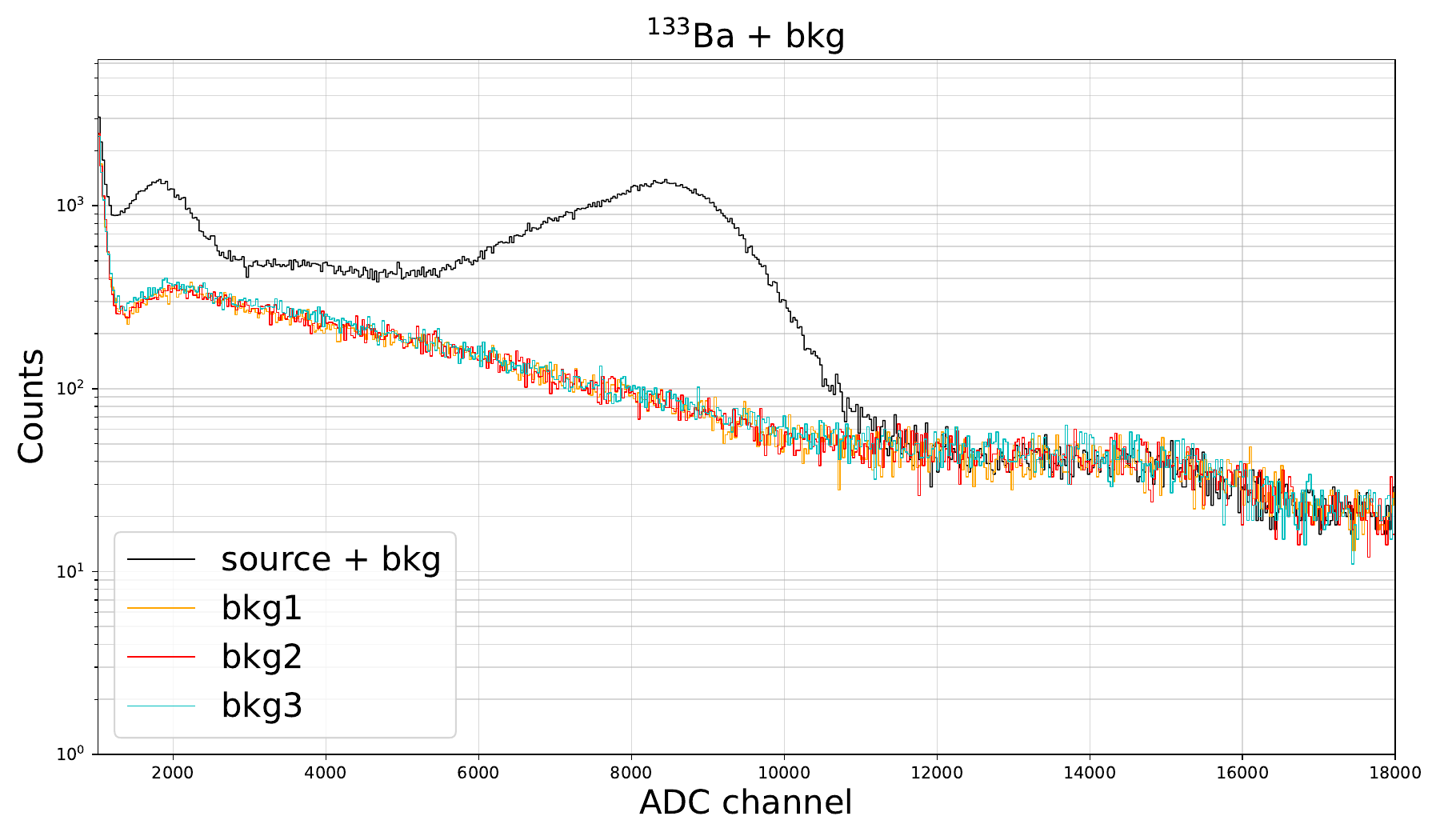}\includegraphics[width=0.33\textwidth]{plots/Na22_bkg_SSL.pdf}\includegraphics[width=0.33\textwidth]{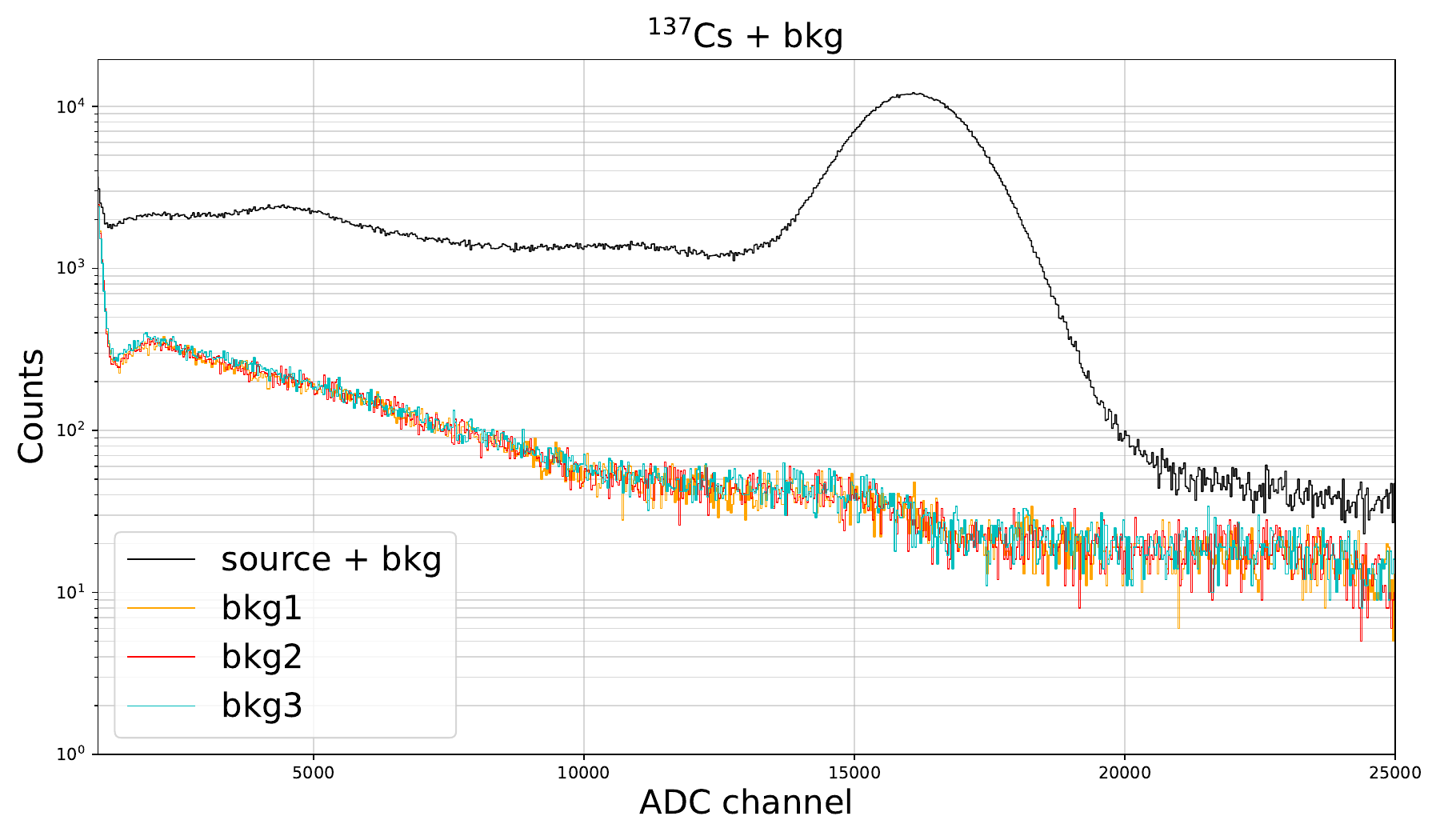}\\\includegraphics[width=0.33\textwidth]{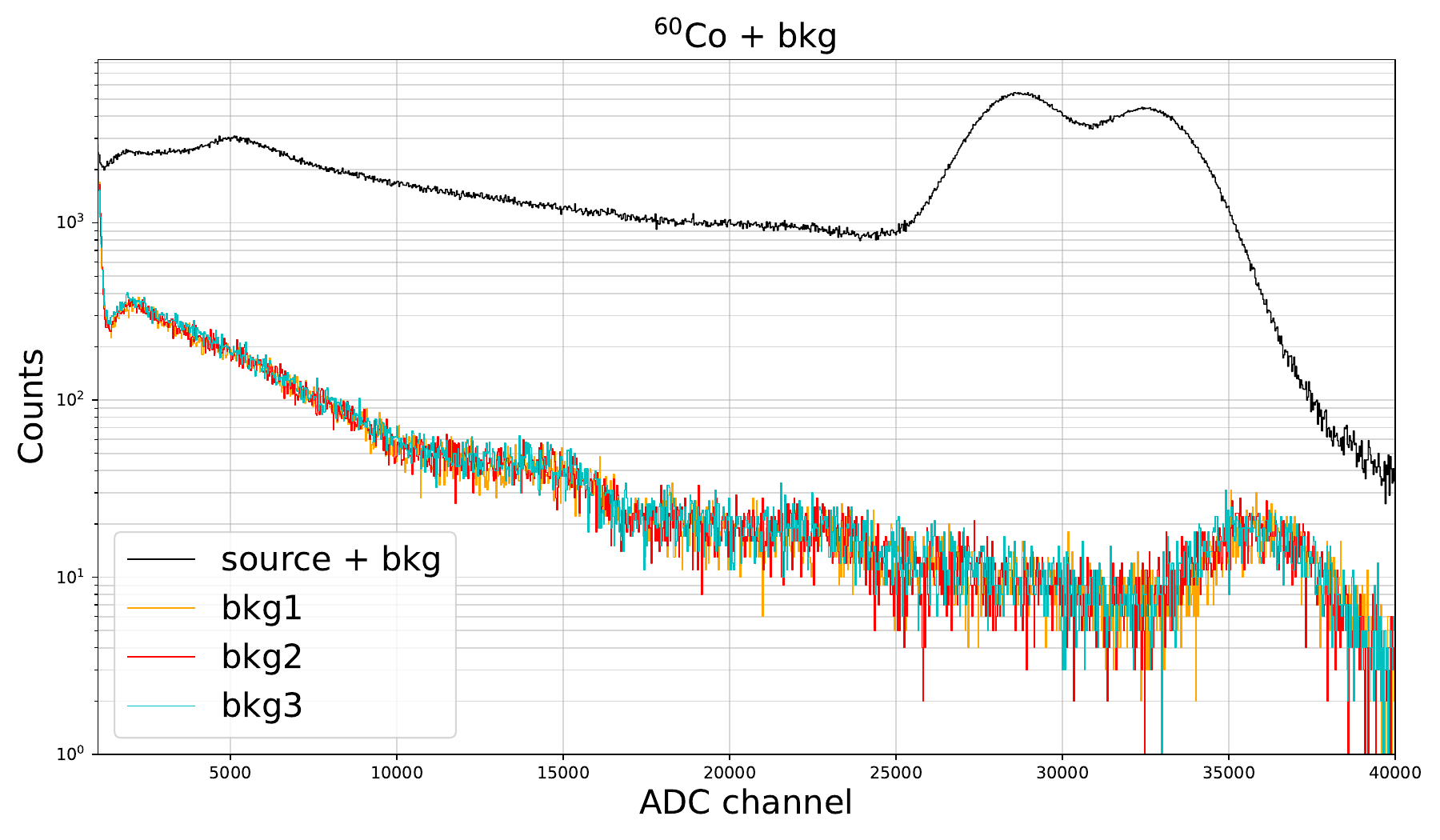}\includegraphics[width=0.33\textwidth]{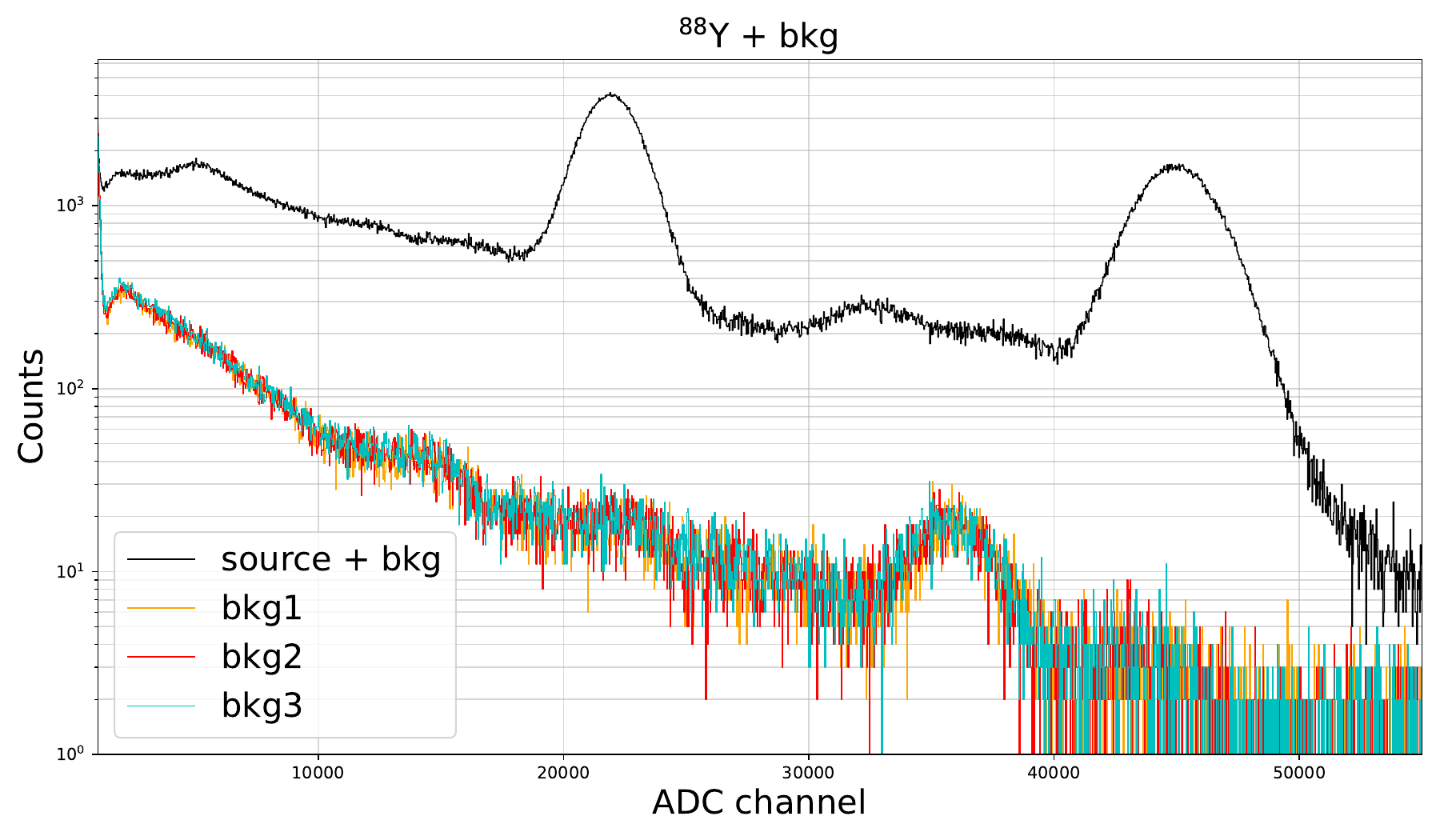}
\caption{Source + background spectra for the ACS X-wall measured at SSL, for each radioactive source}\label{fig:spectra_SSL1}
\end{figure}

We show in Figs. \ref{fig:spectra_NRL2} and \ref{fig:spectra_SSL1} the experimental spectra for each uncollimated source with the corresponding background, for both NRL and SSL experiments.

\section{BGO absorption length and systematic error variations}\label{secA2}
\begin{figure}[h!]
\centering
\includegraphics[width=0.5\textwidth]{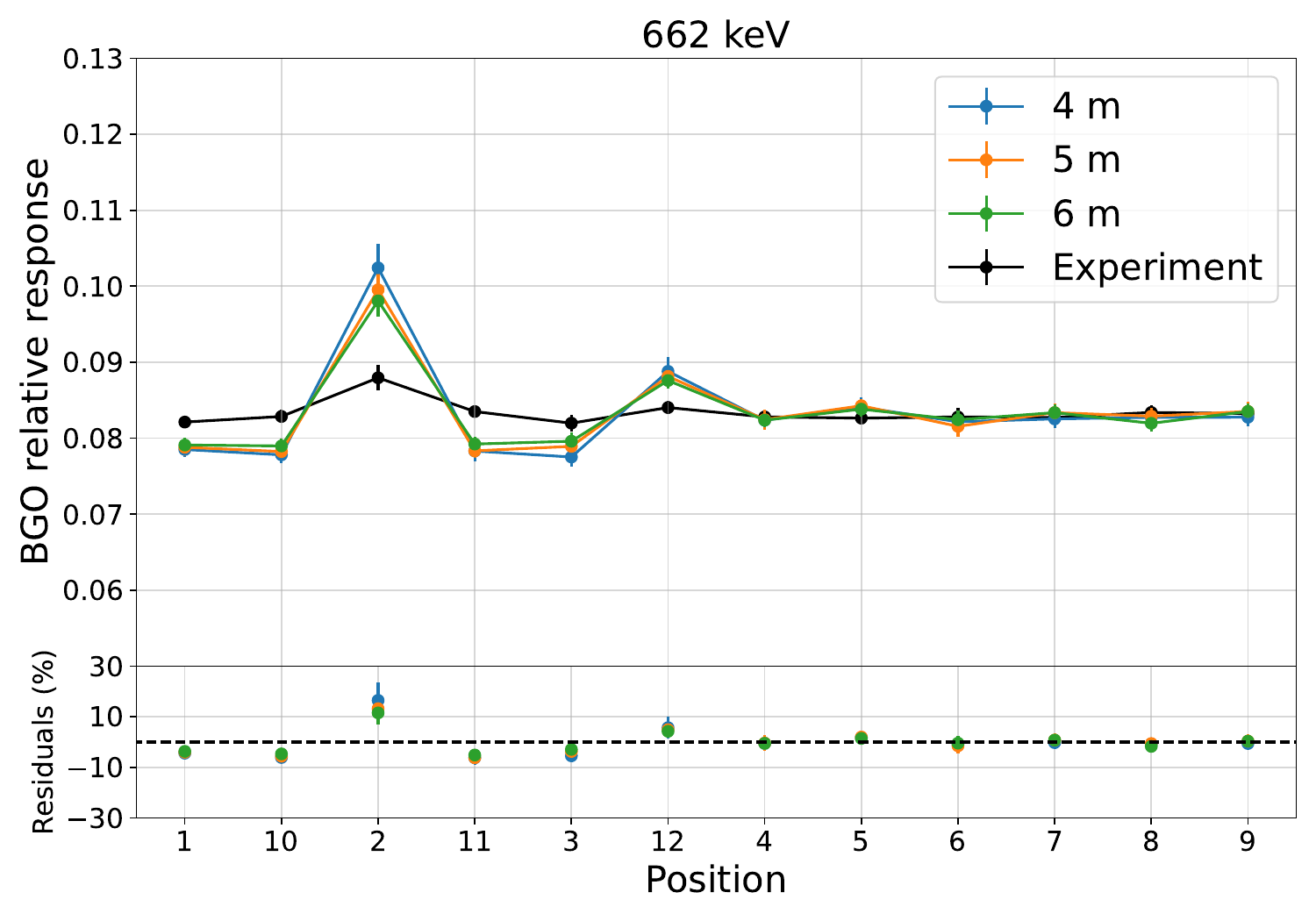}\includegraphics[width=0.5\textwidth]{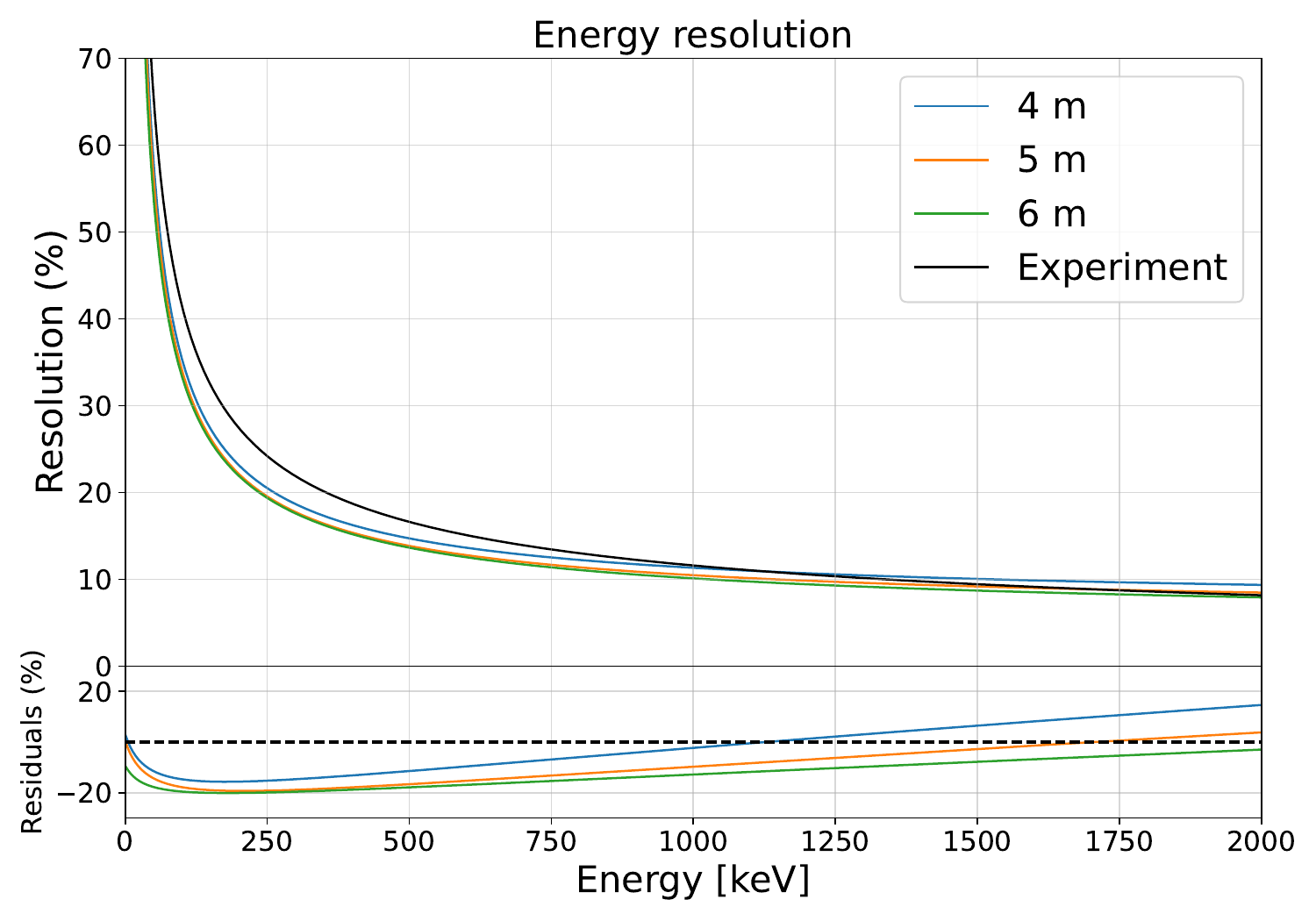}
\caption{On the left, simulated BGO relative response at 662 keV for different BGO absorption lengths (4, 5, 6 m), compared with the experimental one. On the right, simulated best-fit model of the ACS energy resolution for different BGO absorption lengths (4, 5, 6 m), compared with the experimental best-fit model}\label{fig:test_different_absl}
\end{figure}

\begin{figure}[h!]
\centering
\includegraphics[width=0.5\textwidth]{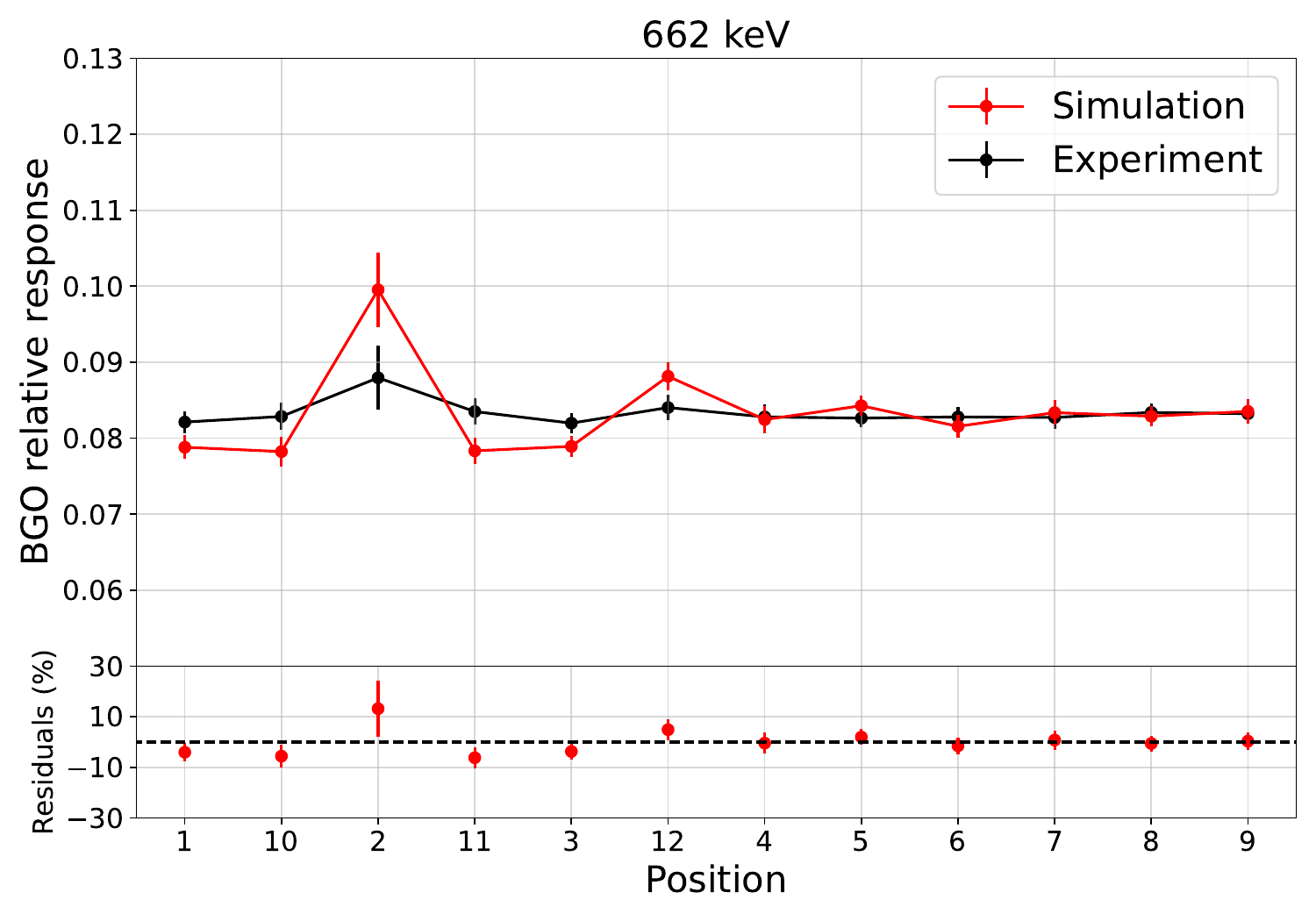}\includegraphics[width=0.5\textwidth]{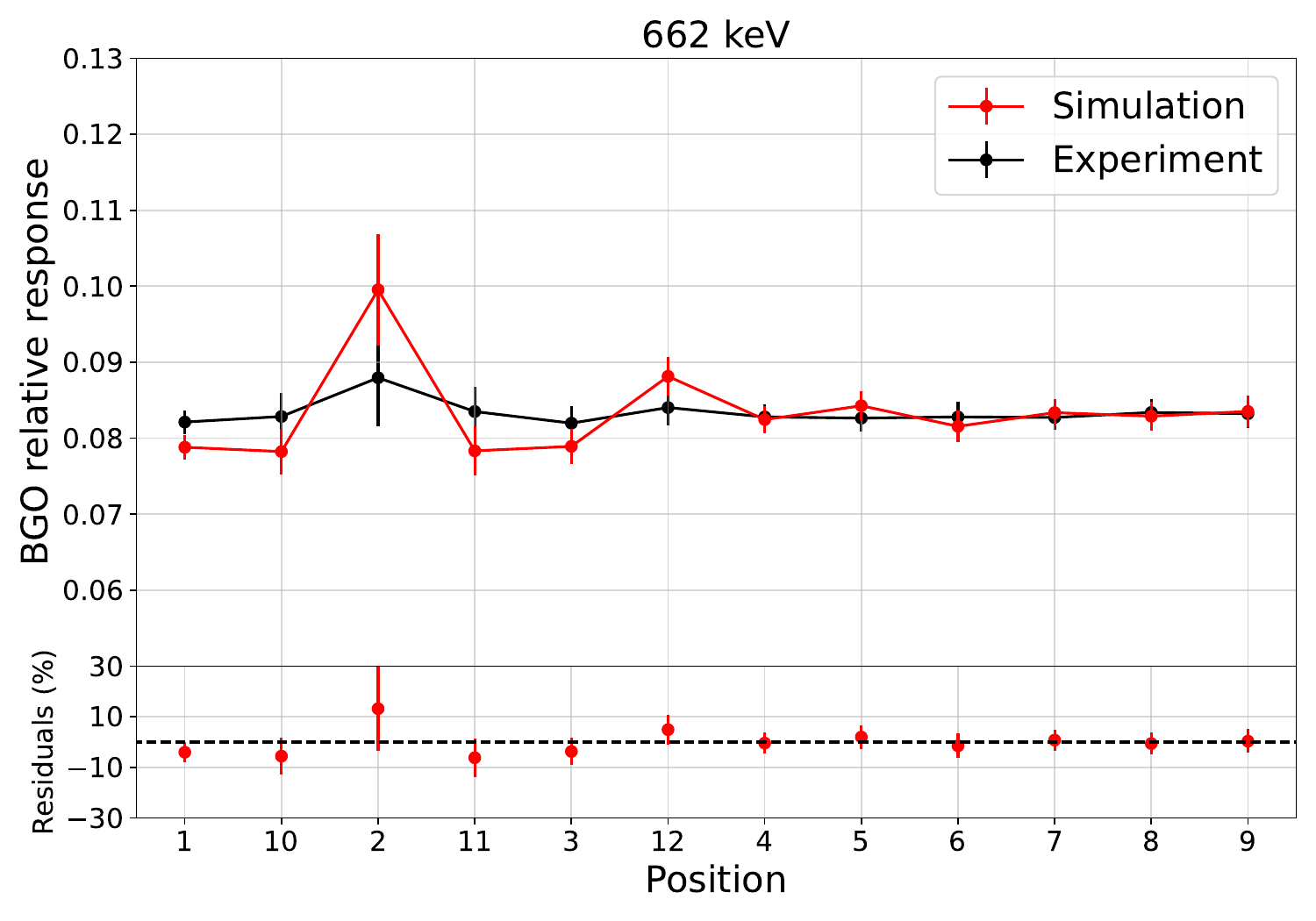}
\caption{Comparison between the experimental and simulated relative response, using a systematic uncertainty on the positions of 5 mm (left) and 8 mm (right)}\label{fig:test_different_sys}
\end{figure}

In Fig. \ref{fig:test_different_absl} we show the simulated relative response and energy resolution for three different values of the BGO absorption length (4, 5, 6 m), compared with the experimental measurements. In Fig. \ref{fig:test_different_sys} we show the comparison between the experimental and simulated relative response, using a systematic uncertainty on the positions of 5 mm and 8 mm.

\end{appendices}

\bibliography{bibliography}% common bib file

\end{document}